\documentclass[12pt, preprint]{aastex}

\usepackage{graphicx}
\usepackage{natbib}
\bibpunct{(}{)}{;}{a}{}{,}
\shorttitle{Light Neutron-Capture Elements in PNe III}
\shortauthors{Sterling et al.}

\begin{document}

\title{The Abundances of Light Neutron-Capture Elements in Planetary Nebulae -- III.\ The Impact of New Atomic Data on Nebular Selenium and Krypton Abundance Determinations}

\author{N.\ C.\ Sterling\altaffilmark{1}, R.\ L.\ Porter\altaffilmark{2}, \& Harriet L.\ Dinerstein\altaffilmark{3}}

\altaffiltext{1}{Department of Physics, University of West Georgia, 1601 Maple Street, Carrollton, GA 30118; nsterlin@westga.edu}
\altaffiltext{2}{Department of Physics and Astronomy and Center for Simulational Physics, University of Georgia, Athens, GA 30602; ryanlporter@gmail.com}
\altaffiltext{3}{Department of Astronomy, University of Texas, 2515 Speedway, C1400, Austin, TX 78712-1205; harriet@astro.as.utexas.edu}

\begin{abstract}

The detection of neutron(\emph{n})-capture elements in several planetary nebulae (PNe) has provided a new means of investigating \emph{s}-process nucleosynthesis in low-mass stars.  However, a lack of atomic data has inhibited accurate trans-iron element abundance determinations in astrophysical nebulae.  Recently, photoionization and recombination data were determined for Se and Kr, the two most widely detected \emph{n}-capture elements in nebular spectra.  We have incorporated these new data into the photoionization code Cloudy.  To test the atomic data, numerical models were computed for 15 PNe that exhibit emission lines from multiple Kr ions.  We found systematic discrepancies between the predicted and observed emission lines that are most likely caused by inaccurate photoionization and recombination data.  These discrepancies were removed by adjusting the Kr$^+$--Kr$^{3+}$ photoionization cross sections within their cited uncertainties and the dielectronic recombination rate coefficients by slightly larger amounts.  From grids of models spanning the physical conditions encountered in PNe, we derive new, broadly applicable ionization correction factor (ICF) formulae for calculating Se and Kr elemental abundances.  The ICFs were applied to our previous survey of near-infrared $[$\ion{Kr}{3}$]$ and $[$\ion{Se}{4}$]$ emission lines in 120 PNe.  The revised Se and Kr abundances are 0.1--0.3 dex lower than former estimates, with average values of $[$Se/(O,~Ar)$]=0.12\pm$0.27 and $[$Kr/(O,~Ar)$]=0.82\pm$0.29, but correlations previously found between their abundances and other nebular and stellar properties are unaffected. We also find a tendency for high-velocity PNe that can be associated with the Galactic thick disk to exhibit larger \emph{s}-process enrichments than low-velocity PNe belonging to the thin disk population.

\end{abstract}


\section{INTRODUCTION} \label{intro}

Despite their low cosmic abundances, neutron(\emph{n})-capture elements (atomic number $Z>30$) can provide significant insight into the nucleosynthetic histories of stellar populations, mixing processes and the internal structure of asymptotic giant branch (AGB) stars, and the chemical evolution of the Universe \citep[e.g.,][]{busso01, herwig05, sneden08, karakas09, simon10, mcwilliam13}.  However, much remains unknown about the production of \emph{n}-capture elements \citep{busso99, sneden08, thielemann11, karakas14, trippella14}, complicating the interpretation of observations of these species.

Our understanding of \emph{n}-capture nucleosynthesis has largely been constructed from stellar abundance patterns \citep[][and references therein]{merrill52, smith90, busso99, travaglio04, sneden08, roederer10}.  In recent years, \emph{n}-capture element emission lines have been detected in a growing number of astrophysical nebulae, including planetary nebulae \citep[PNe;][]{liu04a, zhang05, likkel06, sharpee07, sterling07, sterling08, sterling09, otsuka10, fang11, otsuka11, garcia-rojas12, otsuka13}, \ion{H}{2} regions \citep{aspin94, lumsden96, baldwin00, puxley00, okumura01, blum08, roman-lopes09}, and the interstellar medium of other galaxies \citep{vanzi08}.  The large number of  detections demonstrates the potential of nebular spectroscopy for investigating the abundances and production of \emph{n}-capture elements.  

\defcitealias{sterling08}{Paper~II}

Trans-iron element abundances in PNe are of particular interest, since these nuclei can be produced via slow \emph{n}-capture nucleosynthesis (the \emph{s}-process) in PN progenitor stars.  During the thermally-pulsing AGB evolutionary stage, the \emph{s}-process can occur in the intershell region between the H- and He-burning shells.  Free neutrons are released predominantly by $\alpha$-captures onto $^{13}$C (or $^{22}$Ne in AGB stars with mass $>4$--5~M$_{\odot}$).  Iron-peak nuclei act as seeds from which \emph{s}-process nuclei grow, alternately capturing neutrons and undergoing $\beta$-decays to transform into nuclei of heavier elements \citep{busso99, kappeler11}.  Convective dredge-up events late in the AGB phase transport the \emph{s}-process enriched material to the stellar envelope along with carbon \citep{busso99, herwig05, karakas14}.  Stellar winds and PN ejection supply the surrounding interstellar medium with enriched material, which is eventually incorporated into new generations of stars.

Nebular spectroscopy reveals unique information about \emph{s}-process nucleosynthesis, providing access to \emph{n}-capture elements that are difficult or impossible to detect in AGB stars.  Elemental abundances in PNe also reveal chemical enrichments after the cessation of nucleosynthesis and dredge-up, which is ongoing in thermally-pulsing AGB stars.  In addition, stars that are challenging to study during the AGB phase due to dust obscuration or high mass-loss rates \citep[e.g.,][]{ventura09, dellagli15} produce readily observed PNe \citep[][hereafter Paper~II]{sharpee07, sterling08}.

Selenium and krypton are the two most widely detected trans-iron elements in PNe and other nebulae, owing to their relatively strong near-infrared fine-structure transitions $[$\ion{Kr}{3}$]$~2.199 and $[$\ion{Se}{4}$]$~2.287~$\mu$m first identified by \citet{dinerstein01}.  In \citetalias{sterling08}, Se and Kr abundances were derived in over 80 PNe from observations of these $K$~band transitions, providing the first large-scale investigation of \emph{s}-process enrichments in PNe.

However, abundance determinations of Se and Kr --- and indeed all \emph{n}-capture elements --- in astrophysical nebulae are plagued by uncertainties.  The most important of these uncertainties stems from the absence of reliable atomic data.  Transition probabilities \citep{biemont86b, biemont86a, biemont87, biemont88, biemont95} and effective collision strengths for electron-impact excitation \citep{schoning97, sb98} have been determined for some \emph{n}-capture element ions, allowing ionic abundances to be derived.  However, atomic data needed to reliably correct for the abundances of unobserved ionization stages are unknown.  Such ``ionization correction factors,'' or ICFs, are most reliably derived via numerical simulations of astrophysical nebulae \citep{stasinska78, kb94, kwitter01, rodriguez05, delgado-inglada14}.  However, the simulations are only as accurate as the underlying atomic physics that they include.  Specifically, ionization equilibrium solutions in photoionized nebulae rely on accurate photoionization (PI) cross sections and rate coefficients for radiative recombination (RR), dielectronic recombination (DR), and charge transfer (CT).  Until recently, these data were unknown for nearly all \emph{n}-capture element ions.

\defcitealias{sterling07}{Paper I}

In the absence of these data, \citet[][hereafter Paper~I]{sterling07} used approximations for the necessary atomic parameters to derive ICF prescriptions for Se and Kr.  They utilized hydrogenic PI cross sections and RR rate coefficients self-consistently derived from the PI data, CT rate coefficients of Cu and Zn ions of the same electronic charge, and DR rate coefficients equal to the average rate coefficients of C, N, and O ions of the same charge.  The Kr atomic data was adjusted empirically in order to reproduce emission line strengths in PNe in which multiple Kr ions had been detected.  Monte~Carlo simulations were run to estimate the sensitivity of the derived Se and Kr abundances to atomic data uncertainties.  Those simulations indicated that the poorly-known atomic data can lead to errors in derived Se and Kr abundances by a factor of two or possibly more, if the approximate data are substantially far from the actual values.

\citet{sterling11b} and \citet{sterling11c} recently computed PI cross sections and RR and DR rate coefficients for the first six ions of Se and Kr, to address the lack of reliable atomic data for these elements.  The PI cross sections were benchmarked against experimental measurements \citep{lu06a, lu06b, lu_thesis, bizau11, sterling11a, esteves11, esteves12}, leading to improved accuracy for some ions \citep{sterling11c}.  In addition, \citet{sterling11d} computed CT rate coefficients for low-charge ions of several \emph{n}-capture elements detected in astrophysical nebulae.  Uncertainties were estimated in all of these calculations, enabling a quantitative analysis of the sensitivity of abundance determinations to atomic data uncertainties.  We will present the results of that investigation in a forthcoming paper.

To apply the new atomic data to nebular Se and Kr abundance determinations, we have added the new PI, RR, DR, and CT data to the widely-used photoionization code Cloudy \citep{ferland13}.  The purpose of the present study is twofold: we test the new atomic data against the observed ionization balance in individual PNe, and derive new ICF formulae to account for the abundances of unobserved Se and Kr ions in ionized nebulae.

To address the first goal, we model individual PNe in which multiple Kr ions have been detected and compare the predicted and observed emission line intensities of the detected ions.  Unfortunately the Se atomic data cannot be tested in a similar manner as only one ion, $[$\ion{Se}{4}$]$, has been unambiguously detected in PNe \citep[with the possible exception of Hb~12;][]{sterling13}.  PNe are valuable laboratories for evaluating the accuracy of atomic data, owing to their high surface brightness that makes it possible to detect emission lines with a wide range of intensities.  Indeed, there is a long history of testing new atomic data determinations against observed PN spectra \citep[e.g.,][]{drake72, garstang78, dennefeld83, keenan87, keenan92, biemont95, liu95, kastner96, schoning97, kisielius98, vanhoof00, chen00, wangliu04, bautista09, mcnabb13}.

In order to derive new analytical ICFs for Se and Kr, we computed large grids of models, spanning the range of physical conditions found in most PNe.  Following the methodology of \citetalias{sterling07}, we searched for correlations between the fractional ionic abundances of observed Se and Kr ions and those of routinely detected light elements.  We fitted the correlations with functions to derive ICF prescriptions that can be applied to optical and near-infrared observations of Se and Kr.  Using the new ICFs, we derive more accurate Se and Kr abundances for the PN survey of \citetalias{sterling08}.

The structure of this paper is as follows.  In \S2, we briefly describe our alterations of Cloudy to include the new Se and Kr atomic data.  In \S3, we present detailed models of 15 PNe in which multiple Kr ions have been detected, in order to test the new atomic data's ability to reproduce observed line intensities.  A large grid of models encompassing the range of physical conditions encountered in most PNe is presented in \S4, and we derive new formulae for Se and Kr ICFs from those results.  In \S5, we use the newly-derived ICFs to revise the Se and Kr abundance determinations of \citetalias{sterling08} in 120~PNe, before providing concluding remarks in \S6.

\section{THE INCORPORATION OF SELENIUM AND KRYPTON INTO CLOUDY} \label{cloudy}

The modifications described in this section are relative to the C13 release of Cloudy \citep{ferland13}.  We emphasize the updated atomic database for Se and Kr, but all elements from Ga through Kr were added to the code.  Of the elements we added to Cloudy, only Se and Kr were activated in the models described in the subsequent sections.

Atomic weights are taken from the same source as for lighter elements \citep{coplen01}.  The default abundances are the solar values from \citet{asplund09}, with the exception of interstellar medium abundances and dust depletion factors \citep{savage96, welty99}.  Data for the hydrogenic and He-like isoelectronic sequences were recomputed to extend to $Z=36$.  These calculations are straightforward applications of the analytical treatments described in \citet{porter05}, \citet{porter07}, and \citet{luridiana09}.

We incorporated the direct valence shell PI cross sections of \citet{sterling11b} and \citet{sterling11c} for the neutral and first five ionization stages of Se and Kr.  Following the treatment of other species in Cloudy, resonant photoionization was not included due to uncertainties in the calculated resonance energies \citep{verner96b}.  For all other Se and Kr ions and those of Ga, Ge, As, and Br, we adopt hydrogenic PI cross sections as described in \citetalias{sterling07}.

RR and low-temperature DR rate coefficients for Se$^0$--Se$^{5+}$ and Kr$^0$--Kr$^{5+}$ are taken from \citet{sterling11b} and \citet{sterling11c}.  RR rate coefficients for all other $Z=31$--36 ions are extrapolations of the fits of \citet{verner96a} for $Z\leq 30$ species.  The low-$T$ DR rate coefficients for other $Z>30$ species are treated as described in \citetalias{sterling07}.  High-$T$ DR rate coefficients are taken from \citet{mazzitelli02} for Ga and Ge and from \citet{fournier00} for Kr.  The As, Se, and Br high-$T$ DR rate coefficients are assumed to be equal to the rate coefficient of the Kr ion with the same charge.  While that approximation is crude, uncertainties in the high-$T$ DR rate coefficients are unlikely to be important in models of photoionized nebulae, in which low-$T$ DR is dominant.

We utilize the Landau-Zener and Demkov CT rate coefficient calculations of \citet{sterling11d} for collisions between low-charge (up to five times ionized) Ge, Se, Br, and Kr ions with neutral hydrogen.  CT between H$^+$ and neutral Ge, Se, Br, and Kr are also included.  For the first four ions of Ga and As, the CT rate coefficients are taken to be the average of those for Cu and Ni ions of the same charge.  For all other ionization states of $Z=31$--36 species, CT is treated analytically as described in \citet{ferland97} and \citetalias{sterling07}.  We do not include CT reactions with He, though in some cases these could affect the ionization balance.

We assume that the Ga--Kr electron-impact ionization rate coefficients are equal to the corresponding ion of Zn (or H-like Zn for higher ionization states).  Like high-$T$ DR, collisional ionization does not play a significant role in photoionized plasmas such as PNe and H~II regions.  In addition, Auger ionization is disabled for all six elements.

Ga--Kr have very low cosmic abundances, and should not affect the ionization balance or spectra of lighter elements.  We computed models of the individual PNe in \S\ref{ind_models} with the unaltered version of Cloudy to verify that the light element spectra are unaffected.  We tested the implementation of Ga--Kr in the code by running the ``smoke test'' routine, which showed that the ionization distributions of these species are consistent with those for $Z\leq 30$ elements.  Finally, we ran the full test suite of routines to ensure that our modifications of Cloudy did not affect the code's functionality for lighter species.

\section{MODELS OF INDIVIDUAL NEBULAE} \label{ind_models}

Our first objective is to test whether simulations using the new atomic data reproduce the observed ionization balance of Kr in PNe.  Specifically, we model the intensities of Kr emission lines and compare them to the observed intensities of PNe in which multiple Kr ions have been detected.

We have selected 15 PNe that exhibit emission lines from multiple Kr ions in their optical and near-infrared spectra (Table~\ref{pn_sample}).  Ten of these objects were modeled in \citetalias{sterling07}.  We draw the other objects from optical spectra published since \citetalias{sterling07} \citep{sharpee07, garcia-rojas12}, with the exception of NGC~5315 \citep{peimbert04}.  Intensities for the $K$~band emission lines from \citetalias{sterling08} were determined using the extinction coefficients listed in Table~\ref{pn_sample} and the extinction law adopted by the optical data reference.  We corrected for aperture effects (e.g., different slit widths) by forcing the \ion{H}{1}~Br$\gamma$/H$\beta$ ratio to equal the theoretical value \citep{storey95} for the nebula's temperature and density.  This line ratio is relatively insensitive to temperature and density in the regime of physical conditions found in PNe and H~II regions.

\begin{deluxetable}{lccc}
\tablecolumns{4}
\tabletypesize{\footnotesize}
\tablewidth{0pc} 
\tablecaption{Observational Data References} 
\tablehead{
\colhead{Object} & \colhead{Optical} & \colhead{} & \colhead{Other} \\
\colhead{Name} & \colhead{Data Ref.} & \colhead{$c$(H$\beta$)} & \colhead{Ref.\tablenotemark{a}}}
\startdata
IC 418 & \citet{sharpee03} & 0.34 & \citet{pottasch04} \\
IC 2501 & \citet{sharpee07} & 0.55 & \citet{goharji84, milingo02} \\
IC 4191 & \citet{sharpee07} & 0.77 & \citet{pottasch05a} \\
IC 5117 & \citet{hyung01} & 1.40 & \nodata \\
NGC 2440 & \citet{sharpee07} & 0.55 & \citet{bernard-salas02} \\
NGC 5315 & \citet{peimbert04} & 0.74 & \citet{pottasch02} \\
NGC 6369 & \citet{garcia-rojas12} & 1.93 & \citet{pottasch08} \\
NGC 6572 & \citet{liu04b, liu04a} & 0.48 & \citet{hyung94a}\tablenotemark{b} \\
NGC 6741 & \citet{liu04b, liu04a} & 1.15 & \nodata \\
NGC 6790 & \citet{liu04b, liu04a} & 1.10 & \citet{aller96}\tablenotemark{b} \\
NGC 6826 & \citet{liu04b, liu04a} & 0.06 & \nodata \\
NGC 6884 & \citet{liu04b, liu04a} & 1.00 & \nodata \\
NGC 6886 & \citet{hyung95} & 0.90 & \citet{pottasch05b} \\
NGC 7027 & \citet{zhang05} & 1.37 & \nodata \\
NGC 7662 & \citet{liu04a, liu04b} & 0.18 & \nodata \\
\tableline
\enddata
\label{pn_sample}
\tablecomments{Observational data references and extinction coefficients $c$(H$\beta$) are given.  Near-infrared data are taken from \citetalias{sterling08}, except for IC~5117 \citep{rudy01}, NGC~7027, and NGC~7662 \citep{geballe91}.}
\tablenotetext{a}{Reference for abundances, UV, and mid-IR data if not from optical data reference.}
\tablenotetext{b}{$[$\ion{Kr}{4}$]$~$\lambda$5867.7 is taken from this reference and scaled to $[$\ion{Kr}{4}$]$~$\lambda$5346.0 of \citet{liu04b, liu04a}.  All other abundance, optical, and IR data (except the $K$ band) are from \citet{liu04b, liu04a}.}
\end{deluxetable}

We modeled these 15 PNe with our modified version of Cloudy C13 \citep{ferland13}, adjusting the input parameters to best reproduce the observed optical and (when available) UV and IR spectra.  The methodology is described in detail in \citetalias{sterling07}, and here we provide only a brief synopsis.

Our primary goal in this exercise is to simulate the ionization balance of each nebula, and not necessarily to produce a model that is physically accurate in terms of other nebular and stellar properties.  Given this focus, we have made a number of simplifying assumptions in our models.  We assume that the nebulae are spherical and have uniform densities, with covering and filling factors of unity.  Furthermore, we do not incorporate dust physics, aside from the depletion of refractory elements.  This assumption likely affects the thermal balance and derived central star temperatures of the nebular models \citep[e.g.,][]{dopita00}.  However, we show in \S4 that including dust does not significantly affect the ionization balance of Se or Kr.  Despite these simplifications, the observed spectra are satisfactorily reproduced.  

Several input parameters must be specified in Cloudy models, including the central star temperature, gravity and luminosity; the inner and outer radii of the nebula; the nebular hydrogen density; and abundances of elements.  For the ten PNe modeled in \citetalias{sterling07}, we used those input parameters as the starting values for our new models.  The initial input parameters for the other objects were taken from \citet{sharpee07} for IC~2501, \citet{pottasch05a} for IC~4191, \citet{bernard-salas02} for NGC~2440, \citet{pottasch02} for NGC~5315, and \citet{pottasch08} for NGC~6369.  We obtained initial stellar temperature and gravity data for NGC~5315 and NGC~2440 from \citet{marcolino07} and \citet{heap87}, respectively.  Aside from stellar gravities and nebular inner radii, which were adjusted manually, all other modeled parameters were determined via the subplex optimization routine in Cloudy.  In some cases, parameters were varied manually after optimization to improve the fit to the observed spectra.

For the spectral energy distribution of the PN central stars, we utilized NLTE stellar atmospheres including metal line blanketing \citep{rauch03}.  Since these 15 PNe are Galactic disk objects, we assumed solar metallicities for the central stars.  The \cite{rauch03} grid of models spans the temperature range 50,000--190,000~K, and hence could not be used to model the low-ionization PN IC~418.  For that object, we interpolated on two stellar atmosphere models \citepalias{sterling07} computed by T.\ Rauch (2005, private communication) with $T_{\rm eff}$~=~30,000~K and 40,000~K and log~$g$~=~3.4.  Uncertainties in $T_{\rm eff}$ were estimated by manually adjusting the central star temperature in each model, until diagnostic ratios and/or intensities of high-ionization species were no longer reproduced to within $\sim$50--100\%.

The optimized input parameters to our models are given in Tables~\ref{params1} and \ref{params2}.  Comparisons of modeled and observed line diagnostics and intensities are given in Tables~\ref{lines1}, \ref{lines2}, \ref{lines3}, and \ref{lines4}, with the exception of Kr lines which will be discussed in the following subsection.  Line intensities are given on a normalized scale with $I(\mathrm{H}\beta )=100$.  For the majority of lines, the predicted intensities from the Cloudy models are within 25\% of the observed intensities.  Larger discrepancies exist for a handful of lines in some models, which are likely due to the simplifying assumptions noted above.

\begin{deluxetable}{lccccccccc}
\tablecolumns{10}
\tablewidth{0pc} 
\tabletypesize{\scriptsize}
\tablecaption{Model Input Parameters} 
\tablehead{
\colhead{Parameter} & \colhead{IC 418} & \colhead{IC 2501} & \colhead{IC 4191} & \colhead{IC 5117} & \colhead{NGC 2440} & \colhead{NGC 5315} & \colhead{NGC 6369} & \colhead{NGC 6572} & \colhead{Solar\tablenotemark{a}}}
\startdata
$T_{\rm eff}$ (10$^3$~K) & 38.8$\pm$3.0 & 50.3$\pm$8.0 & 118.8$\pm$10.0 & 125.3$\pm$10.0 & 188.4$\pm$30.0 & 51.0$\pm$7.0 & 62.0$\pm$10.0 & 80.8$\pm$7.0 & \nodata \\
log $g$ & 3.4 & 5.5 & 6.0 & 8.0 & 6.7 & 5.0 & 5.1 & 5.5 & \nodata \\
$L/L_{\odot}$ & 2330 & 3850 & 1680 & 2610 & 920 & 990 & 2620 & 1480 & \nodata \\
$n_{\rm H}$ (10$^3$ cm$^{-3}$) & 8.5 & 11.6 & 9.8 & 59.0 & 6.05 & 30.2 & 3.2 & 19.2 & \nodata \\
log($R_{\rm in}$/pc) & $-2.33$ & $-2.7$ & $-3.0$ & $-2.1$ & $-1.823$ & $-2.0$ &$-2.0$  & $-2.0$ & \nodata \\
log($R_{\rm out}$/pc) & $-1.13$ & $-1.405$ & $-1.524$ & $-1.68$ & $-1.324$ & $-1.234$ & $-1.18$ & $-1.44$ & \nodata \\
He/H & 0.105 & 0.104 & 0.112 & 0.094 & 0.124 & 0.129 & 0.101 & 0.104 & 0.085 \\
C/H & 8.67 & 8.88 & 8.78 & 8.62 & 8.52 & 8.39 & 8.20 & 8.75 & 8.43 \\
N/H & 7.92 & 8.02 & 8.22 & 7.83 & 8.65 & 8.43 & 8.03 & 7.91 & 7.83 \\
O/H & 8.58 & 8.68 & 9.09 & 8.39 & 8.64 & 8.78 & 8.65 & 8.43 & 8.69 \\
Ne/H & 7.90 & 7.97 & 8.40 & 7.70 & 7.93 & 8.14 & 7.95 & 7.70 & 7.93 \\
S/H & 6.63 & 6.85 & 7.24 & 6.77 & 6.59 & 7.16 & 6.84 & 6.44 & 7.12 \\
Cl/H & 4.29 & 5.08 & 5.35 & 5.11 & 5.05 & 5.33 & 4.94 & 4.87 &  5.50 \\
Ar/H & 6.13 & 6.60 & 6.46 & 6.10 & 6.45 & 6.43 & 6.24 & 6.06 & 6.40 \\
Se/H & \nodata & \nodata & \nodata & 3.52$\pm$0.05 & \nodata & \nodata & 3.66$\pm$0.14 & 3.01$\pm$0.03 & 3.34 \\
Kr/H & 3.93$\pm$0.23 & 3.12$\pm$0.13 & 3.22$\pm$0.30 & 3.74$\pm$0.10 & 3.44$\pm$0.10 & 3.42$\pm$0.28 & 3.94$\pm$0.18 & 3.51$\pm$0.15 & 3.25 \\
\tableline
\enddata
\label{params1}
\tablecomments{Input parameters for Cloudy models.  All models were run with Cloudy version C13, modified as described in the text.  The stellar effective temperature, gravity, and luminosity (relative to solar), nebular hydrogen density, and inner and outer nebular radii are given in the first six rows.  Elemental abundances follow, with He/H on a linear scale, and the other elements given as $12+$~log(X/H).}
\tablenotetext{a}{From \citet{asplund09}}
\end{deluxetable}
\clearpage

\begin{deluxetable}{lcccccccc}
\tablecolumns{9}
\tablewidth{0pc} 
\tabletypesize{\footnotesize}
\tablecaption{Model Input Parameters} 
\tablehead{
\colhead{Parameter} & \colhead{NGC 6741} & \colhead{NGC 6790} & \colhead{NGC 6826} & \colhead{NGC 6884} & \colhead{NGC 6886} & \colhead{NGC 7027} & \colhead{NGC 7662} & \colhead{Solar\tablenotemark{a}}}
\startdata
$T_{\rm eff}$ (10$^3$~K) & 144.4$\pm$20.0 & 96.7$\pm$10.0 & 50.1$\pm$5.0 & 114.0$\pm$10.0 & 164.6$\pm$20.0 & 162.7$\pm$20.0 & 126.4$\pm$10.0 & \nodata \\
log $g$ & 5.84 & 5.5 & 5.0 & 5.7 & 7.0 & 6.0 & 5.7 & \nodata \\
$L/L_{\odot}$ & 520 & 5800 & 12,400 & 990 & 780 & 2930 & 1550 & \nodata \\
$n_{\rm H}$ (10$^3$ cm$^{-3}$) & 5.5 & 38.5 & 1.6 & 7.0 & 6.1 & 55.0 & 3.2 & \nodata \\
log($R_{\rm in}$/pc) & $-2.24$ & $-2.3$ & $-1.60$ & $-2.7$ & $-2.79$ & $-2.05$ & $-1.60$ & \nodata \\
log($R_{\rm out}$/pc) & $-1.16$ & $-1.83$ & $-0.82$ & $-1.52$ & $-1.47$ & $-1.68$ & $-1.523$ & \nodata \\
He/H & 0.105 & 0.101 & 0.099 & 0.099 & 0.107 & 0.100 & 0.097 & 0.085 \\
C/H & 8.57 & 8.37 & 8.29 & 8.80 & 8.87 & 9.01 & 8.59 & 8.43 \\
N/H & 8.20 & 7.57 & 7.85 & 8.23 & 8.25 & 8.23 & 7.88 & 7.83 \\
O/H & 8.58 & 8.39 & 8.64 & 8.69 & 8.66 & 8.81 & 8.57 & 8.69 \\
Ne/H & 7.90 & 7.73 & 7.86 & 7.92 & 8.05 & 8.04 & 7.80 & 7.93 \\
S/H & 6.77 & 6.35 & 6.42 & 6.87 & 6.88 & 7.10 & 6.76 & 7.12 \\
Cl/H & 4.96 & 4.84 & 4.96 & 5.12 & 5.13 & 5.33 & 5.19 &  5.50 \\
Ar/H & 6.32 & 5.78 & 6.11 & 6.26 & 6.37 & 6.36 & 6.11 & 6.40 \\
Se/H & 3.31$\pm$0.11 & 2.91$\pm$0.05 & 2.95$\pm$0.13 & 3.40$\pm$0.06 & 3.52$\pm$0.08 & 3.68$\pm$0.08 & 3.53$\pm$0.10 & 3.34 \\
Kr/H & 3.69$\pm$0.20 & 3.26$\pm$0.10 & 3.63$\pm$0.11 & 3.69$\pm$0.15 & 3.70$\pm$0.20 & 4.14$\pm$0.11 & 3.74$\pm$0.10 & 3.25 \\
\tableline
\enddata
\label{params2}
\tablecomments{Input parameters for Cloudy models.  All models were run with Cloudy version C13, modified as described in the text.  The stellar effective temperature, gravity, and luminosity (relative to solar), nebular hydrogen density, and inner and outer nebular radii are given in the first six rows.  Elemental abundances follow, with He/H on a linear scale, and the other elements given as $12+$~log(X/H).}
\tablenotetext{a}{From \citet{asplund09}}
\end{deluxetable}
\clearpage

\begin{deluxetable}{lc|cc|cc|cc|cc|}
\tablecolumns{10}
\tabletypesize{\scriptsize} 
\tablewidth{0pc} 
\tablecaption{Predicted vs. Observed Diagnostics and Line Intensities} 
\tablehead{
\colhead{} & \colhead{} & \multicolumn{2}{|c|}{IC 418} & \multicolumn{2}{|c|}{IC 2501} & \multicolumn{2}{|c|}{IC 4191} & \multicolumn{2}{|c|}{IC 5117} \\ \cline{3-4} \cline{5-6} \cline{7-8} \cline{9-10}
\multicolumn{1}{c}{Ion} & \multicolumn{1}{c}{$\lambda$} & \multicolumn{1}{|c}{Pred.} & \multicolumn{1}{c|}{Obs.} & \multicolumn{1}{|c}{Pred.} & \multicolumn{1}{c|}{Obs.} & \multicolumn{1}{|c}{Pred.} & \multicolumn{1}{c|}{Obs.} & \multicolumn{1}{|c}{Pred.} & \multicolumn{1}{c|}{Obs.}}
\startdata
\ion{Ar}{3}na & \nodata & 2.42E+02 & 2.71E+02 & 1.50E+02\tablenotemark{a} & 1.93E+02\tablenotemark{a} & 1.49E+02\tablenotemark{a} & 8.54E+01\tablenotemark{a} & 7.98E+01 & 1.35E+02 \\
\ion{Ar}{4}nn & \nodata & 1.61E+00 & 1.20E+00 & 1.87E+00 & 1.50E+00 & 1.73E+00 & 1.68E+00 & 4.16E+00 & 4.18E+00 \\
\ion{Cl}{3}nn & \nodata & 6.07E--01 & 5.10E--01 & 5.29E--01 & 5.61E--01 & 5.67E--01 & 4.92E--01 & 2.95E--01 & 3.28E--01 \\
\ion{N}{2}na & \nodata & 8.41E+01 & 7.84E+01 & 5.68E+01 & 5.64E+01 & 6.26E+01 & 4.29E+01 & 1.81E+01 & 2.53E+01 \\
\ion{O}{2}na & \nodata & 8.54E+00 & 6.98E+00 & 5.30E+00 & 4.36E+00 & 6.22E+00 & 2.65E+00 & 9.49E--01 & 6.95E--01 \\
\ion{O}{2}nn & \nodata & 2.18E+00 & 2.37E+00 & 2.26E+00 & 2.38E+00 & 2.23E+00 & 2.31E+00 & 2.55E+00 & 3.10E+00 \\
\ion{O}{3}na & \nodata & 3.19E+02 & 3.08E+02 & 2.45E+02 & 2.39E+02 & 2.34E+02 & 2.13E+02 & 7.16E+01 & 8.83E+01 \\
\ion{S}{2}nn & \nodata & 4.98E--01 & 4.70E--01 & 4.96E--01 & 5.00E--01 & 4.81E--01 & 5.13E--01 & 4.46E--01 & 4.36E--01 \\
\vdots & \vdots & \vdots & \vdots & \vdots & \vdots & \vdots & \vdots & \vdots & \vdots \\
\tableline
\enddata
\label{lines1}
\tablecomments{Table~\ref{lines1} is published in its entirety in the electronic edition of the article.  A portion is shown here for guidance regarding its form and content.  Comparison of predicted and observed line intensities (on the scale $I$(H$\beta)=100$) and diagnostic ratios.  Wavelengths are in \AA\ except where noted.  The diagnostic ratios are defined as follows: \ion{Ar}{3}na = $[$\ion{Ar}{3}$]$ ($7135+7751$)/5192; \ion{Ar}{4}nn = $[$\ion{Ar}{4}$]$ 4740/4711; \ion{Cl}{3}nn = $[$\ion{Cl}{3}$]$ 5518/5538; \ion{N}{2}na = $[$\ion{N}{2}$]$ ($6548+6584$)/5755; \ion{O}{2}na = $[$\ion{O}{2}$]$ ($3726+3729$)/($7320+7330$); \ion{O}{2}nn = $[$\ion{O}{2}$]$ 3726/3729; \ion{O}{3}na = $[$\ion{O}{3}$]$ ($4959+5007$)/4363; and \ion{S}{2}nn = $[$\ion{S}{2}$]$ 6716/6731.  The notation ``na'' indicates a nebular to auroral line ratio, and ``nn'' indicates nebular to nebular.  Nebular lines arise from the first excited term of the ground configuration whereas auroral lines come from the second excited term.}
\tablenotetext{a}{\ion{Ar}{3}na = $[$\ion{Ar}{3}$]$ 7135/5192.}
\end{deluxetable}
\clearpage

\begin{deluxetable}{lc|cc|cc|cc|cc|}
\tablecolumns{10}
\tabletypesize{\scriptsize}
\tablewidth{0pc} 
\tablecaption{Predicted vs. Observed Diagnostics and Line Intensities} 
\tablehead{
\colhead{} & \colhead{} & \multicolumn{2}{|c|}{NGC 2440} & \multicolumn{2}{|c|}{NGC 5315} & \multicolumn{2}{|c|}{NGC 6369} & \multicolumn{2}{|c|}{NGC 6572} \\ \cline{3-4} \cline{5-6} \cline{7-8} \cline{9-10}
\multicolumn{1}{c}{Ion} & \multicolumn{1}{c}{$\lambda$} & \multicolumn{1}{|c}{Pred.} & \multicolumn{1}{c|}{Obs.} & \multicolumn{1}{|c}{Pred.} & \multicolumn{1}{c|}{Obs.} & \multicolumn{1}{|c}{Pred.} & \multicolumn{1}{c|}{Obs.} & \multicolumn{1}{|c}{Pred.} & \multicolumn{1}{c|}{Obs.}}
\startdata
\ion{Ar}{3}na & \nodata & 7.60E+01\tablenotemark{a} & 8.48E+01\tablenotemark{a} & 2.15E+02 & 3.00E+02 & 1.52E+02 & 2.18E+02 & 9.93E+01 & 1.69E+02 \\
\ion{Ar}{4}nn & \nodata & 1.34E+00 & 1.16E+00 & 3.22E+00 & 4.69E+00 & 1.07E+00 & 1.19E+00 & 2.35E+00 & 1.98E+00 \\
\ion{Cl}{3}nn & \nodata & 7.16E--01 & 7.50E--01 & 3.52E--01 & 3.41E--01 & 8.94E--01 & 8.24E--01 & 4.37E--01 & 4.40E--01 \\
\ion{N}{2}na & \nodata & 5.66E+01 & 4.67E+01 & 4.24E+01 & 4.50E+01 & 6.58E+01 & 4.34E+01 & 3.58E+01 & 3.78E+01 \\
\ion{O}{2}na & \nodata & 7.98E+00 & 8.39E+00 & 2.40E+00 & 2.21E+00 & 1.32E+01 & 5.55E+00 & 2.80E+00 & 3.06E+00 \\
\ion{O}{2}nn & \nodata & 1.99E+00 & 1.79E+00 & 2.49E+00 & 2.67E+00 & 1.74E+00 & 1.81E+00 & 2.37E+00 & 2.48E+00 \\
\ion{O}{3}na & \nodata & 8.35E+01 & 7.14E+01 & 2.72E+02 & 2.56E+02 & 1.90E+02 & 1.64E+02 & 2.70E+01\tablenotemark{b} & 4.30E+01\tablenotemark{b} \\
\ion{S}{2}nn & \nodata & 5.60E--01 & 6.45E--01 & 4.58E--01 & 4.76E--01 & 5.75E--01 & 5.98E--01 & 4.76E--01 & 4.70E--01 \\
\vdots & \vdots & \vdots & \vdots & \vdots & \vdots & \vdots & \vdots & \vdots & \vdots \\
\tableline
\enddata
\label{lines2}
\tablecomments{Table~\ref{lines2} is published in its entirety in the electronic edition of the article.  A portion is shown here for guidance regarding its form and content.  Comparison of predicted and observed line intensities (on the scale $I$(H$\beta)=100$) and diagnostic ratios.  Wavelengths are in \AA\ except where noted.  The diagnostic ratios are defined as follows: \ion{Ar}{3}na = $[$\ion{Ar}{3}$]$ ($7135+7751$)/5192; \ion{Ar}{4}nn = $[$\ion{Ar}{4}$]$ 4740/4711; \ion{Cl}{3}nn = $[$\ion{Cl}{3}$]$ 5518/5538; \ion{N}{2}na = $[$\ion{N}{2}$]$ ($6548+6584$)/5755; \ion{O}{2}na = $[$\ion{O}{2}$]$ ($3726+3729$)/($7320+7330$); \ion{O}{2}nn = $[$\ion{O}{2}$]$ 3726/3729; \ion{O}{3}na = $[$\ion{O}{3}$]$ ($4959+5007$)/4363; and \ion{S}{2}nn = $[$\ion{S}{2}$]$ 6716/6731.  The notation ``na'' indicates a nebular to auroral line ratio, and ``nn'' indicates nebular to nebular.  Nebular lines arise from the first excited term of the ground configuration whereas auroral lines come from the second excited term.}
\tablenotetext{a}{\ion{Ar}{3}na = $[$\ion{Ar}{3}$]$ 7135/5192}
\tablenotetext{b}{\ion{O}{3}na = $[$\ion{O}{3}$]$ 4959/4363}
\end{deluxetable}
\clearpage
 
\begin{deluxetable}{lc|cc|cc|cc|cc|}
\tablecolumns{10}
\tabletypesize{\scriptsize}
\tablewidth{0pc} 
\tablecaption{Predicted vs. Observed Diagnostics and Line Intensities} 
\tablehead{
\colhead{} & \colhead{} & \multicolumn{2}{|c|}{NGC 6741} & \multicolumn{2}{|c|}{NGC 6790} & \multicolumn{2}{|c|}{NGC 6826} & \multicolumn{2}{|c|}{NGC 6884} \\ \cline{3-4} \cline{5-6} \cline{7-8} \cline{9-10}
\multicolumn{1}{c}{Ion} & \multicolumn{1}{c}{$\lambda$} & \multicolumn{1}{|c}{Pred.} & \multicolumn{1}{c|}{Obs.} & \multicolumn{1}{|c}{Pred.} & \multicolumn{1}{c|}{Obs.} & \multicolumn{1}{|c}{Pred.} & \multicolumn{1}{c|}{Obs.} & \multicolumn{1}{|c}{Pred.} & \multicolumn{1}{c|}{Obs.}}
\startdata
\ion{Ar}{3}na & \nodata & 9.71E+01 & 1.32E+02 & 7.68E+01 & 9.16E+01 & 2.02E+02 & 2.56E+02 & 1.19E+02 & 1.51E+02 \\
\ion{Ar}{4}nn & \nodata & 1.28E+00 & 1.26E+00 & 3.35E+00 & 4.06E+00 & 9.04E--01 & 9.32E--01 & 1.43E+00 & 1.56E+00 \\
\ion{Cl}{3}nn & \nodata & 7.53E--01 & 7.30E--01 & 3.37E--01 & 4.07E--01 & 1.08E+00 & 1.10E+00 & 6.69E--01 & 7.04E--01 \\
\ion{N}{2}na & \nodata & 5.44E+01 & 6.48E+01 & 2.29E+01 & 2.27E+01 & 8.68E+01 & 7.91E+01 & 5.03E+01 & 5.25E+01 \\
\ion{O}{2}na & \nodata & 8.46E+00 & 9.47E+00 & 1.38E+00 & 1.51E+00 & 2.56E+01 & 1.48E+01 & 6.74E+00 & 7.02E+00 \\
\ion{O}{2}nn & \nodata & 1.93E+00 & 2.04E+00 & 2.50E+00 & 2.45E+00 & 1.43E+00 & 1.48E+00 & 2.09E+00 & 2.17E+00 \\
\ion{O}{3}na & \nodata & 2.33E+01\tablenotemark{a} & 2.67E+01\tablenotemark{a} & 1.87E+01\tablenotemark{a} & 2.24E+01\tablenotemark{a} & 6.43E+01\tablenotemark{a} & 6.71E+01\tablenotemark{a} & 3.41E+01\tablenotemark{a} & 3.78E+01\tablenotemark{a} \\
\ion{S}{2}nn & \nodata & 5.62E--01 & 5.70E--01 & 4.55E--01 & 4.57E--01 & 6.74E--01 & 7.30E--01 & 5.02E--01 & 5.32E--01 \\
\vdots & \vdots & \vdots & \vdots & \vdots & \vdots & \vdots & \vdots & \vdots & \vdots \\
\tableline
\enddata
\label{lines3}
\tablecomments{Table~\ref{lines3} is published in its entirety in the electronic edition of the article.  A portion is shown here for guidance regarding its form and content.  Comparison of predicted and observed line intensities (on the scale $I$(H$\beta)=100$) and diagnostic ratios.  Wavelengths are in \AA\ except where noted.  The diagnostic ratios are defined as follows: \ion{Ar}{3}na = $[$\ion{Ar}{3}$]$ ($7135+7751$)/5192; \ion{Ar}{4}nn = $[$\ion{Ar}{4}$]$ 4740/4711; \ion{Cl}{3}nn = $[$\ion{Cl}{3}$]$ 5518/5538; \ion{N}{2}na = $[$\ion{N}{2}$]$ ($6548+6584$)/5755; \ion{O}{2}na = $[$\ion{O}{2}$]$ ($3726+3729$)/($7320+7330$); \ion{O}{2}nn = $[$\ion{O}{2}$]$ 3726/3729; \ion{O}{3}na = $[$\ion{O}{3}$]$ ($4959+5007$)/4363; and \ion{S}{2}nn = $[$\ion{S}{2}$]$ 6716/6731.  The notation ``na'' indicates a nebular to auroral line ratio, and ``nn'' indicates nebular to nebular.  Nebular lines arise from the first excited term of the ground configuration whereas auroral lines come from the second excited term.}
\tablenotetext{a}{\ion{O}{3}na = $[$\ion{O}{3}$]$ 5007/4363.}
\end{deluxetable}
\clearpage

\begin{deluxetable}{lc|cc|cc|cc|}
\tablecolumns{8}
\tabletypesize{\scriptsize}
\tablewidth{0pc} 
\tablecaption{Predicted vs. Observed Diagnostics and Line Intensities} 
\tablehead{
\colhead{} & \colhead{} & \multicolumn{2}{|c|}{NGC 6886} & \multicolumn{2}{|c|}{NGC 7027} & \multicolumn{2}{|c|}{NGC 7662} \\ \cline{3-4} \cline{5-6} \cline{7-8}
\multicolumn{1}{c}{Ion} & \multicolumn{1}{c}{$\lambda$} & \multicolumn{1}{|c}{Pred.} & \multicolumn{1}{c|}{Obs.} & \multicolumn{1}{|c}{Pred.} & \multicolumn{1}{c|}{Obs.} & \multicolumn{1}{|c}{Pred.} & \multicolumn{1}{c|}{Obs.}}
\startdata
\ion{Ar}{3}na & \nodata & 1.04E+02 & 1.09E+02 & 1.06E+02 & 9.79E+01 & 9.97E+01 & 9.68E+01 \\
\ion{Ar}{4}nn & \nodata & 1.35E+00 & 1.29E+00 & 4.12E+00 & 3.70E+00 & 1.06E+00 & 1.26E+00 \\
\ion{Cl}{3}nn & \nodata & 7.11E--01 & 3.94E--01 & 2.99E--01 & 2.90E--01 & 8.97E--01 & 9.87E--01 \\
\ion{N}{2}na & \nodata & 5.22E+01 & 6.38E+01 & 2.26E+01 & 2.20E+01 & 5.05E+01 & 5.05E+01 \\
\ion{O}{2}na & \nodata & 7.50E+00 & 4.50E+00 & 1.16E+00 & 8.70E--01 & 1.14E+01 & 8.63E+00 \\
\ion{O}{2}nn & \nodata & 2.02E+00 & 1.53E+00 & 2.55E+00 & 2.76E+00 & 1.73E+00 & 1.70E+00 \\
\ion{O}{3}na & \nodata & 1.04E+02 & 1.22E+02 & 9.11E+01 & 8.75E+01 & 2.59E+01\tablenotemark{a} & 2.33E+01\tablenotemark{a} \\
\ion{S}{2}nn & \nodata & 5.15E--01 & 4.84E--01 & 4.47E--01 & 4.40E--01 & 5.79E--01 & 6.58E--01 \\
\vdots & \vdots & \vdots & \vdots & \vdots & \vdots & \vdots & \vdots  \\
\tableline
\enddata
\label{lines4}
\tablecomments{Table~\ref{lines4} is published in its entirety in the electronic edition of the article.  A portion is shown here for guidance regarding its form and content.  Comparison of predicted and observed line intensities (on the scale $I$(H$\beta)=100$) and diagnostic ratios.  Wavelengths are in \AA\ except where noted.  The diagnostic ratios are defined as follows: \ion{Ar}{3}na = $[$\ion{Ar}{3}$]$ ($7135+7751$)/5192; \ion{Ar}{4}nn = $[$\ion{Ar}{4}$]$ 4740/4711; \ion{Cl}{3}nn = $[$\ion{Cl}{3}$]$ 5518/5538; \ion{N}{2}na = $[$\ion{N}{2}$]$ ($6548+6584$)/5755; \ion{O}{2}na = $[$\ion{O}{2}$]$ ($3726+3729$)/($7320+7330$); \ion{O}{2}nn = $[$\ion{O}{2}$]$ 3726/3729; \ion{O}{3}na = $[$\ion{O}{3}$]$ ($4959+5007$)/4363; and \ion{S}{2}nn = $[$\ion{S}{2}$]$ 6716/6731.  The notation ``na'' indicates a nebular to auroral line ratio, and ``nn'' indicates nebular to nebular.  Nebular lines arise from the first excited term of the ground configuration whereas auroral lines come from the second excited term.}
\tablenotetext{a}{\ion{O}{3}na = $[$\ion{O}{3}$]$ 4959/4363.}
\end{deluxetable}
\clearpage

We determined uncertainties in the modeled Se and Kr abundances (Tables~\ref{params1} and \ref{params2}) by considering errors in line intensities and central star temperature.  For the upper and lower limits to the stellar temperature, we re-optimized the Se and Kr abundances to estimate their uncertainty.  The effects of uncertainties in line intensities on the abundance determinations were treated differently for Se and Kr.  In the case of Kr, we manually adjusted each model's Kr abundance until the most discrepant Kr line intensity was reproduced.  We adopted the difference with the best model as the uncertainty, or 25\% (the assumed error bars in the optical line intensities), whichever value was larger.  In contrast, only one line of Se was detected (if at all) in the modeled nebulae, and we utilized the percentage uncertainty in the intensity as that of the abundance \citepalias[this underestimates the magnitude of the actual error bars, as discussed in][]{sterling07}.  The overall abundance uncertainties were calculated by adding those from the intensities and variations in the central star temperature in quadrature.

\subsection{The Ionization Balance of Krypton} \label{krionbal}

In Table~\ref{krlines}, we compare the observed Kr line intensities with those predicted by our models.  We find that the ionization balance of Kr is not reproduced well by our models that incorporate the new atomic data.  Specifically, the models predict values for the ionic ratio Kr$^{3+}$/Kr$^{2+}$ that are lower than the observed values.  $[$\ion{Kr}{3}$]$ line strengths are predicted to be larger than observed by about 50\% up to factors of 2--3, while the $[$\ion{Kr}{4}$]$ lines are predicted to be weaker than observed by similar amounts.  This is a \textit{systematic} effect, in that it is seen in all of the 15 modeled PNe independent of ionization level, central star temperature, morphology, extinction, density, and other nebular properties.  This result suggests that the errors in the modeled Kr ionic ratios are due to errors in the new atomic data.  To verify this, we first investigate whether observational uncertainties or modeling biases can explain the poorly-reproduced Kr ionization equilibrium.

\begin{deluxetable}{lc|ccc|ccc|ccc}
\tablecolumns{11}
\tabletypesize{\scriptsize}
\tablewidth{0pc} 
\tablecaption{Predicted vs.\ Observed Krypton Line Intensities} 
\tablehead{
\multicolumn{1}{c}{Ion} & \multicolumn{1}{c}{$\lambda$} & \multicolumn{1}{|c}{Pred.\tablenotemark{a}} & \multicolumn{1}{c}{Pred.\tablenotemark{b}} & \multicolumn{1}{c|}{Obs.} & \multicolumn{1}{|c}{Pred.\tablenotemark{a}} & \multicolumn{1}{c}{Pred.\tablenotemark{b}} & \multicolumn{1}{c|}{Obs.} & \multicolumn{1}{|c}{Pred.\tablenotemark{a}} & \multicolumn{1}{c}{Pred.\tablenotemark{b}} & \multicolumn{1}{c|}{Obs.}}
\startdata
\multicolumn{2}{c}{} & \multicolumn{3}{c}{IC 418} & \multicolumn{3}{c}{IC 2501} & \multicolumn{3}{c}{IC 4191} \\ \hline
$[$\ion{Kr}{3}$]$ & 6826 & 4.71E--02 & 2.83E--02 & 2.90E--02 & 8.08E--03 & 5.15E--03 & 4.60E--03 & 6.38E--03 & 4.60E--03 & 1.50E--03 \\
$[$\ion{Kr}{3}$]$ & 2.199$\mu$m & 1.44E--01 & 8.75E--02 & 6.99E--02 & 2.19E--02 & 1.40E--02 & \nodata & 1.72E--02 & 1.23E--02 & \nodata \\
$[$\ion{Kr}{4}$]$ & 5346 & 8.55E--04 & 1.88E--03 & 3.50E--03 & 4.04E--03 & 5.54E--03 & 6.10E--03 & 9.78E--03 & 1.24E--02 & 2.40E--02 \\
$[$\ion{Kr}{4}$]$ & 5868 & 1.30E--03 & 2.85E--03 & 3.50E--03 & 6.08E--03 & 8.34E--03 & 8.50E--03 & 1.43E--02 & 1.81E--02 & 1.90E--02 \\
$[$\ion{Kr}{5}$]$ & 6256 & 1.54E--10 & 1.74E--12 & \nodata & 1.31E--08 & 4.35E--10 & \nodata & 5.40E--04 & 1.03E--04 & \nodata \\
$[$\ion{Kr}{5}$]$ & 8243 & 1.40E--10 & 1.59E--12 & \nodata & 1.19E--08 & 3.97E--10 & \nodata & 4.93E--04 & 9.37E--05 & \nodata \\
$[$\ion{Kr}{6}$]$ & 1.233$\mu$m & \nodata & \nodata & \nodata & 8.03E--17 & \nodata & \nodata & 9.96E--05 & 3.72E--05 & \nodata \\
\hline\hline
\multicolumn{2}{c}{} & \multicolumn{3}{c}{IC 5117} & \multicolumn{3}{c}{NGC 2440} & \multicolumn{3}{c}{NGC 5315} \\ \hline
$[$\ion{Kr}{3}$]$ & 6826 & 4.13E--02 & 3.05E--02 & \nodata & 4.20E--02 & 1.03E--02 & \nodata & 1.75E--02 & 1.17E--02 & 6.20E--03 \\
$[$\ion{Kr}{3}$]$ & 2.199$\mu$m & 7.16E--02 & 5.26E--02 & 4.23E--02 & 8.82E--02 & 2.08E--02 & \nodata & 5.08E--02 & 3.44E--02 & \nodata \\
$[$\ion{Kr}{4}$]$ & 5346 & 8.76E--02 & 1.07E--01 & 1.29E--01 & 4.65E--02 & 3.48E--02 & 3.50E--02 & 3.33E--03 & 4.99E--03 & 9.40E--03 \\
$[$\ion{Kr}{4}$]$ & 5868 & 1.23E--01 & 1.49E--01 & 1.45E--01 & 6.25E--02 & 4.62E--02 & 4.60E--02 & 5.19E--03 & 7.78E--03 & \nodata \\
$[$\ion{Kr}{5}$]$ & 6256 & 1.75E--03 & 2.20E--04 & \nodata & 2.78E--02 & 4.77E--03 & 4.40E--02 & 1.48E--09 & 9.74E--14 & \nodata \\
$[$\ion{Kr}{5}$]$ & 8243 & 1.59E--03 & 2.01E--04 & \nodata & 2.53E--02 & 4.35E--03 & \nodata & 1.35E--09 & 8.88E--14 & \nodata \\
$[$\ion{Kr}{6}$]$ & 1.233$\mu$m & 5.24E--05 & 6.84E--06 & \nodata & 5.22E--02 & 1.18E--02 & \nodata & \nodata & \nodata & \nodata \\
\hline\hline
\multicolumn{2}{c}{} & \multicolumn{3}{c}{NGC 6369} & \multicolumn{3}{c}{NGC 6572} & \multicolumn{3}{c}{NGC 6741} \\ \hline
$[$\ion{Kr}{3}$]$ & 6826 & 5.58E--02 & 3.92E--02 & 2.10E--02 & 2.60E--02 & 1.87E--02 & 1.80E--02 & 4.20E--02 & 2.45E--02 & 3.70E--02 \\
$[$\ion{Kr}{3}$]$ & 2.199$\mu$m & 1.36E--01 & 9.50E--02 & \nodata & 5.07E--02 & 3.62E--02 & 2.49E--02 & 8.37E--02 & 4.81E--02 & 5.40E--02 \\
$[$\ion{Kr}{4}$]$ & 5346 & 4.20E--02 & 5.68E--02 & 8.80E--02 & 2.97E--02 & 3.90E--02 & 4.70E--02 & 4.23E--02 & 6.61E--02 & 6.90E--02 \\
$[$\ion{Kr}{4}$]$ & 5868 & 6.08E--02 & 8.22E--02 & 1.23E--01 & 4.14E--02 & 5.44E--02 & 6.40E--02 & 5.68E--02 & 8.81E--02 & 5.50E--02 \\
$[$\ion{Kr}{5}$]$ & 6256 & 2.52E--06 & 1.49E--07 & \nodata & 8.69E--06 & 5.85E--07 & \nodata & 1.81E--02 & 5.71E--03 & \nodata \\
$[$\ion{Kr}{5}$]$ & 8243 & 2.30E--06 & 1.36E--07 & \nodata & 7.92E--06 & 5.34E--07 & \nodata & 1.65E--02 & 5.20E--03 & \nodata \\
$[$\ion{Kr}{6}$]$ & 1.233$\mu$m & 5.14E--04 & 2.82E--15 & \nodata & 6.34E--13 & 4.08E--14 & \nodata & 1.08E--02 & 5.10E--03 & \nodata \\
\hline\hline
\multicolumn{2}{c}{} & \multicolumn{3}{c}{NGC 6790} & \multicolumn{3}{c}{NGC 6826} & \multicolumn{3}{c}{NGC 6884} \\ \hline
$[$\ion{Kr}{3}$]$ & 6826 & 1.34E--02 & 9.91E--03 & 1.20E--02 & 2.67E--02 & 1.57E--02 & 1.60E--02 & 2.76E--02 & 1.99E--02 & 1.50E--02 \\
$[$\ion{Kr}{3}$]$ & 2.199$\mu$m & 1.96E--02 & 1.69E--02 & \nodata & 7.51E--02 & 4.41E--02 & \nodata & 5.95E--02 & 4.26E--02 & \nodata \\
$[$\ion{Kr}{4}$]$ & 5346 & 3.33E--02 & 4.70E--02 & 5.30E--02 & 1.80E--02 & 2.38E--02 & 2.30E--02 & 5.37E--02 & 6.58E--02 & 9.50E--02 \\
$[$\ion{Kr}{4}$]$ & 5868 & 4.58E--02 & 6.47E--02 & 5.20E--02 & 2.66E--02 & 3.50E--02 & 4.50E--02 & 7.45E--02 & 9.13E--02 & 1.14E--01 \\
$[$\ion{Kr}{5}$]$ & 6256 & 1.94E--04 & 1.79E--05 & \nodata & 5.15E--08 & 1.20E--09 & \nodata & 1.39E--03 & 1.97E--04 & \nodata \\
$[$\ion{Kr}{5}$]$ & 8243 & 1.77E--04 & 1.63E--05 & \nodata & 4.70E--08 & 1.09E--09 & \nodata & 1.26E--03 & 1.80E--04 & \nodata \\
$[$\ion{Kr}{6}$]$ & 1.233$\mu$m & 2.75E--06 & 2.85E--07 & \nodata & 4.36E--17 & \nodata & \nodata & 1.40E--04 & 3.48E--05 & \nodata \\
\hline\hline
\multicolumn{2}{c}{} & \multicolumn{3}{c}{NGC 6886} & \multicolumn{3}{c}{NGC 7027} & \multicolumn{3}{c}{NGC 7662} \\ \hline
$[$\ion{Kr}{3}$]$ & 6826 & 4.34E--02 & 2.10E--02 & \nodata & 7.59E--02 & 4.59E--02 & 4.10E--02 & 2.06E--02 & 1.25E--02 & 1.30E--02 \\
$[$\ion{Kr}{3}$]$ & 2.199$\mu$m & 8.89E--02 & 4.28E--02 & 5.89E--02 & 1.53E--01 & 9.20E--02 & 8.74E--02 & 4.18E--02 & 2.54E--02 & \nodata \\
$[$\ion{Kr}{4}$]$ & 5346 & 4.88E--02 & 6.25E--02 & 8.00E--02 & 9.05E--02 & 1.48E--01 & 1.48E--01 & 1.07E--01 & 1.30E--01 & 1.59E--01 \\
$[$\ion{Kr}{4}$]$ & 5868 & 6.64E--02 & 8.43E--02 & 5.40E--02 & 1.30E--01 & 2.10E--01 & 2.22E--01 & 1.45E--01 & 1.75E--01 & 1.60E--01 \\
$[$\ion{Kr}{5}$]$ & 6256 & 2.45E--02 & 6.87E--03 & \nodata & 4.47E--02 & 2.11E--02 & 2.50E--02 & 1.61E--02 & 2.44E--03 & \nodata \\
$[$\ion{Kr}{5}$]$ & 8243 & 2.24E--02 & 6.27E--03 & \nodata & 4.08E--02 & 1.93E--02 & 1.40E--02 & 1.47E--02 & 2.23E--03 & \nodata \\
$[$\ion{Kr}{6}$]$ & 1.233$\mu$m & 2.64E--02 & 1.10E--02 & \nodata & 1.11E--01 & 6.50E--02 & \nodata & 1.42E--03 & 2.37E--04 & \nodata \\
\tableline
\enddata
\label{krlines}
\tablecomments{Comparison of predicted and observed Kr line intensities (on the scale $I$(H$\beta)=100$).  Wavelengths are in \AA\ unless noted.}
\tablenotetext{a}{Predicted Kr line intensities, using the photoionization and recombination data from \cite{sterling11c}.}
\tablenotetext{b}{Predicted Kr line intensities, after adjusting the PI cross sections and DR rate coefficients within their uncertainties (see text).}
\end{deluxetable}

The fluxes of weak, low signal-to-noise emission lines are often overestimated compared to stronger lines \citep[e.g.,][]{rolapelat94}.  Because of the low cosmic abundance of Kr, even ``strong'' Kr emission lines are comparable in flux to weak permitted features of more abundant elements.  As seen in Table~\ref{krlines}, $[$\ion{Kr}{3}$]$ and $[$\ion{Kr}{5}$]$ lines tend to be weaker than $[$\ion{Kr}{4}$]$ lines for PNe of moderate to high excitation.  Therefore, if line measurement errors substantially affect the Kr ionization equilibrium results, one would expect the observed $[$\ion{Kr}{3}$]$ and $[$\ion{Kr}{5}$]$ lines to be stronger than predicted by models.  However, the \textit{opposite} effect is seen in our models, where the predicted $[$\ion{Kr}{3}$]$ line intensities are too large compared to observational measurements.  In NGC~7027, the observed $[$\ion{Kr}{5}$]$ intensities are also much weaker than predicted.  Therefore bias in the measurement of weak emission lines does not explain the inconsistency between the predicted and observed Kr ionization balance.

Calibration uncertainties can also potentially be important, particularly since many of the optical data we utilize are cross-dispersed echelle spectra, for which flux calibrations are often difficult and uncertain.  However, since the data we utilize were taken with different instruments and reduced by independent teams of researchers, one would expect that calibration errors would be random and should not produce systematic effects.  Even if systematic effects were introduced by improper calibrations, then these should in principle be seen for other species.  However, no such trends are seen for other elements.

Assumptions and/or uncertainties in our numerical models are another potential explanation.  Indeed, we make several simplifying assumptions in our models: we assume that the nebulae are uniform density and isotropic, we ignore the effects of dust on the ionized gas, and we assume that the filling and covering factors are equal to unity.  These assumptions certainly affect the predicted emission line spectra to some extent.  Discrepancies in the predicted S, Ar, and Cl ionization balance for some objects are likely due in part to our simplifications.  However, it is important to note that \textit{there are no systematic trends seen in the predicted ionization balance of elements other than Kr.}  This suggests that the simplifying assumptions in our models are unlikely to be the source of the Kr ionization equilibrium disparity.  The fact that the inclusion or exclusion of dust in our models does not substantially affect the Kr ionization balance (\S\ref{icfs}) further supports this argument.

Therefore it appears to be unlikely that observational or calibration errors and simplifications in our models can explain the disagreement between the predicted and observed Kr ionization balance.  On the other hand, it is possible that errors and uncertainties in the atomic data may be responsible for the ionization balance discrepancy.  \citet{sterling11c} cite error bars in their computed PI cross sections and RR and DR rate coefficients, based on comparisons to experimental measurements, the electronic configurations included in the calculations, and the effects of variations in the radial scaling parameters of the orbital wavefunctions.  For example, the computed Kr$^{3+}$ direct PI cross section is a factor of two larger in magnitude than the experimental PI cross section of \citet{lu06b} \citep[see Fig.\ 5 of][]{sterling11c}, and efforts to resolve this discrepancy were not successful.  For other ions, the PI cross sections were cited to be accurate to within 30--50\%.  DR rate coefficients for near-neutral heavy element ions have large uncertainties regardless of the theoretical method utilized, due to the unknown energies of autoionizing resonances near the ionization threshold.  In the ``low'' temperature environments of photoionized nebulae, electrons are captured into these levels, and hence whether they lie just above the ionization threshold or just below can substantially affect the computed DR rate coefficient \citep{savin99, ferland03}.  \citet{sterling11c} estimated uncertainties of factors of 2--3 for DR of Kr ions, though the actual errors could be larger.

Uncertainties in DR rate coefficients are likely to be more important than those for RR rate coefficients, since DR is the dominant recombination mechanism for Kr ions at electron temperatures near 10$^4$~K \citep{sterling11c}.  Moreover, the RR rate coefficients have a relatively small uncertainty ($\leq 10$\% for all but singly-ionized Kr) compared to DR.  The CT calculations of \citet{sterling11d} are another potential source of error, as Landau-Zener calculations are only accurate to within a factor of $\sim$3 for systems with large CT rate coefficients.  However, the importance of CT decreases with increasing charge of the reactant ion, as less neutral hydrogen is generally intermixed with more highly-charged species.

Therefore, we consider the DR and PI data to be the most potentially important sources of uncertainty in the Kr ionization balance.  In Cloudy, we manually adjusted the PI cross sections and DR rate coefficients for various Kr ions to investigate how they affect the relative strengths of $[$\ion{Kr}{3}$]$ and $[$\ion{Kr}{4}$]$ emission lines.  We found that altering only the PI cross sections (by up to a factor of two) or only the DR rate coefficients (by up to a factor of 10) did not eliminate the systematic discrepancy in the predicted Kr ionization balance.  However, by adjusting both the PI and DR data for Kr$^+$, Kr$^{2+}$, and Kr$^{3+}$ by modest amounts, systematic trends in the relative strengths of $[$\ion{Kr}{3}$]$ and $[$\ion{Kr}{4}$]$ line intensities could be eliminated.  

\begin{deluxetable}{lcc}
\tablecolumns{3}
\tablewidth{0pc} 
\tablecaption{Adjustments made to Kr atomic data of \citet{sterling11c}} 
\tablehead{
\multicolumn{1}{l}{} & \multicolumn{1}{c}{$\sigma_{\rm pi}$(adjusted)/} & \multicolumn{1}{c}{$\alpha_{\rm DR}$(adjusted)/} \\
\multicolumn{1}{l}{Ion} & \multicolumn{1}{c}{$\sigma_{\rm pi}$(S11)} & \multicolumn{1}{c}{$\alpha_{\rm DR}$(S11)}}
\startdata
Kr$^+$    & 0.67 & \nodata \\
Kr$^{2+}$ & 1.5  & 10.0 \\
Kr$^{3+}$ & 0.5  & 0.1 \\
Kr$^{4+}$ & \nodata & 10.0 \\
\tableline
\enddata
\label{kradjust}
\tablecomments{The adjustment factors made to the Kr atomic data of \citet{sterling11c}.  The first column lists the ion before the indicated reaction occurs.  Therefore, the PI cross section ($\sigma_{\rm pi}$) of Kr$^+$ refers to photoionization of Kr$^+$ to form Kr$^{2+}$.  Likewise, the DR rate coefficient ($\alpha_{\rm DR}$) for Kr$^{2+}$ refers to Kr$^{2+}$ forming Kr$^+$.}
\end{deluxetable}

The PI cross sections were altered within the uncertainties cited by \citet{sterling11c}, while the DR rate coefficients were modified by an order of magnitude (larger than the cited -- but approximate -- uncertainties; see Table~\ref{kradjust}).  Specifically, we adjusted the PI cross sections in the following manner: Kr$^+$ was reduced by 50\%, Kr$^{2+}$ was increased by 50\%, and Kr$^{3+}$ was reduced by a factor of two.  The larger change to the Kr$^{3+}$ PI cross section is motivated by the experimental cross section \citep{lu06a}, which has a magnitude half that of the cross section computed by \citet{sterling11c}.  The DR rate coefficient of Kr$^{3+}$ (forming Kr$^{2+}$) was reduced by a factor of ten, while those of Kr$^{2+}$ and Kr$^{4+}$ were increased by a factor of 10.  These changes reduced the Kr$^{2+}$ ionic fractions in our numerical simulations, while increasing those of Kr$^{3+}$.  The alterations to the atomic data also reduced the (initially overestimated) $[$\ion{Kr}{5}$]$~$\lambda \lambda$6256, 8243 line strengths in NGC~7027, though the agreement worsened for $\lambda$6256 in NGC~2440.\footnote{\citet{sharpee07} note that this line is affected by telluric contamination, and hence its flux may be uncertain.}

With these changes to the Kr PI and DR atomic data, there is no systematic disagreement between the modeled and observed Kr ionization balance.  The remaining inconsistencies in the predicted Kr line intensities are similar in magnitude to those seen for S and Ar, and can be attributed to observational uncertainties and/or model simplifications.

This exercise highlights the value of testing theoretical calculations against experimental and observational measurements, particularly for complex systems such as near-neutral ions of heavy elements.  As discussed in the introduction, PNe have served as natural laboratories to evaluate the accuracy of atomic data determinations for many ions.  Observations, paired with increasingly sophisticated theoretical and experimental methods to determine atomic properties, have led to more reliable atomic data for many species.  Therefore it should come as little surprise that a similar scenario occurs for Kr. 

Because the atomic data governing the ionization balance of Se cannot currently be tested in the absence of detections of ions other than Se$^{3+}$, derived Se abundances will have inherently larger uncertainties than Kr.  However, since Se$^{3+}$ is expected to be the dominant Se ion in moderate to high-excitation PNe \citep{dinerstein01}, the corrections for unobserved ions should be relatively modest and thus the effects of atomic data uncertainties mitigated.

\subsection{Comparison of Modeled Se and Kr Abundances with Previous Results}

We compare the derived Se and Kr abundances from our new Cloudy models with those of \citetalias{sterling07} in Table~\ref{sekrcomp}.  For Kr, we show the abundances determined both before and after we adjusted the PI and DR data to better reproduce the Kr ionization balance.  The abundances found using the new atomic data show remarkably good agreement with those found using the approximate atomic data of \citetalias{sterling07}.

\begin{deluxetable}{lcc|ccc}
\tablecolumns{6}
\tablewidth{0pc} 
\tabletypesize{\footnotesize}
\tablecaption{Comparison of Modeled Se and Kr Abundances} 
\tablehead{
\multicolumn{1}{l}{} & \multicolumn{2}{c}{12~+~log(Se/H)} & \multicolumn{3}{c}{12~+~log(Kr/H)} \\
\cline{2-3} \cline{4-6}
\multicolumn{1}{l}{PN Name} & \multicolumn{1}{c}{This work} & \multicolumn{1}{c}{Paper~I} & \multicolumn{1}{c}{Adopted} & \multicolumn{1}{c}{No Adjustment} & \multicolumn{1}{c}{Paper~I}}
\startdata
IC 418 & \nodata & \nodata & 3.93$\pm$0.23 & 3.95 & 3.90$\pm$0.21 \\
IC 2501 & \nodata & \nodata & 3.12$\pm$0.13 & 3.12 & \nodata \\
IC 4191 & \nodata & \nodata & 3.22$\pm$0.30 & 3.20 & \nodata \\
IC 5117 & 3.52$\pm$0.05 & 3.67$\pm$0.05 & 3.74$\pm$0.10 & 3.72 & 3.76$\pm$0.11 \\
NGC 2440 & \nodata & \nodata & 3.44$\pm$0.10 & 3.90 & \nodata \\
NGC 5315 & \nodata & \nodata & 3.42$\pm$0.28 & 3.43 & \nodata \\
NGC 6369 & 3.66$\pm$0.14 & \nodata & 3.94$\pm$0.18 & 3.92 & \nodata \\
NGC 6572 & 3.01$\pm$0.03 & 3.11$\pm$0.08 & 3.51$\pm$0.15 & 3.47 & 3.54$\pm$0.18 \\
NGC 6741 & 3.31$\pm$0.11 & 3.39$\pm$0.12 & 3.69$\pm$0.20 & 3.75 & 3.75$\pm$0.13 \\
NGC 6790 & 2.91$\pm$0.05 & 3.10$\pm$0.05 & 3.34$\pm$0.10 & 3.23 & 3.24$\pm$0.14 \\
NGC 6826 & 2.95$\pm$0.13 & 2.89$\pm$0.13 & 3.63$\pm$0.11 & 3.67 & 3.57$\pm$0.20 \\
NGC 6884 & 3.40$\pm$0.06 & 3.45$\pm$0.10 & 3.69$\pm$0.15 & 3.67 & 3.65$\pm$0.16 \\
NGC 6886 & 3.70$\pm$0.08 & 3.59$\pm$0.11 & 3.70$\pm$0.20 & 3.87 & 3.81$\pm$0.18 \\
NGC 7027 & 3.68 $\pm$0.08& 3.78$\pm$0.12 & 4.14$\pm$0.11 & 4.23 & 4.26$\pm$0.12 \\
NGC 7662 & 3.53$\pm$0.10 & 3.50$\pm$0.10 & 3.74$\pm$0.10 & 3.79 & 3.84$\pm$0.14 \\
\tableline
\enddata
\label{sekrcomp}
\tablecomments{Comparison of Se and Kr abundances derived from models of 15 individual PNe with those from \citetalias{sterling07}.  For Kr, we show the adopted Kr abundances (after the atomic data was altered to improve the modeled Kr ionization balance in each PN), and those found before the atomic data were adjusted.  The adjusted atomic data do not significantly impact Kr abundances when multiple Kr ions are detected, but will likely have a larger effect when a single ionization stage is detected.}
\end{deluxetable}

To explore the reason for this, in Table~\ref{sekr_vs_sdk} we compare the PI cross sections near the ionization threshold and DR rate coefficients at 10$^4$~K from \citet{sterling11b} and \citet{sterling11c} to the approximate values of \citetalias{sterling07} (we do not consider RR, which at typical PN temperatures is dominated by DR for these species).  For both Se and Kr, the PI cross sections are generally smaller (with the exception of singly-ionized species) while the DR rate coefficients are generally larger (with the exception of doubly-ionized species).  

\begin{deluxetable}{lcc}
\tablecolumns{3}
\tablewidth{0pc} 
\tablecaption{Comparison of Se and Kr Atomic Data with \citetalias{sterling07}} 
\tablehead{
\multicolumn{1}{l}{} & \multicolumn{1}{c}{$\sigma_{\rm pi}$(thresh)/} & \multicolumn{1}{c}{$\alpha_{\rm DR}$(10$^4$~K)/} \\
\multicolumn{1}{l}{Ion} & \multicolumn{1}{c}{$\sigma_{\rm pi}$(\citetalias{sterling07})} & \multicolumn{1}{c}{$\alpha_{\rm DR}$(\citetalias{sterling07})}}
\startdata
Se$^+$    & 2.5 & 27.1 \\
Se$^{2+}$ & 0.9 & 0.3 \\
Se$^{3+}$ & 0.3 & 1.1 \\
Se$^{4+}$ & 0.3 & 5.5 \\
Kr$^+$    & 1.6 & 1.6 \\
Kr$^{2+}$ & 0.6 & 0.2 \\
Kr$^{3+}$ & 0.7 & 7.9 \\
Kr$^{4+}$ & 0.3 & 24.4 \\
\tableline
\enddata
\label{sekr_vs_sdk}
\tablecomments{Ratio of PI cross sections (just above the ionization threshold) and DR rate coefficients of \citet{sterling11b} and \citet{sterling11c} relative to the approximate data of \citetalias{sterling07}.  The first column lists the ion before the indicated reaction occurs.  Therefore, the PI cross section ($\sigma_{\rm pi}$) of Se$^+$ refers to photoionization of Se$^+$ to form Se$^{2+}$.  Likewise, the DR rate coefficient ($\alpha_{\rm DR}$) for Se$^{2+}$ refers to Se$^{2+}$ forming Se$^+$.}
\end{deluxetable}

In the case of Kr, the DR rate coefficients of Kr$^{3+}$ (forming Kr$^{2+}$) and Kr$^{4+}$ are significantly larger than used in \citetalias{sterling07}.  Combined with the smaller PI cross sections of these species, it is expected that the overall Kr ionization balance will shift to lower charge states.  In Table~\ref{kr_ions}, we compare the modeled Kr ionic fractions (averaged over volume) with those of \citetalias{sterling07} for each PN common to both studies.  In all models, the Kr ionic fractions are indeed shifted to less ionized species.  The Kr$^{2+}$ fractional abundances are larger in our new models with the exception of IC~418, in which Kr$^+$ is now predicted to be more abundant than Kr$^{2+}$.   Likewise, the Kr$^{3+}$ fraction is reduced compared to \citetalias{sterling07} in all but the most highly-ionized nebulae, for which the Kr$^{3+}$ abundance is larger at the expense of Kr$^{4+}$ and/or Kr$^{5+}$ in our new models.

\begin{deluxetable}{l|cc|cc|cc|cc|cc|}
\rotate
\tablecolumns{11}
\tablewidth{0pc} 
\tabletypesize{\footnotesize}
\tablecaption{Comparison of Kr Ionic Fractions With \citetalias{sterling07}} 
\tablehead{
\multicolumn{1}{l}{PN} & \multicolumn{2}{|c|}{Kr$^+$} & \multicolumn{2}{c|}{Kr$^{2+}$} & \multicolumn{2}{c|}{Kr$^{3+}$} & \multicolumn{2}{c|}{Kr$^{4+}$} & \multicolumn{2}{c|}{Kr$^{5+}$} \\
\cline{2-3} \cline{4-5} \cline{6-7} \cline{8-9} \cline{10-11}
\multicolumn{1}{l}{Name} & \multicolumn{1}{|c}{This work} & \multicolumn{1}{c}{Paper I} & \multicolumn{1}{|c}{This work} & \multicolumn{1}{c}{Paper I} & \multicolumn{1}{|c}{This work} & \multicolumn{1}{c}{Paper I} & \multicolumn{1}{|c}{This work} & \multicolumn{1}{c}{Paper I} & \multicolumn{1}{|c}{This work} & \multicolumn{1}{c|}{Paper I}}
\startdata
IC 418   & 0.56 & 0.40 & 0.41 & 0.55 & 0.02 & 0.03 & \nodata & \nodata & \nodata & \nodata \\
IC 5117  & 0.10 & 0.12 & 0.26 & 0.17 & 0.61 & 0.68 & \nodata & 0.02 & \nodata & \nodata \\
NGC 6572 & 0.16 & 0.13 & 0.32 & 0.21 & 0.49 & 0.64 & \nodata & \nodata & \nodata & \nodata \\
NGC 6741 & 0.25 & 0.29 & 0.25 & 0.20 & 0.35 & 0.30 & 0.06 & 0.09 & 0.03 & 0.09 \\
NGC 6790 & 0.08 & 0.09 & 0.21 & 0.15 & 0.69 & 0.74 & \nodata & 0.02 & \nodata & \nodata \\
NGC 6826 & 0.08 & 0.01 & 0.40 & 0.31 & 0.52 & 0.68 & \nodata & \nodata & \nodata & \nodata \\
NGC 6884 & 0.06 & 0.01 & 0.28 & 0.19 & 0.66 & 0.74 & \nodata & 0.04 & \nodata & 0.02 \\
NGC 6886 & 0.14 & \nodata & 0.26 & 0.11 & 0.43 & 0.47 & 0.09 & 0.16 & 0.08 & 0.17 \\
NGC 7027 & 0.11 & 0.12 & 0.20 & 0.14 & 0.39 & 0.33 & 0.09 & 0.10 & 0.16 & 0.19 \\
NGC 7662 & 0.02 & \nodata & 0.13 & 0.10 & 0.82 & 0.77 & 0.03 & 0.09 & \nodata & 0.04 \\
\tableline
\enddata
\label{kr_ions}
\tablecomments{Comparison of modeled Kr ionic fractions (averaged over volume) between this work and \citetalias{sterling07}.  Ionic fractions less than 1\% are omitted.}
\end{deluxetable}

The excellent agreement between the Kr abundances in our new models and previous results can be understood given that we selected PNe in which multiple Kr ions have been detected.  In most objects, the detected species are Kr$^{2+}$ and Kr$^{3+}$.  These ions have an ionization potential range of 24.4--50.9~eV, and hence are expected to be the dominant Kr ions in all but the lowest and highest excitation PNe.  Indeed, according to our Cloudy models the combined ionic fractions of Kr$^{2+}$ and Kr$^{3+}$ exceed 70--80\% of the total Kr elemental abundance in all but IC~418, NGC~2440, NGC~6741, and NGC~7662.  These values are within 5--10\% of the combined ionic fractions from \citetalias{sterling07} (Table~\ref{kr_ions}).  Therefore, the corrections for unobserved Kr ions are small due to the detection of the dominant ionization stages. 

The agreement is not expected to be as strong when only one ionization stage is detected, as is the case for most of the PNe in the sample of \citetalias{sterling08}.  Indeed, the larger Kr$^{2+}$ ionic fractions in our modeled objects suggest that previous ICF schemes systematically overestimated total elemental Kr abundances in PNe.  In \S\ref{new_abund}, we see that this is indeed the case.

It is also worth noting that our adjustment of the Kr atomic data has very little effect on the derived Kr elemental abundance in our models, due to the detection of both Kr$^{2+}$ and Kr$^{3+}$ (Table~\ref{sekrcomp}).  The difference in the derived Kr abundance is greater than 0.1~dex only in the case of NGC~2440 and NGC~6886.  For NGC~2440, we were unable to fit $[$\ion{Kr}{5}$]$~$\lambda$6256.1 simultaneously with $[$\ion{Kr}{4}$]$~$\lambda \lambda$5346.0, 5867.7 after the atomic data adjustments.  \citet{sharpee07} note that $\lambda$6256.1 is blended with telluric features in their NGC~2440 spectrum, and hence its flux may be uncertain.  Regardless, a larger Kr abundance was required to reproduce the Kr$^{3+}$ emission lines in NGC~2440 after the adjustment to the atomic data.  For NGC~6886, our adjustment of the atomic data shifted a significant amount of Kr$^{2+}$ and Kr$^{4+}$ to Kr$^{3+}$ in our models.  Along with the uncertainties in the $[$\ion{Kr}{4}$]$ line intensities (see Table~\ref{krlines}) this required a larger Kr abundance to reproduce the observed Kr emission lines.

Our Se abundances also agree well with those derived in \citetalias{sterling07}.  At face value, this seems more surprising, given that only a single ion of Se was detected.  In many of the 15 modeled PNe in which it was detected, Se$^{3+}$ comprises at least 50\% of the total gaseous Se abundance, with the exceptions of NGC~6741, NGC~6886, NGC~7027, and NGC~7662.  But even in those four PNe, the modeled Se$^{3+}$ fractional abundances agree with those of \citetalias{sterling07} to within 6\%.  For other objects, the ionic fraction differs from the \citetalias{sterling07} models by up to 25\%.  

However, the reason for this seeming coincidence emerges when the new PI and DR data for Se ions are compared to the previous approximate values.  Note in Table~\ref{sekr_vs_sdk} that the PI cross section of Se$^{2+}$ and DR rate coefficient of Se$^{3+}$ forming Se$^{2+}$ are quite similar to those adopted in \citetalias{sterling07}.  On the other hand, the PI cross section of Se$^{3+}$ is considerably smaller than \citetalias{sterling07}'s while the DR rate coefficient of Se$^{4+}$ forming Se$^{3+}$ is much larger.  

Therefore, it may be expected that the Se abundances derived with the newly determined atomic data will be lower for high-ionization objects (i.e., the Se$^{3+}$ ionic fraction will be larger at the expense of Se$^{4+}$, and hence the correction for unobserved ions will be smaller) but quite similar for lower-excitation PNe.  We contrast the Se$^+$--Se$^{5+}$ fractional abundances from our current models with those of \citetalias{sterling07} in Table~\ref{se_ions}.  In spite of the differing model input parameters, for most objects it can be seen that the Se$^{3+}$ ionic fraction is larger than found in \citetalias{sterling07}, typically due to a smaller Se$^{4+}$ fraction.  This follows the expected trend given the differences between the Se atomic data from \citet{sterling11b} and those adopted in \citetalias{sterling07}.  The primary reason that the overall Se abundances are not drastically changed is that the Se$^{4+}$ fractional abundances found in \citetalias{sterling07} were relatively small (no larger than 0.32), thereby changing the Se$^{3+}$ fraction by a modest amount.

\begin{deluxetable}{l|cc|cc|cc|cc|cc|}
\rotate
\tablecolumns{11}
\tablewidth{0pc} 
\tabletypesize{\footnotesize}
\tablecaption{Comparison of Se Ionic Fractions With \citetalias{sterling07}} 
\tablehead{
\multicolumn{1}{l}{PN} & \multicolumn{2}{|c|}{Se$^+$} & \multicolumn{2}{c|}{Se$^{2+}$} & \multicolumn{2}{c|}{Se$^{3+}$} & \multicolumn{2}{c|}{Se$^{4+}$} & \multicolumn{2}{c|}{Se$^{5+}$} \\
\cline{2-3} \cline{4-5} \cline{6-7} \cline{8-9} \cline{10-11}
\multicolumn{1}{l}{Name} & \multicolumn{1}{|c}{This work} & \multicolumn{1}{c}{Paper I} & \multicolumn{1}{|c}{This work} & \multicolumn{1}{c}{Paper I} & \multicolumn{1}{|c}{This work} & \multicolumn{1}{c}{Paper I} & \multicolumn{1}{|c}{This work} & \multicolumn{1}{c}{Paper I} & \multicolumn{1}{|c}{This work} & \multicolumn{1}{c|}{Paper I}}
\startdata
IC 418   & 0.34 & 0.20 & 0.57 & 0.67 & 0.09 & 0.12 & \nodata & \nodata & \nodata & \nodata \\
IC 5117  & 0.11 & 0.13 & 0.20 & 0.20 & 0.58 & 0.40 & 0.11 & 0.26 & \nodata & 0.01 \\
NGC 6572 & 0.17 & 0.14 & 0.27 & 0.24 & 0.54 & 0.45 & 0.02 & 0.17 & \nodata & \nodata \\
NGC 6741 & 0.28 & 0.29 & 0.19 & 0.24 & 0.24 & 0.19 & 0.26 & 0.14 & 0.02 & 0.09 \\
NGC 6790 & 0.08 & 0.10 & 0.17 & 0.17 & 0.64 & 0.41 & 0.10 & 0.32 & \nodata & \nodata \\
NGC 6826 & 0.08 & \nodata & 0.39 & 0.30 & 0.52 & 0.61 & \nodata & 0.08 & \nodata & \nodata \\
NGC 6884 & 0.05 & 0.01 & 0.24 & 0.24 & 0.58 & 0.50 & 0.12 & 0.22 & \nodata & 0.03 \\
NGC 6886 & 0.11 & 0.01 & 0.18 & 0.14 & 0.29 & 0.31 & 0.36 & 0.27 & 0.04 & 0.18 \\
NGC 7027 & 0.11 & 0.12 & 0.14 & 0.17 & 0.28 & 0.22 & 0.23 & 0.15 & 0.11 & 0.17 \\
NGC 7662 & 0.02 & \nodata & 0.13 & 0.12 & 0.44 & 0.50 & 0.40 & 0.31 & \nodata & 0.06 \\
\tableline
\enddata
\label{se_ions}
\tablecomments{Comparison of Se ionic fractions between this work and that of \citetalias{sterling07}.  Ionic fractions less than 1\% are omitted.}
\end{deluxetable}

Some of the PNe (NGC~6741, NGC~6826, NGC~6886, and NGC~7662) do not show the expected shift of Se$^{4+}$ from our previous models to Se$^{3+}$ in those we present here.  This is largely due to the different model input parameters that we adopt for these objects.  For example, we utilize a lower $T_{\rm eff}$ for the central stars of NGC~6741 and NGC~6826, shifting the Se ionization balance to lower charge states (the larger Se$^{4+}$ fraction in NGC~6741 is due to a lower amount of higher-charge states than the previous model).  The differing nebular radii of the NGC~6886 and NGC~7662 models from those of \citetalias{sterling07} account for the Se ionic fractions in those objects.  The larger outer radius of the current NGC~6886 model shifts the Se ionization balance to lower charge states, while the smaller outer radius of the NGC~7662 model has the opposite effect (leading to a larger Se$^{4+}$ fraction).  Finally, in our new IC~418 model we note that the much larger Se$^{2+}$ to Se$^+$ DR rate coefficient produces a larger Se$^+$/Se$^{2+}$ ratio than found in \citetalias{sterling07}.

\subsection{Remaining Atomic Data Needs}

The imperfect agreement between the modeled and observed Kr ionization balance demonstrates that improvements can be made to the atomic data for Se and Kr.  Experimental PI cross sections of these ions \citep{lu06a, lu06b, lu_thesis, bizau11, sterling11a, esteves11, esteves12, hinojosa12} are invaluable for benchmarking PI cross section calculations, particularly for complex systems such as near-neutral \emph{n}-capture elements.  Comparison of the experimental and calculated PI cross sections suggest that more computationally-intensive calculations --- such as R-matrix and/or fully relativistic treatments --- would be valuable for Kr$^{3+}$.

Low-temperature DR cannot be accurately treated theoretically without knowledge of the near-threshold autoionizing state energies.  Unfortunately these energies have not been measured experimentally for any element beyond the second row of the Periodic Table.  We emphasize the importance of experimental DR rate coefficient measurements conducted at facilities such as the Test Storage Ring \citep{savin99, savin06, schippers10} and particularly the Cryogenic Storage Ring \citep{wolf06}, which can be used to measure DR rate coefficients for low charge-to-mass species observed in photoionized plasmas.

Finally, CT rate coefficients derived through the Landau-Zener or Demkov approximations \citep{bd80b, demkov64, swartz94, sterling11d} are accurate to within a factor of three for systems with large rate coefficients, but can have larger errors in systems with small rate coefficients.  Detailed quantal calculations \citep[e.g.,][]{kimura89, zygelman92, wang04} can significantly improve the accuracy of these data, but are time- and computationally-intensive.  In addition, we have not considered CT reactions with He, which can also affect the ionization balance.

In a forthcoming paper, we will conduct a Monte~Carlo analysis of the sensitivity of Se and Kr abundance determinations to atomic data uncertainties, which will highlight the systems and processes that most urgently require further theoretical and/or experimental exploration.  That study will help to prioritize atomic data needs for these two elements, which is critical given the intensive nature of the theoretical methods and experimental measurements discussed.

\section{SELENIUM AND KRYPTON IONIZATION CORRECTION FACTORS} \label{icfs}

In this section, we derive analytical ICF formulae that can be used to estimate the abundances of unobserved Se and Kr ions.  We consider detected (or detectable) Se and Kr ions in the optical and $K$~band portions of the spectrum.

We intend for these corrections to be widely applicable, and therefore we have constructed a large grid of Cloudy models whose properties span a wide range of central star temperatures and luminosities, as well as nebular densities.  Specifically, we consider central star temperatures $T_{\mathrm{eff}}$ ranging from 50,000~K to 190,000~K (the range of temperatures of the Rauch~2003 grid), with step sizes of 5,000~K.  Stellar gravities were assumed to be $\mathrm{log}$~$ g=5.5$ in all models.  Hydrogen densities $n_{\mathrm{H}}=10^3$--10$^5$~cm$^{-3}$ are considered, with step sizes of 3,000~cm$^{-3}$.  Central star luminosities are controlled through the ionization parameter $U=Q(\mathrm{H})/(4\pi R_{\mathrm{in}}^2 n_{\mathrm{H}} c)$, where $Q$(H) is the number of hydrogen-ionizing photons emitted per second, $R_{\mathrm{in}}$ is the inner radius of the nebula, and $c$ is the speed of light.  Given the hydrogen density and inner radius, $U$ is directly related to the ionizing luminosity of the central star.  In our grids, we vary log($U$) from $-3.0$ to 2.0, in step sizes of 0.5.  A model is computed for each combination of $T_{\mathrm{eff}}$, $n_{\mathrm{H}}$, and log($U$), for a total of 10,471 individual models in the grid.

Each model is assumed to have spherical geometry and uniform density, with an inner radius of 10$^{-2.5}$~pc.  The models are stopped when the electron temperature decreases to 4,000~K.  We adopt a solar composition \citep{asplund09} with O-rich dust grains \citep[the ``grain Orion silicate'' command, with grain properties from][]{vanhoof04}.  In \S \ref{robust}, we show that our derived ICF formulae are valid over a wide range of nebular compositions, metallicity, and assumptions regarding dust composition.

We derive ICFs based on correlations between the Se and Kr ionic fractions and those of routinely detected species in the optical spectra of PNe.  For each model we extracted the fractional abundance of all He, N, O, and Ne ions, and the first six ions of S, Cl, Ar, Se, and Kr.  Combining the optical and $K$~band spectra of PNe, it is possible to detect transitions of $[$\ion{Se}{3}$]$, $[$\ion{Se}{4}$]$, $[$\ion{Kr}{3}$]$, $[$\ion{Kr}{4}$]$, and $[$\ion{Kr}{5}$]$.  We searched for correlations between the fractional abundances of these species (and combinations thereof) and ionic fractions of He, N, O, Ne, S, Cl, and Ar.  

\begin{figure}
\epsscale{0.8}
\plotone{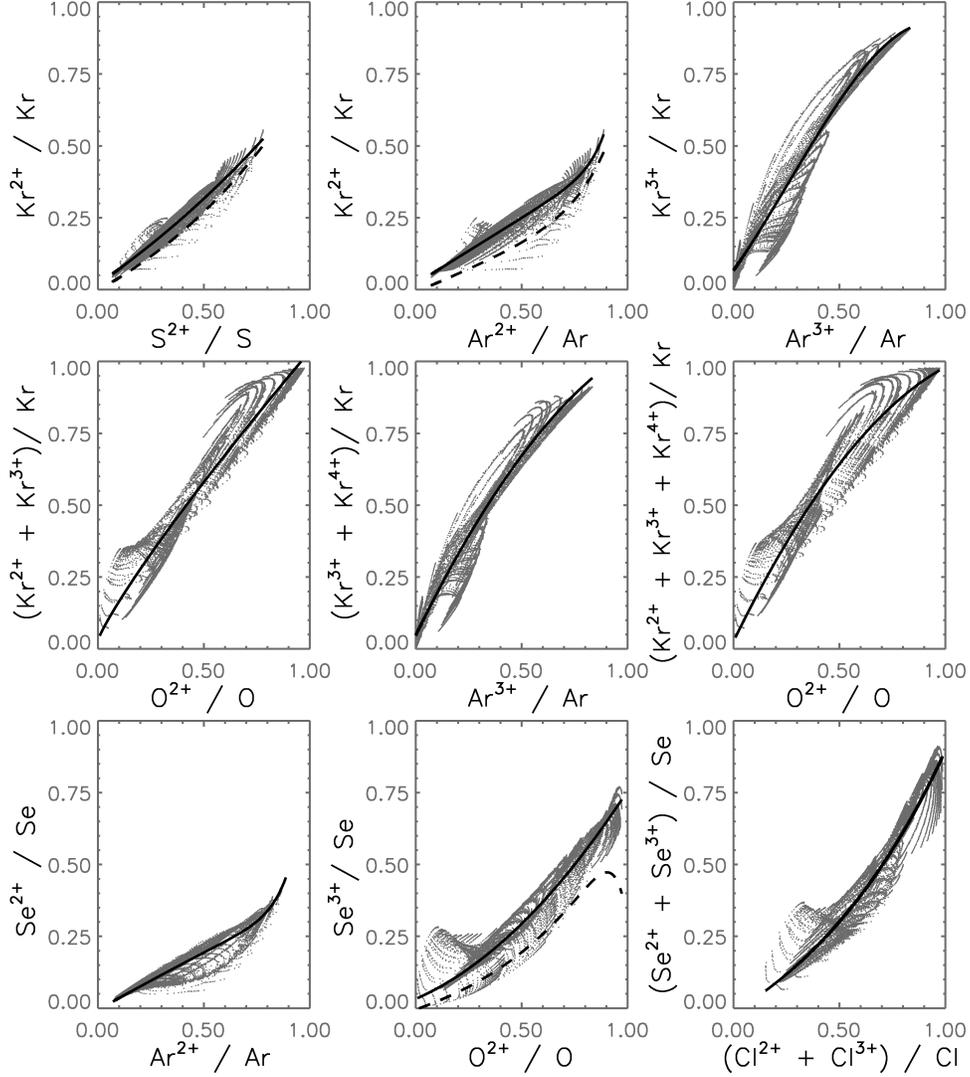} 
\caption{The strongest correlations between the fractional abundances of observable Se and Kr ions and those of commonly detected light element ions are plotted.  These data are from a grid of models with solar abundances and oxygen-rich dust grains.  Each gray dot represents the ionic fractions from one of the 10,471 models in our grid.  Analytical fits to these correlations (see text) are shown as solid black curves, while fits from \citetalias{sterling07} are shown as dashed curves if the correlation is between the same ions.  Families of curves are seen in some of the correlations.  Each of these curves generally corresponds to a single value of $T_{\rm eff}$, with variations in  $n_{\rm H}$ and $U$ producing the changes along the curve.}
\label{icfs_solar}
\end{figure}

Figure~\ref{icfs_solar} depicts the strongest correlations that we found for Se and Kr ionic fractions.  Each gray dot corresponds to the ionic fractions from a single Cloudy model.  In some of the correlation plots, families of curves can be distinguished.  These curves usually correspond to models with the same central star temperature, with the ionic fractions changing with the density and ionization parameter.

To determine ICFs, we fit each correlation in Fig.~\ref{icfs_solar} with an analytical function, using a $\chi^2$ minimization routine written in IDL.  We considered polynomial, exponential, and logarithmic functional forms, attempting several types of functions (and adjusting their orders) to provide the best fit to each correlation.  The fits are displayed as solid curves in the panels of Fig.~\ref{icfs_solar}.  The inverses of these functions can be utilized as ICF formulae to correct for the abundances of unobserved Se and Kr ions in PNe.

As in \citetalias{sterling07}, we find that the Kr$^{2+}$ fractional abundance correlates strongly with the S$^{2+}$ and Ar$^{2+}$ ionic fractions (top left and middle panels of Fig.~\ref{icfs_solar}).  We derive the following expressions for the ICFs:
\begin{equation}
\mathrm{ICF}(\mathrm{Kr}) = \mathrm{Kr}/ \mathrm{Kr}^{2+} = (-0.8774 + 0.8963e^{0.5735x})^{-1}, \label{kr2s2}
\end{equation}
\begin{displaymath}
x = \mathrm{S}^{2+}/\mathrm{S};
\end{displaymath}
and
\begin{equation}
\mathrm{ICF}(\mathrm{Kr}) = \mathrm{Kr}/ \mathrm{Kr}^{2+} = (0.02038 + 0.4577x + 4.752\times10^{-7}e^{13.96x})^{-1}, \label{kr2ar2}
\end{equation}
\begin{displaymath}
x = \mathrm{Ar}^{2+}/\mathrm{Ar}.
\end{displaymath}

Equation~\ref{kr2ar2} is likely to result in the most reliable correction for unobserved Kr ions, given the ``sulfur anomaly'' observed in Galactic and Magellanic Cloud PNe \citep{henry04, shaw12}.  A variety of causes have been sought to explain the deficit of the S abundance in PNe relative to \ion{H}{2}~regions and stars, but the most likely explanation is the inability of current ICF schemes to accurately account for unobserved S ions \citep{henry12, shingles13}.  If this is correct, then S abundances in PNe are less accurate than those of Ar, which is the reason for our preference for Equation~\ref{kr2ar2}.

We display the fits from \citetalias{sterling07} as dashed lines in Figure~\ref{icfs_solar}.  \citetalias{sterling07} found lower Kr$^{2+}$ fractional abundances for given S$^{2+}$ and Ar$^{2+}$ ionic fractions, meaning that their (Kr/Kr$^{2+}$) ICFs systematically overestimate Kr elemental abundances in PNe.  This is in agreement with the higher Kr$^{2+}$ ionic fractions found in our models of individual PNe (\S3.2) compared to those of \citetalias{sterling07}.  In the case of Equation~\ref{kr2s2}, the disagreement with the corresponding ICF from \citetalias{sterling07} is less than 10\% at the highest values of S$^{2+}$/S ($>0.64$) in our grid, but reaches a factor of two at the lowest values.  Equation~\ref{kr2ar2} shows starker disagreement with the ICF from \citetalias{sterling07}: from $\sim25$\% for large Ar$^{2+}$/Ar values to an order of magnitude at the lowest fractional abundances.

The top right panel of Fig.~\ref{icfs_solar} shows the correlation used to derive ICFs when Kr$^{3+}$ is the only Kr ion detected, while the middle panels show correlations for combinations of Kr ions.  The corresponding ICF formulae are:
\begin{equation}
\mathrm{ICF}(\mathrm{Kr}) = \mathrm{Kr}/ \mathrm{Kr}^{3+} = (0.06681 + 1.05x + 0.7112x^2 - 0.907x^3)^{-1}, \label{kr3ar3}
\end{equation}
\begin{displaymath}
x = \mathrm{Ar}^{3+}/\mathrm{Ar};
\end{displaymath}
\begin{equation}
\mathrm{ICF}(\mathrm{Kr}) = \mathrm{Kr}/ (\mathrm{Kr}^{2+} + \mathrm{Kr}^{3+}) = (-2.648 + 1.047x^{0.8749} + 2.678e^{-0.01552x})^{-1}, \label{kr23o2}
\end{equation}
\begin{displaymath}
x = \mathrm{O}^{2+}/\mathrm{O};
\end{displaymath}
\begin{equation}
\mathrm{ICF}(\mathrm{Kr}) = \mathrm{Kr}/ (\mathrm{Kr}^{3+} + \mathrm{Kr}^{4+}) = (0.04747 + 1.512x - 0.5257x^2)^{-1}, \label{kr34ar3}
\end{equation}
\begin{displaymath}
x = \mathrm{Ar}^{3+}/\mathrm{Ar} \leq 0.932;
\end{displaymath}
and
\begin{equation}
\mathrm{ICF}(\mathrm{Kr}) = \mathrm{Kr}/ (\mathrm{Kr}^{2+} + \mathrm{Kr}^{3+} + \mathrm{Kr}^{4+}) = (0.02699 + 1.511x - 0.5571x^2)^{-1}, \label{kr234o2}
\end{equation}
\begin{displaymath}
x = \mathrm{O}^{2+}/\mathrm{O}.
\end{displaymath}
The upper limit to $x$~=~Ar$^{3+}$/Ar in Equation~\ref{kr34ar3} provides the range of validity for this ICF (above this value, the ICF is smaller than unity).  None of the models in our grid achieved such high fractional abundances of Ar$^{3+}$, and hence this equation should be generally applicable.  \citetalias{sterling07} used different reference ions than we utilize for the $(\mathrm{Kr}^{2+} + \mathrm{Kr}^{3+})$ ICF, and hence a direct comparison is not possible.

The bottom three panels of Fig.~\ref{icfs_solar} show correlations for Se ionic fractions, including that of Se$^{2+}$, whose abundance can in principle be derived from $[$\ion{Se}{3}$]$~$\lambda$8854.0.  The Se$^{2+}$ ionic abundance has not yet been reliably determined in PNe, due to the lack of collisional excitation data and because of blending with a weak \ion{He}{1} line at nearly the same wavelength \citep{pb94, sharpee07}.  From these correlations we find the following ICF formulae:
\begin{equation}
\mathrm{ICF}(\mathrm{Se}) = \mathrm{Se}/ \mathrm{Se}^{2+} = (2.411 + 0.5146x^{12.29} - 2.417e^{-0.1723x})^{-1}, \label{se2ar2}
\end{equation}
\begin{displaymath}
x = \mathrm{Ar}^{2+}/\mathrm{Ar} \geq 0.0344;
\end{displaymath}
\begin{equation}
\mathrm{ICF}(\mathrm{Se}) = \mathrm{Se}/ \mathrm{Se}^{3+} = (0.03238 + 0.3225x + 0.4013x^2)^{-1}, \label{se3o2}
\end{equation}
\begin{displaymath}
x = \mathrm{O}^{2+}/\mathrm{O};
\end{displaymath}
and
\begin{equation}
\mathrm{ICF}(\mathrm{Se}) = \mathrm{Se}/ (\mathrm{Se}^{2+} + \mathrm{Se}^{3+}) = (-0.3703 + 0.3558e^{1.27x})^{-1}, \label{se23cl23}
\end{equation}
\begin{displaymath}
x = (\mathrm{Cl}^{2+} + \mathrm{Cl}^{3+}) / \mathrm{Cl} \geq 0.0315.
\end{displaymath}

The Se$^{3+}$ correlation shows the largest dispersion of these three.  The situation is exacerbated at low O$^{2+}$ fractional abundances ($\leq 0.25$), where the Se$^{3+}$ ionic fraction has a bimodal distribution in our models.  The bimodality of this correlation is a result of variations in the ionization parameter $U$.  For small values (log~($U$)~$\leq -1.5$), the ``upper branch'' tends to be populated, while models with larger log($U$) generally populate the ``lower branch.''

\citetalias{sterling07} also found O$^{2+}$ to be the best tracer of the Se$^{3+}$ fractional abundance.  Our models show that Se$^{3+}$ comprises a larger fraction of the Se abundance than found in \citetalias{sterling07}.  Equation~\ref{se3o2} produces ICFs smaller than the \citetalias{sterling07} values by 30--50\% when O$^{2+}$/O$>0.5$, and by more than an order of magnitude at the smallest O$^{2+}$/O values.

Originally, we planned to extend our grid of models to lower central star temperatures (down to $\sim$30,000~K).  Since the \citet{rauch03} grid of PN central star atmospheres does not extend to such low temperatures, we utilized LTE plane-parallel ATLAS models \citep{castelli04} for the range $T_{\mathrm{eff}}=30,000$~K to 50,000~K.  However, we found that the ionic fractions from these models do not follow the correlations depicted in Figure~\ref{icfs_solar}.  This suggests that either the ATLAS stellar atmosphere models do not accurately describe the ionizing radiation from low-temperature PN central stars, or that our ICFs break down for objects with sufficiently low central star temperatures.  Therefore, Equations~\ref{kr2s2}--\ref{se23cl23} should be used with caution for PNe with cool central stars or for \ion{H}{2} regions.  However, we note that the Kr abundance derived with these ICFs for IC~418, the only object in our sample with $T_{\rm eff}<$~50,000~K (see Table~\ref{params1}), agrees well with the model results (see \S \ref{icfcomp}).

\subsection{Robustness of the Derived ICFs} \label{robust}

The grid described in the previous section was recomputed with different nebular abundances and dust properties, in order to test the robustness of the derived Se and Kr ICFs for various compositions.

\begin{figure}
\epsscale{0.8}
\plotone{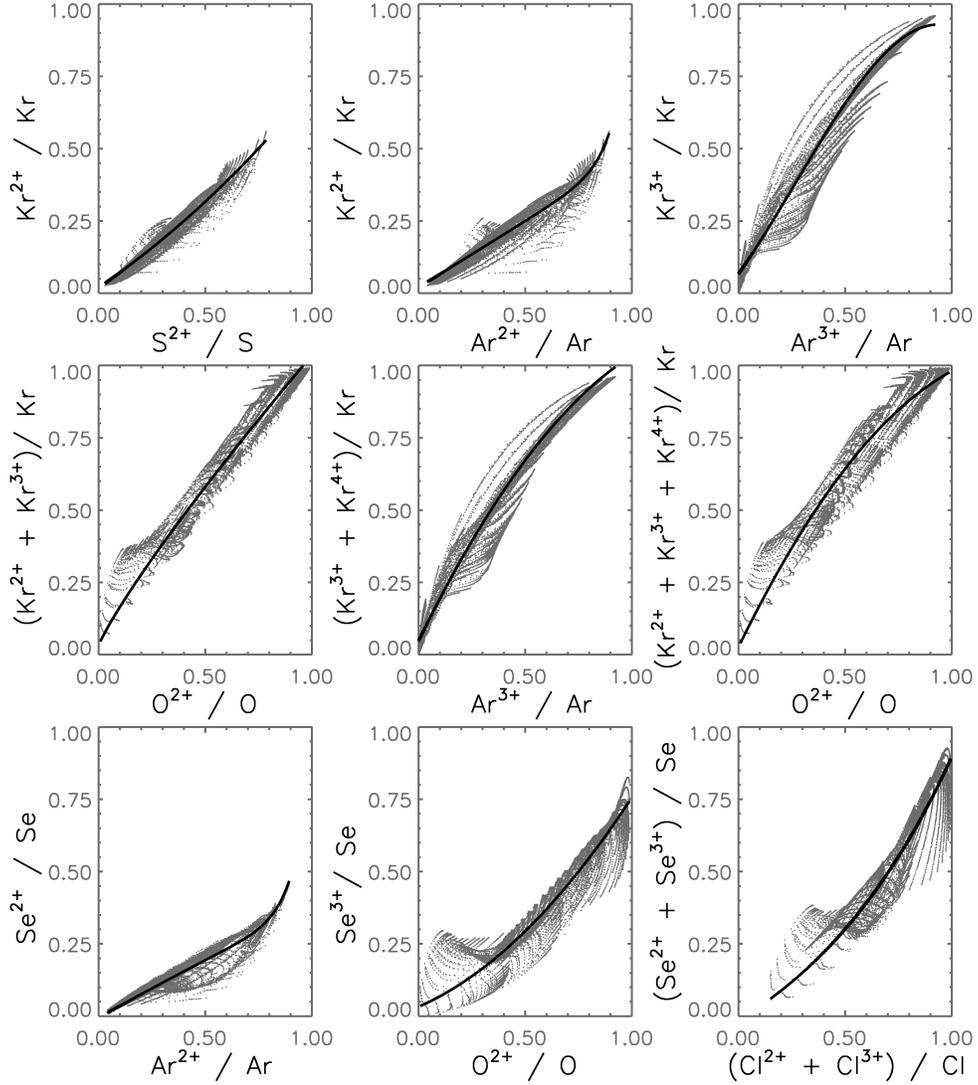}
\caption{The same as Figure~\ref{icfs_solar}, except in this grid dust is not included.  The analytical fits to the correlations from Figure~\ref{icfs_solar} are overplotted in each panel.} 
\label{icfs_solarnodust}
\end{figure}

In Figure~\ref{icfs_solarnodust}, we display the same correlations as shown in Figure~\ref{icfs_solar} for a grid of models that has a solar composition, but dust is not included.  There is not as pronounced a bimodality in the Se$^{3+}$ correlation at small O$^{2+}$ fractional abundances as seen in the grid including dust.  Specifically, the ``lower branch'' of this correlation, at which the Se$^{3+}$ fractional abundance is small compared to O$^{2+}$, is not as highly populated in the dust-free models.  That being said, Equation~\ref{se3o2} falls well within the dispersion of this correlation.  For other ions, the fits describe all of the correlations in the dust-free grid to good approximation, with a slightly larger dispersion in some cases.  We conclude that our ICFs are robust regardless of whether dust is included in the models.

\begin{figure}
\epsscale{0.8}
\plotone{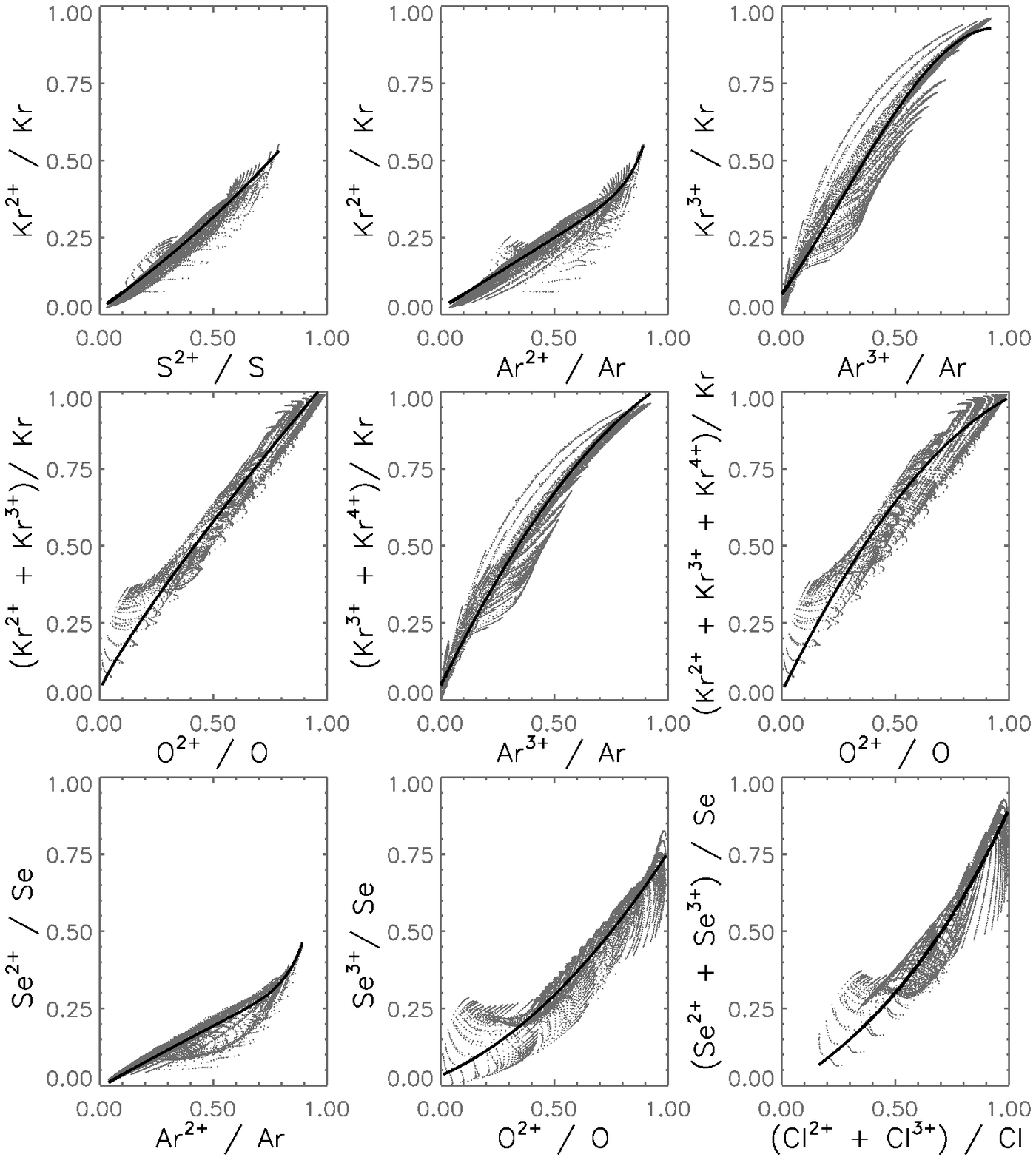}
\caption{The same as Figure~\ref{icfs_solar}, but from a grid of models with the default Cloudy PN abundances \citep{ac83}, in which C/O~$>1$, and no dust.  The analytical fits to the correlations from Figure~\ref{icfs_solar} are overplotted in each panel.  The correlations show excellent agreement with those seen in the dust-free solar composition models (Figure~\ref{icfs_solarnodust}), indicating that the ICFs do not depend significantly on the carbon/oxygen chemistry.  Similarly, results for C-rich nebular compositions with C-rich dust grains agree well with solar composition models with O-rich dust (Figure~\ref{icfs_solar}).} 
\label{icfs_default}
\end{figure}

The correlations are shown in Figure~\ref{icfs_default} for a grid with Cloudy's default (C-rich) PN abundances \citep{ac83} that does not include dust.  The correlations are virtually identical to those found in the dust-free solar composition grid, indicating that the ICFs do not depend strongly on nebular composition at near-solar metallicities.  Similarly, we find that models with Cloudy's PN abundances that include carbon-rich dust grains \citep[the ``grain Orion graphite'' command, where the grain properties are taken from][]{baldwin91} agree well with the correlations depicted in Figure~\ref{icfs_solar}.

\begin{figure}
\epsscale{0.8}
\plotone{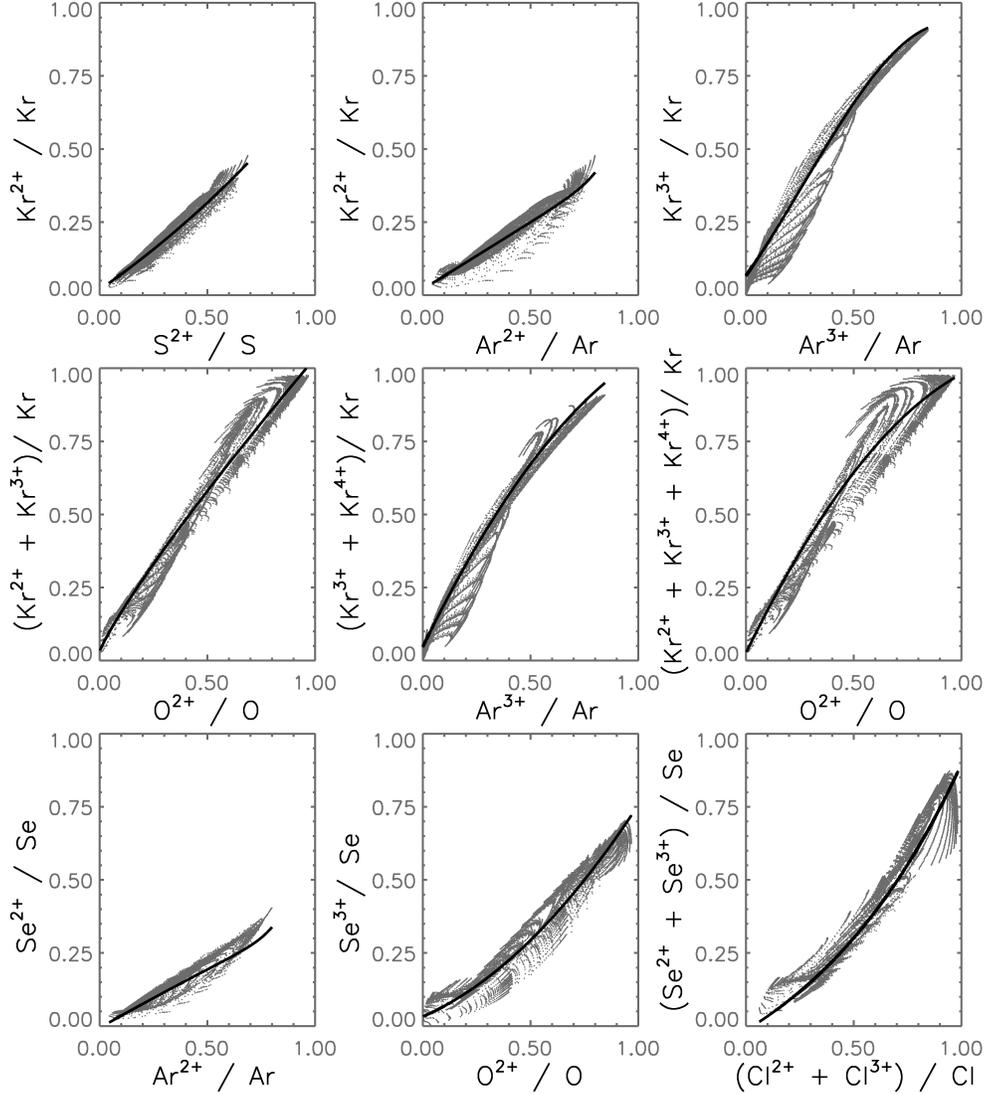}
\caption{The same as Figure~\ref{icfs_solar}, but with metal abundances reduced by a factor of ten from a solar composition.  The analytical fits to the correlations from Figure~\ref{icfs_solar} are overplotted in each panel.  Small discrepancies with the fits derived from higher-metallicity grids are apparent for some of the correlations in the middle and bottom panels, though the Kr$^{2+}$, Kr$^{3+}$, and Se$^{3+}$ correlations appear to be robust at various metallicities.  The plotted ionic fractions are similar to those found in a grid with metal abundances reduced by two orders of magnitude.} 
\label{icfs_z0.1}
\end{figure}

At lower metallicities, Equations~\ref{kr2s2}--\ref{se23cl23} also describe the ionic fraction correlations well.  Figure~\ref{icfs_z0.1} shows the fractional abundance plots of Figure~\ref{icfs_solar} for a scaled solar composition (without dust) in which the abundances of all elements heavier than He have been reduced by a factor of ten.  NLTE stellar atmosphere models with halo abundances \citep{rauch03} were used to describe the central star radiation in this grid.  The correlations in the top panels, for single Kr ions, show excellent agreement with the solar metallicity grid.  For the correlations depicted in the middle panels, the fits tend to slightly overestimate the fractional Kr abundances when the sum of the Kr$^{2+}$ and Kr$^{3+}$ ionic fractions is small ($\leq0.1$--0.5, depending on the correlation).  This will cause elemental Kr abundances derived from multiple ions to be slightly underestimated for some low-metallicity PNe.  The Se$^{2+}$ fractional abundance can be underestimated by Equation~\ref{se2ar2} at low metallicity when Ar$^{2+}$/Ar~$>0.5$, which would lead to overestimated Se abundances.

The correlations shown in Figure~\ref{icfs_z0.1} are very similar to those found from a grid of models with metal abundances reduced to 1\% solar (not shown), though the discrepancies with the solar metallicity models described above are slightly more pronounced.  Likewise, grids with supersolar abundances (metal abundances increased by 50\% from the solar values) show negligible deviation from the solar abundance models with O-rich dust shown in Figure~\ref{icfs_solar}.

These tests show that our derived ICFs exhibit only a slight dependence on dust physics and metallicity.  The ICFs for Kr$^{2+}$, Kr$^{3+}$, and Se$^{2+}$ appear to be quite robust, while those for combinations of Kr ions show discrepancies only in low-metallicity objects when their sum represents a small fraction of the total Kr abundance.  We conclude that our analytical ICFs are valid for a wide range of nebular compositions, with errors becoming significant only for low-metallicity, low-ionization nebulae.

\subsection{Comparison of ICF- and Model-Derived Se and Kr Abundances} \label{icfcomp}

We have computed empirical Se and Kr abundances from the optical and near-infrared spectra of the 15 PNe that we individually modeled (Table~\ref{pn_sample}), using all possible ICFs (Equations~\ref{kr2s2}--\ref{se23cl23}).  The O, S, and Ar ionic fractions needed for the ICFs are taken from the references listed in Table~\ref{pn_sample}.  In cases where error bars were not provided for the ionic and/or elemental O, S, and Ar abundances, we assume a 30\% uncertainty in the ionic fractions.

Kr ionic abundances were determined using a five-level model atom calculation with effective collision strengths computed by \citet{schoning97} and transition probabilities from \citet{biemont86b, biemont86a}.  Electron temperatures and densities were adopted from the references in Table~\ref{pn_sample}, assuming uncertainties of 1000~K in $T_{\rm e}$ and 20\% in $n_{\rm e}$ if values were not provided in those references.  The Kr ionic abundances for each of the 15 modeled PNe are given in Table~\ref{krionab}, where the error bars take into account uncertainties in line intensities (assumed to be 25\%), electron temperature, and electron density.

Ionic Kr abundances derived from different lines of the same emitting species agree within the uncertainties with the exception of Kr$^{2+}$ in IC~418 (perhaps due to aperture differences in the optical and near-infrared spectra, or incomplete removal of telluric OH contamination in $\lambda$6826.7).  \citetalias{sterling07} discussed the identities of optical and near-infrared Se and Kr emission lines, refuting previously proposed alternative identifications.  The good agreement in ionic abundances derived from different emission lines of the same Kr ion provides further evidence that these lines indeed arise from Kr fine-structure transitions.  In particular, note the excellent agreement in Kr$^{4+}$ abundances derived from the $\lambda$6256.1 and $\lambda$8243.4 features in NGC~7027.  These two lines arise from the same upper level, and therefore are independent of temperature and density.  The relative intensities validate the accuracy of the transition probabilities of \citet{biemont86b}, with the small difference likely due to uncertainties in the line fluxes.  To our knowledge, the detection of $[$\ion{Kr}{5}$]$~$\lambda$8243.4 by \citet{sharpee07} is the first in a high-resolution PN spectrum.  

\begin{deluxetable}{l|cc|cc|cc|}
\tablecolumns{7}
\tablewidth{0pc} 
\tablecaption{Empirical Kr Ionic Abundances in the Modeled PNe} 
\tablehead{
\colhead{PN} & \multicolumn{2}{|c|}{Kr$^{2+}$/H$^+$ $\times 10^9$} & \multicolumn{2}{c|}{Kr$^{3+}$/H$^+$ $\times 10^9$} &  \multicolumn{2}{c|}{Kr$^{4+}$/H$^+$ $\times 10^9$} \\
\colhead{Name} & \multicolumn{1}{|c}{$\lambda$6826} & \multicolumn{1}{c|}{$\lambda$2.199~$\mu$m} & \multicolumn{1}{c}{$\lambda$5346} & \multicolumn{1}{c|}{$\lambda$5868} & \multicolumn{1}{c}{$\lambda$6256} & \multicolumn{1}{c|}{$\lambda$8243}}
\startdata
IC 418   & 3.96$\pm$0.88   & 2.89$\pm$0.05 & 0.35$\pm$0.10 & 0.24$\pm$0.07 & \nodata       & \nodata       \\
IC 2501  & 0.49$\pm$0.14   & \nodata       & 0.53$\pm$0.16 & 0.50$\pm$0.15 & \nodata       & \nodata       \\
IC 4191  & 0.14$\pm$0.04   & \nodata       & 1.80$\pm$0.62 & 0.98$\pm$0.29 & \nodata       & \nodata       \\
IC 5117  & \nodata         & 1.35$\pm$0.34 & 5.92$\pm$1.50 & 4.90$\pm$1.19 & \nodata       & \nodata       \\
NGC 2440 & \nodata         & \nodata       & 0.91$\pm$0.20 & 0.88$\pm$0.23 & 0.25$\pm$0.05 & \nodata       \\
NGC 5315 & 0.90$\pm$0.26   & \nodata       & 1.19$\pm$0.33 & \nodata       & \nodata       & \nodata       \\
NGC 6369 & 1.39$\pm$0.38   & \nodata       & 4.15$\pm$1.24 & 4.13$\pm$1.21 & \nodata       & \nodata       \\
NGC 6572 & 1.45$\pm$0.55   & 0.88$\pm$0.21 & 2.89$\pm$0.93 & 2.74$\pm$0.97 & \nodata       & \nodata       \\
NGC 6741 & 2.03$\pm$0.35   & 1.57$\pm$0.40 & 2.60$\pm$0.74 & 1.73$\pm$0.56 & \nodata       & \nodata       \\
NGC 6790 & 0.64$\pm$0.09   & \nodata       & 1.99$\pm$0.56 & 1.36$\pm$0.32 & \nodata       & \nodata       \\
NGC 6826 & 1.77$\pm$0.43   & \nodata       & 2.05$\pm$0.75 & 2.77$\pm$0.69 & \nodata       & \nodata       \\
NGC 6884 & 1.12$\pm$0.24   & \nodata       & 5.22$\pm$1.62 & 5.07$\pm$0.96 & \nodata       & \nodata       \\
NGC 6886 & \nodata         & 1.80$\pm$0.37 & 3.45$\pm$1.01 & 2.77$\pm$0.79 & \nodata       & \nodata       \\
NGC 7027 & 2.55$\pm$0.67   & 2.36$\pm$0.41 & 5.68$\pm$1.28 & 5.43$\pm$1.21 & 1.07$\pm$0.29 & 0.98$\pm$0.27 \\
NGC 7662 & 0.63$\pm$0.13   & \nodata       & 5.16$\pm$1.42 & 4.63$\pm$0.63 & \nodata       & \nodata       \\
\tableline
\enddata
\label{krionab}
\tablecomments{Comparison of Kr ionic abundances derived from optical and near-infrared emission lines.  See Table~\ref{pn_sample} for references to the spectroscopic data utilized.}
\end{deluxetable}

We converted the ionic abundances to elemental Kr abundances using all of the relevant ICFs (Table~\ref{krcomp}).  The Kr abundances found from our Cloudy models of each object are given in the last column of the table.  Some variation is expected in the abundances derived with different ICFs, due to observational uncertainties (e.g., line fluxes, electron temperatures, and extinction coefficients).  In addition, the abundances will generally be more accurate (i.e., the ICFs smaller) when more ions are detected.  The empirical Kr abundances derived using the maximum number of detected Kr ions (Kr$^{2+}$ and Kr$^{3+}$ for most objects, with the addition of Kr$^{4+}$ in NGC~7027 and NGC~2440) agree to within $\sim0.25$~dex -- and often better -- with the modeled Kr abundances.  We refer to the Kr abundances derived from the maximum number of detected Kr ions, using Equations~\ref{kr23o2}--\ref{kr234o2}, as the ``best empirical'' Kr abundances since the corresponding ICFs are the smallest and least sensitive to uncertainties.  Elemental abundances derived from a single Kr ion show more scatter (Table~\ref{krcomp}), as expected, though in many cases there is good agreement with the best empirical values.

\begin{deluxetable}{lccccccc}
\rotate
\tablecolumns{8}
\tablewidth{0pc} 
\tabletypesize{\scriptsize}
\tablecaption{Comparison of ICF- and Model-Derived Kr Abundances} 
\tablehead{
\colhead{PN} & \colhead{$12+$ Log(Kr/H)} & \colhead{$12+$ Log(Kr/H)} & \colhead{$12+$ Log(Kr/H)} & \colhead{$12+$ Log(Kr/H)} & \colhead{$12+$ Log(Kr/H)} & \colhead{$12+$ Log(Kr/H)} & \colhead{$12+$ Log(Kr/H)}\\
\colhead{Name} & \colhead{ICF Eqn.\ref{kr2ar2}} & \colhead{ICF Eqn.\ref{kr2s2}} & \colhead{ICF Eqn.\ref{kr3ar3}} & \colhead{ICF Eqn.\ref{kr23o2}} & \colhead{ICF Eqn.\ref{kr34ar3}} & \colhead{ICF Eqn.\ref{kr234o2}} & \colhead{Model}}
\startdata
IC 418   & 4.20$\pm$0.15 & 3.87$\pm$0.20 & \nodata\tablenotemark{a}       & 4.06$\pm$0.18 & \nodata       & \nodata       & 3.93$\pm$0.23 \\
IC 2501  & 3.01$\pm$0.33 & 3.02$\pm$0.24 & 3.55$\pm$0.12 & 3.02$\pm$0.09 & \nodata       & \nodata       & 3.12$\pm$0.13 \\
IC 4191  & 2.83$\pm$0.19 & 2.81$\pm$0.23 & 3.53$\pm$0.17 & 3.20$\pm$0.09 & \nodata       & \nodata       & 3.22$\pm$0.30 \\
IC 5117  & 3.64$\pm$0.21 & 3.92$\pm$0.22 & 4.07$\pm$0.15 & 3.90$\pm$0.12 & \nodata       & \nodata       & 3.74$\pm$0.10 \\
NGC 2440 & \nodata       & \nodata       & 3.44$\pm$0.18 & \nodata       & 3.44$\pm$0.11 & \nodata       & 3.44$\pm$0.10 \\
NGC 5315 & 3.37$\pm$0.30 & 3.24$\pm$0.13 & 3.68$\pm$0.17 & 3.36$\pm$0.10 & \nodata       & \nodata       & 3.42$\pm$0.28 \\
NGC 6369 & 3.54$\pm$0.31 & 3.73$\pm$0.23 & 4.14$\pm$0.15 & 3.77$\pm$0.10 & \nodata       & \nodata       & 3.94$\pm$0.18 \\
NGC 6572 & 3.48$\pm$0.29 & 3.66$\pm$0.22 & 3.99$\pm$0.16 & 3.60$\pm$0.08 & \nodata       & \nodata       & 3.51$\pm$0.15 \\
NGC 6741 & 3.78$\pm$0.21 & 3.81$\pm$0.23 & 3.81$\pm$0.16 & 3.95$\pm$0.19 & \nodata       & \nodata       & 3.69$\pm$0.20 \\
NGC 6790 & 3.35$\pm$0.16 & 3.42$\pm$0.19 & 3.50$\pm$0.15 & 3.37$\pm$0.08 & \nodata       & \nodata       & 3.34$\pm$0.10 \\
NGC 6826 & 3.53$\pm$0.34 & 3.61$\pm$0.23 & 4.16$\pm$0.13 & 3.64$\pm$0.09 & \nodata       & \nodata       & 3.63$\pm$0.11 \\
NGC 6884 & 3.61$\pm$0.17 & 3.86$\pm$0.20 & 3.98$\pm$0.16 & 3.82$\pm$0.10 & \nodata       & \nodata       & 3.69$\pm$0.15 \\
NGC 6886 & 3.82$\pm$0.20 & 3.94$\pm$0.16 & 3.94$\pm$0.16 & 3.82$\pm$0.20 & \nodata       & \nodata       & 3.70$\pm$0.20 \\
NGC 7027 & 3.94$\pm$0.16 & 4.10$\pm$0.20 & 4.02$\pm$0.15 & 3.97$\pm$0.20 & 4.09$\pm$0.11 & 4.01$\pm$0.10 & 4.14$\pm$0.11 \\
NGC 7662 & 3.49$\pm$0.16 & 4.03$\pm$0.21 & 3.87$\pm$0.15 & 3.96$\pm$0.19 & \nodata       & \nodata       & 3.74$\pm$0.11 \\
\cline{1-8}
\multicolumn{1}{c}{PN} & \multicolumn{1}{c}{ICF} & \multicolumn{1}{c}{ICF} & \multicolumn{1}{c}{ICF} & \multicolumn{1}{c}{ICF} & \multicolumn{1}{c}{ICF} & \multicolumn{1}{c}{ICF} & \multicolumn{1}{c}{}\\
\multicolumn{1}{c}{Name} & \multicolumn{1}{c}{Eqn.\ref{kr2ar2}}  & \multicolumn{1}{c}{Eqn.\ref{kr2s2}} & \multicolumn{1}{c}{Eqn.\ref{kr3ar3}} & \multicolumn{1}{c}{Eqn.\ref{kr23o2}} & \multicolumn{1}{c}{Eqn.\ref{kr34ar3}} & \multicolumn{1}{c}{Eqn.\ref{kr234o2}} & \multicolumn{1}{c}{} \\
\cline{1-8}
IC 418   & 4.60$\pm$1.36 & 2.14$\pm$0.18 & \nodata       & 3.12$\pm$1.22 & \nodata       & \nodata       & \\
IC 2501  & 2.10$\pm$1.21 & 3.02$\pm$0.24 & 6.97$\pm$1.19 & 1.04$\pm$0.22 & \nodata       & \nodata       & \\
IC 4191  & 4.80$\pm$1.41 & 4.63$\pm$1.79 & 2.44$\pm$0.69 & 1.05$\pm$0.22 & \nodata       & \nodata       & \\
IC 5117  & 3.21$\pm$1.11 & 6.21$\pm$2.37 & 2.18$\pm$0.62 & 1.17$\pm$0.30 & \nodata       & \nodata       & \\
NGC 2440 & \nodata       & \nodata       & 3.10$\pm$0.83 & \nodata       & 2.82$\pm$0.72 & \nodata       & \\
NGC 5315 & 2.58$\pm$1.36 & 1.92$\pm$0.07 & 4.01$\pm$0.98 & 1.09$\pm$0.25 & \nodata       & \nodata       & \\
NGC 6369 & 2.48$\pm$1.36 & 3.83$\pm$1.50 & 3.36$\pm$0.88 & 1.08$\pm$0.24 & \nodata       & \nodata       & \\
NGC 6572 & 2.58$\pm$1.36 & 3.93$\pm$1.53 & 3.50$\pm$0.90 & 1.00$\pm$0.19 & \nodata       & \nodata       & \\
NGC 6741 & 3.37$\pm$1.11 & 3.56$\pm$1.40 & 2.99$\pm$0.81 & 2.26$\pm$0.92 & \nodata       & \nodata       & \\
NGC 6790 & 3.54$\pm$1.13 & 4.14$\pm$1.61 & 1.88$\pm$0.54 & 1.00$\pm$0.19 & \nodata       & \nodata       & \\
NGC 6826 & 1.90$\pm$1.15 & 2.29$\pm$0.94 & 6.01$\pm$1.16 & 1.04$\pm$0.22 & \nodata       & \nodata       & \\
NGC 6884 & 3.66$\pm$1.14 & 6.45$\pm$2.46 & 1.84$\pm$0.52 & 1.07$\pm$0.24 & \nodata       & \nodata       & \\
NGC 6886 & 3.60$\pm$1.13 & 5.23$\pm$2.01 & 2.78$\pm$0.76 & 1.34$\pm$0.57 & \nodata       & \nodata       & \\
NGC 7027 & 4.00$\pm$1.22 & 5.78$\pm$2.21 & 2.24$\pm$0.64 & 1.35$\pm$0.58 & 2.09$\pm$0.54 & 1.27$\pm$0.27 & \\
NGC 7662 & 4.91$\pm$1.44 & 16.9$\pm$6.8  & 1.52$\pm$0.41 & 1.64$\pm$0.69 & \nodata       & \nodata       & \\
\tableline
\enddata
\label{krcomp}
\tablecomments{Comparison of Kr elemental abundances (\textit{top}) derived with different ICFs and from Cloudy models.  The solar Kr abundance is $12$~+~Log(Kr/H)~=~$3.26\pm0.06$ \citep{asplund09}.  The values of the ICFs are given in the lower portion of the table.}
\tablenotetext{a}{Equation (3) is not used to determine the Kr abundance of IC~418, since Ar$^{3+}$ was not detected in the optical spectrum used by \citet{pottasch04}.}
\end{deluxetable}

\begin{figure}
\epsscale{0.8}
\plotone{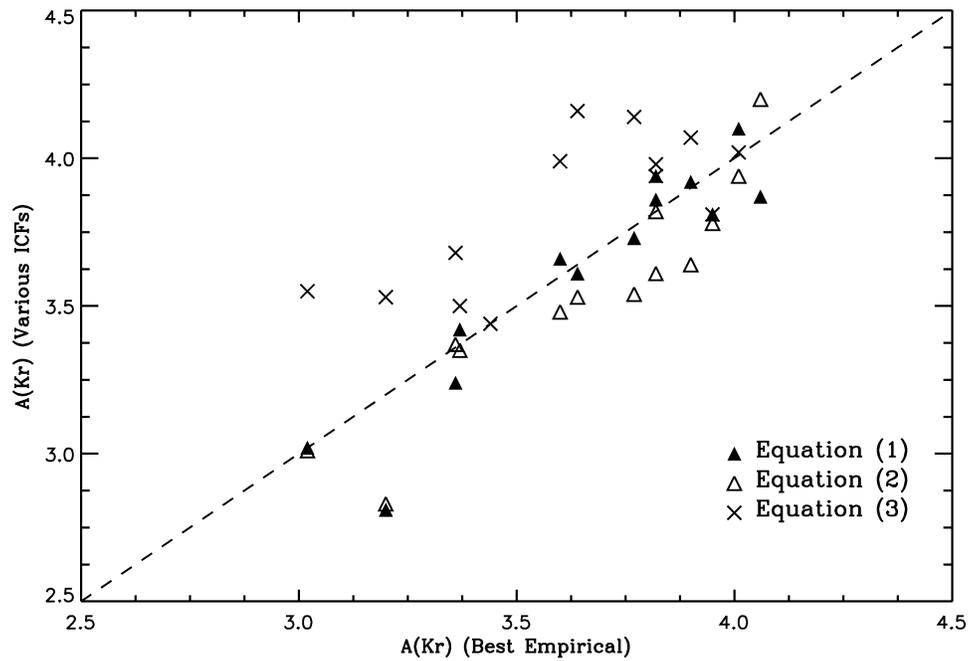}
\caption{A comparison of the Kr abundances derived from Kr$^{2+}$ lines (Equations~\ref{kr2s2} and \ref{kr2ar2}; filled and open triangles respectively) and from Kr$^{3+}$ lines (Equation~\ref{kr3ar3}; $\times$ symbols) with the best empirical Kr abundances (derived from all detected Kr ions).  The dashed line denotes perfect agreement, and is not a fit to the data.} 
\label{kricf_fig}
\end{figure}

In Figure~\ref{kricf_fig}, we plot the Kr abundances determined from single Kr ions (using Equations~\ref{kr2s2} and \ref{kr2ar2} for Kr$^{2+}$, and Equation~\ref{kr3ar3} for Kr$^{3+}$) against the best empirical Kr abundance.  Abundances derived with Equation~\ref{kr2s2} agree very well with the best empirical values.  This is surprising in light of the uncertainty regarding total elemental S abundances in PNe \citep{henry04, henry12}, which can affect the value of the ICF.  Equation~\ref{kr2ar2} underestimates the gaseous Kr abundance in some of the PNe, though the agreement with the best empirical value is typically quite good (less than 0.2~dex).  We nevertheless regard Equation~\ref{kr2ar2} to be the most reliable ICF when only Kr$^{2+}$ emission is detected, both because of the sulfur abundance anomaly in PNe and because the S$^{2+}$ abundance (needed for Equation~\ref{kr2s2}) is often derived from the strongly temperature-sensitive $\lambda$6312.1 line.  

There is a tendency for Equation~\ref{kr3ar3} to produce larger Kr abundances than the best empirical estimate.  This is not a systematic trend, as in some cases the elemental abundances are in good agreement (e.g., NGC~6741, NGC~6790, NGC~6886, NGC~7027, NGC~7662).  Equation~\ref{kr3ar3} is largest (and hence most sensitive to uncertainties) when the Ar$^{3+}$/Ar fraction is small.  Indeed, many of the larger disparities are seen for objects in which Ar$^{3+}$/Ar~$\lesssim 0.20$ (IC~2501, NGC~5315, NGC~6369, NGC~6572, NGC~6826).  Fortunately, these objects are low- or moderate-excitation PNe (with $T_{\rm eff}$ typically between 50,000~K and 60,000~K) in which Kr$^{2+}$ can be readily detected.  Therefore, the ICF from Equation~\ref{kr3ar3} should be used with caution in such objects, in which Equations~\ref{kr2ar2} and \ref{kr23o2} are preferable if $[$\ion{Kr}{3}$]$ emission is detected.  

The only other object with a discrepant Kr abundance from Equation~\ref{kr3ar3} is IC~4191.  This is most likely due to observational error, as \citet{sharpee07} flag $[$\ion{Kr}{3}$]$~$\lambda$6826.7 as a marginal detection.  For this object, the disagreement in Kr abundances from the different ICFs is likely due to the uncertain Kr$^{2+}$ abundance.

Observational errors in the O, S, and Ar ionic fractions also affect the Kr abundances.  For example, the NGC~7662 Kr abundances derived from Equations~\ref{kr2ar2} and \ref{kr3ar3} differ by 0.38~dex when using ionic Ar fractions from \citet{liu04b}, but only by 0.15~dex (3.68 and 3.83, respectively) if those from \citet{hyung97} are used.  In addition, various ICF schema exist for computing Ar abundances \citep[e.g.,][]{kb94, kwitter01, delgado-inglada14}, which can also lead to differences in Ar ionic fractions.  These effects also contribute to the scatter in Kr abundances derived with the various ICFs.

Se$^{3+}$/H$^+$ abundances were derived from the 2.287~$\mu$m line, using data from the near-infrared references in Table~\ref{pn_sample}.  Since the ground configuration of Se$^{3+}$ consists of a $^2$P term, a two-level model atom is sufficient for deriving the ionic abundance (the first excited term has an energy $\sim$10~eV higher than the ground configuration).  We utilized the transition probability of \citet{biemont87} and the collision strength calculated by K.\ Butler (2007, private communication).

Table~\ref{secomp} displays the calculated Se$^{3+}$ abundances, as well as the empirical and modeled elemental abundances.  The ICF- and model-derived abundances agree to within the uncertainties for all but NGC~6886, and are within 0.15~dex of each other with the exceptions of NGC~6886 and NGC~7027.  The level of agreement seen is consistent with observational and modeling uncertainties, and with the dispersion in the ionic fraction correlations of Fig.~\ref{icfs_solar}.  

\begin{deluxetable}{lcccc}
\tablecolumns{5}
\tablewidth{0pc} 
\tablecaption{Comparison of ICF- and Model-Derived Se Abundances} 
\tablehead{
\colhead{PN} & \colhead{$n$(Se$^{3+}$)/} & \colhead{} & \colhead{$12+$ Log(Se/H)} & \colhead{$12+$ Log(Se/H)} \\
\colhead{Name} & \colhead{$n$(H$^+$)$\times 10^9$} & \colhead{ICF(Se)} & \colhead{ICF} & \colhead{Model}}
\startdata
IC 418 & \nodata         & \nodata       & \nodata       & \nodata \\
IC 2501 & \nodata        & \nodata       & \nodata       & \nodata \\
IC 4191 & \nodata        & \nodata       & \nodata       & \nodata \\
IC 5117 & 2.15$\pm$0.23  & 1.83$\pm$0.81 & 3.59$\pm$0.22 & 3.52$\pm$0.05 \\
NGC 2440 & \nodata       & \nodata       & \nodata       & \nodata \\
NGC 5315 & \nodata       & \nodata       & \nodata       & \nodata \\
NGC 6369 & 2.05$\pm$0.42 & 1.60$\pm$0.64 & 3.51$\pm$0.21 & 3.66$\pm$0.14 \\
NGC 6572 & 0.68$\pm$0.04 & 1.40$\pm$0.50 & 2.98$\pm$0.16 & 3.01$\pm$0.03 \\
NGC 6741 & 0.58$\pm$0.14 & 4.99$\pm$1.75 & 3.46$\pm$0.20 & 3.31$\pm$0.11 \\
NGC 6790 & 0.55$\pm$0.06 & 1.38$\pm$0.48 & 2.88$\pm$0.17 & 2.91$\pm$0.05 \\
NGC 6826 & 0.47$\pm$0.14 & 1.50$\pm$0.57 & 2.85$\pm$0.23 & 2.95$\pm$0.13 \\
NGC 6884 & 1.50$\pm$0.22 & 1.57$\pm$0.62 & 3.37$\pm$0.20 & 3.40$\pm$0.06 \\
NGC 6886 & 0.99$\pm$0.11 & 2.29$\pm$1.01 & 3.35$\pm$0.21 & 3.70$\pm$0.08 \\
NGC 7027 & 1.26$\pm$0.08 & 2.33$\pm$1.03 & 3.47$\pm$0.21 & 3.68$\pm$0.08 \\
NGC 7662 & 1.44$\pm$0.32 & 3.16$\pm$1.29 & 3.66$\pm$0.22 & 3.53$\pm$0.10 \\
\tableline
\enddata
\label{secomp}
\tablecomments{Comparison of Se abundances derived from ICFs and from Cloudy models.  The solar Se abundance is $12$~+~Log(Se/H)~=~$3.34\pm0.03$ \citep{asplund09}.}
\end{deluxetable}

This analysis shows that our model-derived ICF formulae produce Se and Kr elemental abundances that are consistent within the uncertainties with abundances derived from multiple Kr ions and with Cloudy models of the individual PNe.  We now apply our ICFs to the $K$~band PN survey of \citetalias{sterling08}.

\section{APPLICATION OF ICFS TO PLANETARY NEBULA SE AND KR ABUNDANCES} \label{new_abund}

In this section, we utilize Equations~\ref{kr2s2}, \ref{kr2ar2}, and \ref{se3o2} to re-evaluate Se and Kr enrichments in PNe.  We consider the $K$~band spectra of 120~PNe from \citetalias{sterling08}, and examine whether the correlations found in that study are affected by the changes to the Se and Kr abundances due to using updated atomic data.

We aim to make the comparison with \citetalias{sterling08} as direct as possible, and therefore utilize the same ionic abundances, electron temperatures, and densities as before.  We change only the Se and Kr ICFs in order to isolate the effects of the updated atomic data.  The solar abundances of \citet{asplund05} are adopted, following \citetalias{sterling08}.  More recent compilations of the solar composition \citep[e.g.,][]{asplund09} adopt a significantly larger Ar abundance, in better agreement with the studies of \citet{lodders03, lodders08} than \citet{asplund05}.  We discuss how the revised solar abundances would affect the Se and Kr enrichment factors in the following text and tables.

We present the Se and Kr abundances relative to the reference elements O and Ar, in order to suppress effects due to intrinsic dispersion in metallicity as well as systematic effects such as abundance gradients with radial Galactocentric distance or vertical distance from the Galactic plane \citep{maciel06, henry10, stanghellini10}.  Following \citetalias{sterling08}, we determine Se and Kr abundances relative to Ar in Type~I PNe and to O in other objects.  Type~I PNe are believed to be descendants of more massive progenitor stars ($>3$--4~M$_{\odot}$), based on their He and N enrichments \citep{peimbert78, kb94}, small Galactic scale height, high nebular and stellar masses, and peculiar velocities \citep{torres-peimbert97, corradi95, gorny97, pena13}.  The primary reason for using different metallicity indicators is that the Ar/O ratio is a factor of two larger in Type~I PNe than in non-Type~I objects of \citetalias{sterling08}'s sample.  Similar trends were found for S/O and Cl/O, suggesting that uncertainties in the Ar abundances are not to blame.  In \citetalias{sterling08}, we speculated that this disparity in Ar/O ratios for the two populations could be due to ON-burning at the base of the convective envelope during the thermally-pulsing AGB stage of evolution \citep[i.e., hot bottom burning or HBB;][]{ventura05a, ventura05b, karakas06}.  However, subsequent studies of AGB nucleosynthesis showed that the overall O abundance decreases only modestly (by $\sim0.06$--0.15~dex, depending on the assumed solar abundances) as a result of HBB for solar-metallicity stars with initial masses 6--6.5~M$_{\odot}$ \citep{karakas09, karakas14}.  Furthermore, S, Cl, and Ar are not significantly affected by AGB nucleosynthesis \citep{karakas09, henry12, shingles13}.  Therefore the cause of the difference in the Ar/O (and S/O and Cl/O) ratios for Type~I and non-Type~I PNe is unresolved at this time.  Whether it is due to systematic uncertainties in the abundance determinations of Type~I PNe or to an unknown nucleosynthetic or mixing process in intermediate-mass AGB stars, we adopt Ar as the reference element in Type~I PNe.  Oxygen abundances are generally more accurate in PNe, and hence we use O as the metallicity reference in non-Type~I PNe; the results below are not significantly affected if Ar is used as the reference element for the full sample.

The Se and Kr ionic abundances used in our abundance determinations were derived previously and can be found in Table~9 of \citetalias{sterling08}.  Likewise, the electron temperatures, densities, and ionic and elemental O, S, and Ar abundances for the sample are given in their Tables~7 and 8.  We use our new Equation~\ref{kr2ar2} to derive Kr abundances, except when Ar ionic or elemental abundances are not available for a PN, or if Equation~\ref{kr2ar2} gives an anomalously large ICF compared to Equation~\ref{kr2s2}.   Se and Kr abundances derived with the new ICFs are given in Table~\ref{newsekr}, relative to H, O, and Ar.  The $[$Kr/O$]$ and $[$Se/O$]$ values for Type~I PNe are enclosed in parentheses to indicate that they may not be reliable.  If the solar abundances of \citet{asplund09} are adopted, the listed $[$Kr/O$]$ and $[$Kr/Ar$]$ values would increase by 0.05 and 0.24~dex, respectively, while $[$Se/O$]$ and $[$Se/Ar$]$ would increase by 0.02 and 0.21~dex.

\begin{deluxetable}{lccccccc}
\tablecolumns{8}
\tablewidth{0pc} 
\tabletypesize{\scriptsize}
\tablecaption{Se and Kr Abundances Relative to H, O, and Ar} 
\tablehead{
\colhead{Object} & \colhead{} & \colhead{} & \colhead{} & \colhead{} & \colhead{} & \colhead{} & \colhead{}\\
\colhead{Name} & \colhead{[Kr/H]} & \colhead{[Kr/O]\tablenotemark{a,b}} & \colhead{[Kr/Ar]\tablenotemark{b}} & \colhead{[Se/H]} & \colhead{[Se/O]\tablenotemark{a,b}} & \colhead{[Se/Ar]\tablenotemark{b}} & \colhead{Ref.\tablenotemark{c}}}
\startdata
BD+30$^{\rm o}$3639 & $<$0.96\tablenotemark{f} & $<$   0.96 & $<$   0.43 & $<$0.55 & $<$   0.54 & $<$   0.01 & IR2, 7 \\
  & $<$0.89\tablenotemark{f} & $<$   0.89 & $<$   0.35 & \nodata & \nodata & \nodata & IR4, 7 \\
  & 0.22$\pm$0.21\tablenotemark{f} &    0.22$\pm$0.23 &   -0.32$\pm$0.23 & \nodata & \nodata & \nodata & IR6, 7 \\
Cn 3-1 & $<$0.73 & $<$   0.76 & $<$   0.82 & $<$0.71 & $<$   0.74 & $<$   0.80 & IR1, 46 \\
DdDm 1 & $<$1.06 & $<$   1.67 & $<$   2.08 & $<$0.24 & $<$   0.85 & $<$   1.26 & IR1, 46 \\
Hb 4 & $<$0.58 & ($<$   0.56) & $<$   0.26 & 0.14$\pm$0.27 &    (0.12$\pm$0.30) &   -0.18$\pm$0.34 & IR1,  2 \\
Hb 5 & 0.36$\pm$0.18 &    (0.32$\pm$0.20) &   -0.24$\pm$0.20 & -0.48$\pm$0.24\tablenotemark{e} &   (-0.52$\pm$0.26)\tablenotemark{e} &   -1.08$\pm$0.26\tablenotemark{e} & IR1, 37 \\
Hb 6 & $<$0.70 & ($<$   0.65) & $<$   0.18 & 0.12$\pm$0.34 &    (0.07$\pm$0.41) &   -0.40$\pm$0.41 & IR1,  2 \\
Hb 7 & \nodata & \nodata & \nodata & $<$0.42 & $<$   0.43 & $<$   0.58 & IR1, 41 \\
Hb 12 & 0.39$\pm$0.15 &    0.71$\pm$0.18 &    0.27$\pm$0.18 & -0.38$\pm$0.18 &   -0.06$\pm$0.21 &   -0.50$\pm$0.21 & IR5, 18 \\
He 2-459 & $<$1.25\tablenotemark{f} & ($<$   2.06) & \nodata & $<$0.68 & ($<$   1.49) & \nodata & IR1, 14 \\
Hu 1-1 & $<$1.12\tablenotemark{f} & $<$   1.22 & $<$   2.10 & $<$0.36 & $<$   0.46 & $<$   1.34 & IR1, 46 \\
Hu 1-2 & $<$0.87 & ($<$   1.45) & $<$   1.09 & $<$0.06 & ($<$   0.64) & $<$   0.27 & IR1, 29 \\
Hu 2-1 & 0.51$\pm$0.30 &    0.66$\pm$0.32 &    0.91$\pm$0.32 & $<$-0.71 & $<$  -0.56 & $<$  -0.31 & IR1, 46 \\
IC 351 & $<$1.19 & $<$   1.45 & $<$   1.42 & 0.10$\pm$0.28\tablenotemark{e} &    0.36$\pm$0.31\tablenotemark{e} &    0.33$\pm$0.31\tablenotemark{e} & IR1, 46 \\
IC 418 & 0.50$\pm$0.20\tablenotemark{f} &    0.62$\pm$0.22 &    0.43$\pm$0.22 & $<$-0.32 & $<$  -0.21 & $<$  -0.40 & IR1, 39 \\
  & 0.44$\pm$0.46\tablenotemark{e,f} &    0.55$\pm$0.49\tablenotemark{e} &    0.36$\pm$0.49\tablenotemark{e} & \nodata & \nodata & \nodata & IR3, 39 \\
IC 1747 & $<$1.06 & $<$   1.14 & $<$   1.16 & 0.37$\pm$0.22 &    0.45$\pm$0.24 &    0.46$\pm$0.24 & IR1, 46 \\
IC 2003 & $<$0.95 & $<$   1.17 & $<$   1.18 & 0.33$\pm$0.21 &    0.55$\pm$0.24 &    0.56$\pm$0.24 & IR1, 46 \\
  & $<$0.95 & $<$   1.17 & $<$   1.18 & 0.45$\pm$0.22 &    0.67$\pm$0.24 &    0.68$\pm$0.24 & IR2, 46 \\
  & \nodata & \nodata & \nodata & 0.39$\pm$0.31\tablenotemark{e} &    0.61$\pm$0.34\tablenotemark{e} &    0.62$\pm$0.34\tablenotemark{e} & IR3, 46 \\
IC 2149 & $<$-0.06 & $<$   0.14 & $<$   0.22 & $<$-0.51 & $<$  -0.30 & $<$  -0.23 & IR1, 13 \\
IC 2165 & $<$0.50 & $<$   0.73 & $<$   0.60 & -0.10$\pm$0.23 &    0.13$\pm$0.25 &   -0.00$\pm$0.25 & IR1, 39 \\
  & $<$0.59 & $<$   0.82 & $<$   0.69 & 0.04$\pm$0.24 &    0.27$\pm$0.26 &    0.14$\pm$0.26 & IR2, 39 \\
IC 3568 & $<$0.28 & $<$   0.55 & $<$   0.57 & -0.56$\pm$0.22\tablenotemark{e} &   -0.29$\pm$0.24\tablenotemark{e} &   -0.27$\pm$0.24\tablenotemark{e} & IR1, 29 \\
IC 4593 & $<$0.51 & $<$   0.48 & $<$   0.43 & $<$-0.06 & $<$  -0.10 & $<$  -0.14 & IR1, 16 \\
IC 4634 & $<$0.17 & $<$   0.26 & $<$   0.35 & -0.41$\pm$0.18 &   -0.32$\pm$0.21 &   -0.23$\pm$0.21 & IR1, 21 \\
IC 4732 & $<$0.82 & $<$   1.06 & $<$   1.02 & 0.02$\pm$0.39\tablenotemark{e} &    0.26$\pm$0.47\tablenotemark{e} &    0.22$\pm$0.47\tablenotemark{e} & IR1,  2 \\
IC 4846 & $<$0.56 & $<$   0.72 & $<$   0.78 & $<$-0.38 & $<$  -0.22 & $<$  -0.16 & IR1, 46 \\
IC 4997\tablenotemark{g} & $<$-0.11 & $<$  -0.53 & $<$   0.46 & $<$-0.67 & $<$  -1.09 & $<$  -0.10 & IR1, 20 \\
  & $<$-0.22 & $<$  -0.64 & $<$   0.36 & $<$-0.90 & $<$  -1.32 & $<$  -0.32 & IR2, 20 \\
IC 5117 & 0.38$\pm$0.16 &    0.45$\pm$0.19 &    0.35$\pm$0.19 & 0.26$\pm$0.15 &    0.33$\pm$0.18 &    0.23$\pm$0.18 & IR7, 22 \\
  & 0.48$\pm$0.28\tablenotemark{e} &    0.55$\pm$0.30\tablenotemark{e} &    0.45$\pm$0.30\tablenotemark{e} & 0.32$\pm$0.16 &    0.39$\pm$0.19 &    0.30$\pm$0.19 & IR2, 22 \\
  & 0.60$\pm$0.14 &    0.67$\pm$0.17 &    0.57$\pm$0.17 & \nodata & \nodata & \nodata & IR4, 22 \\
IC 5217 & $<$0.50 & $<$   0.66 & $<$   0.58 & -0.09$\pm$0.20 &    0.07$\pm$0.23 &   -0.01$\pm$0.23 & IR1, 46 \\
J 320 & $<$0.78 & $<$   1.04 & $<$   0.98 & $<$-0.11 & $<$   0.16 & $<$   0.10 & IR1, 31 \\
J 900 & 0.51$\pm$0.30\tablenotemark{e} &    0.62$\pm$0.31\tablenotemark{e} &    0.61$\pm$0.32\tablenotemark{e} & 0.32$\pm$0.25 &    0.43$\pm$0.26 &    0.42$\pm$0.27 & IR1, 28 \\
K 3-17 & $>$0.25 & $>$   0.05 & \nodata & 0.43$\pm$0.23 &    0.24$\pm$0.28 & \nodata & IR1, 25 \\
K 3-55 & $>$0.20 & $>$  -0.04 & \nodata & 0.24$\pm$0.28 &    0.01$\pm$0.34 & \nodata & IR1, 24 \\
K 3-60 & 1.16$\pm$0.39 &    1.19$\pm$0.75 &    0.92$\pm$0.75 & 0.58$\pm$0.60 &    0.61$\pm$0.80 &    0.34$\pm$0.80 & IR1,  2 \\
  & 1.13$\pm$0.39 &    1.16$\pm$0.72 &    0.89$\pm$0.72 & 0.58$\pm$0.64 &    0.61$\pm$0.84 &    0.34$\pm$0.84 & IR6,  2 \\
K 3-61 & $<$1.42 & ($<$   1.39) & $<$   1.08 & 0.36$\pm$0.34 &    (0.33$\pm$0.36) &    0.02$\pm$0.36 & IR1,  2 \\
K 3-62 & $>$0.04 & \nodata & \nodata & $>$-0.36 & \nodata & \nodata & IR6 \\
K 3-67 & \nodata & \nodata & \nodata & -0.35$\pm$0.33 &    0.05$\pm$0.40 &   -0.17$\pm$0.40 & IR6,  2 \\
K 4-48 & $<$0.95 & $<$   0.94 & $<$   1.40 & $<$0.76 & $<$   0.75 & $<$   1.20 & IR1, 9 \\
  & 0.73$\pm$0.34 &    0.71$\pm$0.35 &    1.17$\pm$0.38 & \nodata & \nodata & \nodata & IR6, 9 \\
M 1-1 & $<$1.84 & $<$   1.80 & $<$   1.69 & $<$1.09 & $<$   1.05 & $<$   0.95 & IR1,  3 \\
M 1-4 & $<$0.94 & $<$   1.10 & $<$   0.94 & 0.30$\pm$0.34 &    0.46$\pm$0.41 &    0.30$\pm$0.41 & IR1,  2 \\
  & 0.75$\pm$0.21 &    0.91$\pm$0.28 &    0.75$\pm$0.28 & 0.31$\pm$0.37 &    0.47$\pm$0.45 &    0.31$\pm$0.45 & IR6,  2 \\
M 1-5 & 0.41$\pm$0.38 &    0.96$\pm$0.44 &    0.59$\pm$0.44 & $<$-0.78 & $<$  -0.23 & $<$  -0.60 & IR1, 31 \\
M 1-6 & 0.12$\pm$0.24\tablenotemark{f} &    0.43$\pm$0.29 &   -0.32$\pm$0.29 & $<$0.09 & $<$   0.41 & $<$  -0.34 & IR1, 8 \\
  & 0.05$\pm$0.24\tablenotemark{f} &    0.37$\pm$0.29 &   -0.38$\pm$0.29 & \nodata & \nodata & \nodata & IR6, 8 \\
M 1-9 & $<$0.94 & $<$   1.28 & $<$   1.12 & $<$-0.06 & $<$   0.28 & $<$   0.12 & IR1, 8 \\
M 1-11 & 0.79$\pm$0.34 &    1.71$\pm$0.38 &    1.86$\pm$0.39 & 0.40$\pm$0.26\tablenotemark{e} &    1.32$\pm$0.70\tablenotemark{e} &    1.47$\pm$0.75\tablenotemark{e} & IR1, 9 \\
  & 0.79$\pm$0.35 &    1.71$\pm$0.39 &    1.85$\pm$0.39 & -0.09$\pm$0.37 &    0.83$\pm$0.41 &    0.97$\pm$0.41 & IR6, 9 \\
M 1-12 & 0.46$\pm$0.37 &    0.68$\pm$0.38 &    1.32$\pm$0.41 & $<$0.63 & $<$   0.85 & $<$   1.49 & IR1, 27 \\
  & 0.41$\pm$0.37 &    0.63$\pm$0.38 &    1.27$\pm$0.41 & \nodata & \nodata & \nodata & IR6, 27 \\
M 1-14 & $<$0.18\tablenotemark{f} & $<$   0.54 & $<$   0.05 & $<$0.04 & $<$   0.40 & $<$  -0.09 & IR1, 8 \\
M 1-16 & $<$0.63 & ($<$   0.91) & $<$   0.55 & $<$-0.08 & ($<$   0.20) & $<$  -0.16 & IR1, 32 \\
M 1-17 & 0.94$\pm$0.39 &    0.74$\pm$0.45 &    0.60$\pm$0.45 & 0.25$\pm$0.30 &    0.05$\pm$0.35 &   -0.09$\pm$0.35 & IR1, 8 \\
M 1-20 & \nodata & \nodata & \nodata & -0.10$\pm$0.34 &   -0.06$\pm$0.40 &    0.03$\pm$0.40 & IR6, 40 \\
M 1-25 & 0.63$\pm$0.40 &    0.54$\pm$0.46 &    0.15$\pm$0.46 & $<$-0.02 & $<$  -0.11 & $<$  -0.50 & IR1, 14 \\
M 1-31 & $<$0.57 & $<$   0.74\tablenotemark{d} & $<$   0.33\tablenotemark{d} & 0.04$\pm$0.44\tablenotemark{e} &    0.21$\pm$0.51\tablenotemark{d,e} &   -0.20$\pm$0.64\tablenotemark{d,e} & IR1, 14 \\
M 1-32 & 1.17$\pm$0.34\tablenotemark{f} &    1.17$\pm$0.35 &    0.46$\pm$0.35 & 0.51$\pm$0.23 &    0.51$\pm$0.28 &   -0.20$\pm$0.28 & IR1, 14 \\
M 1-35 & $<$1.09 & ($<$   1.27) & $<$   0.64 & 0.14$\pm$0.48 &    (0.32$\pm$0.59) &   -0.31$\pm$0.59 & IR1,  2 \\
M 1-40 & 0.22$\pm$0.24 &    (0.39$\pm$0.24) &   -0.20$\pm$0.29 & -0.02$\pm$0.08\tablenotemark{e} &    (0.15$\pm$0.08)\tablenotemark{e} &   -0.45$\pm$0.16\tablenotemark{e} & IR1, 15 \\
M 1-46 & $<$1.08 & $<$   0.87 & $<$   0.73 & $<$1.17 & $<$   0.96 & $<$   0.82 & IR1, 14 \\
M 1-50 & $<$0.86 & $<$   0.70 & $<$   0.67 & -0.04$\pm$0.40\tablenotemark{e} &   -0.21$\pm$0.46\tablenotemark{e} &   -0.23$\pm$0.46\tablenotemark{e} & IR1, 31 \\
M 1-51 & 1.02$\pm$0.36 &    0.76$\pm$0.37 &    0.66$\pm$0.40 & 0.53$\pm$0.36 &    0.27$\pm$0.42 &    0.16$\pm$0.56 & IR1, 14 \\
M 1-54 & $<$1.05\tablenotemark{f} & ($<$   0.94) & \nodata & $<$0.44 & ($<$   0.33) & \nodata & IR1, 31 \\
M 1-57 & 0.54$\pm$0.28 &    (0.39$\pm$0.33) &    0.05$\pm$0.33 & -0.04$\pm$0.37 &   (-0.19$\pm$0.43) &   -0.53$\pm$0.43 & IR1, 31 \\
M 1-58 & $<$1.03 & $<$   0.93 & $<$   0.74 & 0.34$\pm$0.30 &    0.24$\pm$0.35 &    0.05$\pm$0.83 & IR1, 9 \\
M 1-60 & $<$0.74 & ($<$   0.62) & $<$   0.24 & 0.24$\pm$0.51 &    (0.12$\pm$0.60) &   -0.25$\pm$0.59 & IR1, 14 \\
M 1-61 & $<$0.31 & $<$   0.48 & $<$   0.32 & -0.21$\pm$0.39 &   -0.04$\pm$0.45 &   -0.20$\pm$0.45 & IR1, 14 \\
M 1-71 & 0.56$\pm$0.26 &    0.46$\pm$0.31 &    0.24$\pm$0.31 & -0.09$\pm$0.43 &   -0.19$\pm$0.49 &   -0.41$\pm$0.49 & IR1, 14 \\
M 1-72 & $<$0.17 & $<$   0.90 & $<$   0.78 & $<$0.33 & $<$   1.06 & $<$   0.94 & IR1, 27 \\
M 1-74 & $<$0.38 & $<$   0.44 & $<$   0.32 & -0.64$\pm$0.44\tablenotemark{e} &   -0.58$\pm$0.46\tablenotemark{e} &   -0.70$\pm$0.46\tablenotemark{e} & IR1, 46 \\
  & \nodata & \nodata & \nodata & -0.41$\pm$0.23 &   -0.35$\pm$0.25 &   -0.47$\pm$0.25 & IR6, 46 \\
M 1-75 & $<$1.41\tablenotemark{f} & ($<$   1.28) & $<$   0.75 & $<$0.93 & ($<$   0.80) & $<$   0.27 & IR1,  2 \\
M 1-80 & $<$1.09\tablenotemark{f} & $<$   1.16 & $<$   1.14 & 0.45$\pm$0.29 &    0.52$\pm$0.37 &    0.51$\pm$0.36 & IR1,  2 \\
M 2-2 & $<$1.30 & $<$   1.53 & $<$   1.53 & $<$0.22 & $<$   0.45 & $<$   0.45 & IR1,  2 \\
M 2-31 & $<$0.75 & $<$   0.73 & $<$   0.56 & 0.05$\pm$0.43 &    0.03$\pm$0.50 &   -0.14$\pm$0.50 & IR1, 40 \\
M 2-43 & 0.81$\pm$0.35 &    0.84$\pm$0.39 &    0.82$\pm$0.39 & 0.06$\pm$0.32 &    0.09$\pm$0.37 &    0.08$\pm$0.37 & IR1, 14 \\
M 2-48 & $<$1.42 & ($<$   1.53) & $<$   0.73 & $<$0.33 & ($<$   0.44) & $<$  -0.36 & IR1, 30 \\
M 3-15 & $<$0.50 & $<$   0.80 & $<$   0.54 & -0.29$\pm$0.38 &    0.01$\pm$0.44 &   -0.25$\pm$0.58 & IR1, 14 \\
M 3-25 & $<$0.56 & ($<$   0.39) & $<$   0.22 & 0.39$\pm$0.39 &    (0.22$\pm$0.45) &    0.05$\pm$0.45 & IR1, 9 \\
M 3-28 & $<$2.10\tablenotemark{f} & ($<$   2.03) & \nodata & $<$0.42 & ($<$   0.35) & \nodata & IR1, 25 \\
M 3-35 & \nodata & \nodata & \nodata & $>$-0.75 & \nodata & \nodata & IR1,  5 \\
M 3-41 & $<$1.53 & ($<$   2.10) & $<$   1.72 & $<$1.50 & ($<$   2.07) & $<$   1.69 & IR1, 40 \\
M 4-18 & \nodata & \nodata & \nodata & \nodata & \nodata & \nodata & IR1, 10 \\
Me 1-1 & $<$0.81 & ($<$   0.91) & $<$   0.63 & $<$-0.07 & ($<$   0.03) & $<$  -0.25 & IR1, 42 \\
Me 2-1 & \nodata & \nodata & \nodata & 0.07$\pm$0.21 &    0.01$\pm$0.23 &    0.05$\pm$0.23 & IR6, 43 \\
Me 2-2 & $<$0.43 & ($<$   0.76) & $<$   0.84 & $<$-0.42 & ($<$  -0.09) & $<$  -0.01 & IR1, 46 \\
NGC 40 & 1.09$\pm$0.19\tablenotemark{f} &    1.03$\pm$0.21 &    0.74$\pm$0.21 & \nodata & \nodata & \nodata & IR4, 38 \\
  & 1.13$\pm$0.20\tablenotemark{f} &    1.07$\pm$0.23 &    0.78$\pm$0.23 & \nodata & \nodata & \nodata & IR3, 38 \\
NGC 1501 & $<$1.76 & ($<$   1.89) & $<$   1.83 & $<$0.95 & ($<$   1.07) & $<$   1.02 & IR1, 12 \\
NGC 2392 & $<$0.92 & $<$   1.00 & $<$   0.88 & $<$0.17 & $<$   0.26 & $<$   0.14 & IR1, 16 \\
NGC 2440 & $<$0.52 & ($<$   0.61) & $<$   0.20 & $<$-0.30 & ($<$  -0.22) & $<$  -0.62 & IR2, 6 \\
NGC 3242 & $<$0.73 & $<$   0.87 & $<$   0.92 & 0.01$\pm$0.21 &    0.15$\pm$0.23 &    0.20$\pm$0.23 & IR1, 44 \\
  & $<$0.16 & $<$   0.30 & $<$   0.34 & 0.01$\pm$0.19 &    0.15$\pm$0.21 &    0.20$\pm$0.21 & IR2, 44 \\
  & \nodata & \nodata & \nodata & 0.05$\pm$0.23 &    0.19$\pm$0.25 &    0.24$\pm$0.25 & IR3, 44 \\
  & \nodata & \nodata & \nodata & -0.14$\pm$0.25 &   -0.00$\pm$0.27 &    0.04$\pm$0.27 & IR3, 44 \\
NGC 6210 & $<$0.18 & $<$   0.20 & $<$   0.24 & -0.21$\pm$0.18 &   -0.20$\pm$0.21 &   -0.15$\pm$0.21 & IR2, 29 \\
  & \nodata & \nodata & \nodata & -0.27$\pm$0.24 &   -0.25$\pm$0.27 &   -0.21$\pm$0.27 & IR3, 29 \\
NGC 6302 & $<$0.66 & ($<$   0.95) & $<$   0.06 & $<$-0.14 & ($<$   0.16) & $<$  -0.74 & IR2, 33 \\
NGC 6369 & $<$1.25\tablenotemark{f} & $<$   1.45 & $<$   0.81 & 0.43$\pm$0.34 &    0.63$\pm$0.42 &   -0.01$\pm$0.42 & IR1,  2 \\
NGC 6439 & $<$1.20\tablenotemark{f} & ($<$   1.21) & $<$   0.93 & 0.25$\pm$0.27 &    (0.26$\pm$0.30) &   -0.02$\pm$0.30 & IR1,  2 \\
NGC 6445 & 1.09$\pm$0.28\tablenotemark{e} &    0.70$\pm$0.33\tablenotemark{e} &    0.77$\pm$0.33\tablenotemark{e} & 1.02$\pm$0.26 &    0.63$\pm$0.31 &    0.70$\pm$0.31 & IR1, 27 \\
NGC 6537 & $<$0.25\tablenotemark{f} & ($<$   0.64) & $<$  -0.19 & -0.31$\pm$0.20 &    (0.09$\pm$0.22) &   -0.74$\pm$0.22 & IR1, 34 \\
  & $<$0.45\tablenotemark{f} & ($<$   0.85) & $<$   0.02 & $<$-0.47 & ($<$  -0.08) & $<$  -0.90 & IR2, 34 \\
NGC 6543 & \nodata & \nodata & \nodata & -0.16$\pm$0.28 &   -0.36$\pm$0.30 &   -0.49$\pm$0.30 & IR3, 45 \\
NGC 6567 & $<$0.34 & $<$   0.70 & $<$   0.85 & -0.08$\pm$0.18 &    0.28$\pm$0.21 &    0.43$\pm$0.21 & IR1, 19 \\
NGC 6572 & 0.08$\pm$0.28 &    0.18$\pm$0.30 &    0.07$\pm$0.30 & -0.35$\pm$0.14 &   -0.24$\pm$0.17 &   -0.36$\pm$0.17 & IR1, 29 \\
  & 0.19$\pm$0.31\tablenotemark{e} &    0.29$\pm$0.33\tablenotemark{e} &    0.18$\pm$0.33\tablenotemark{e} & -0.39$\pm$0.16 &   -0.29$\pm$0.18 &   -0.40$\pm$0.18 & IR2, 29 \\
  & 0.15$\pm$0.49\tablenotemark{e} &    0.25$\pm$0.53\tablenotemark{e} &    0.14$\pm$0.53\tablenotemark{e} & -0.39$\pm$0.18 &   -0.29$\pm$0.21 &   -0.40$\pm$0.21 & IR3, 29 \\
  & 0.31$\pm$0.39\tablenotemark{e} &    0.41$\pm$0.41\tablenotemark{e} &    0.30$\pm$0.41\tablenotemark{e} & -0.28$\pm$0.16 &   -0.18$\pm$0.19 &   -0.30$\pm$0.19 & IR3, 29 \\
NGC 6578 & $<$1.03 & $<$   0.94 & $<$   0.71 & 0.31$\pm$0.28 &    0.22$\pm$0.92 &   -0.01$\pm$0.92 & IR1,  2 \\
NGC 6629 & 0.96$\pm$0.26 &    1.02$\pm$0.33 &    0.54$\pm$0.33 & -0.32$\pm$0.68\tablenotemark{e} &   -0.26$\pm$1.00\tablenotemark{e} &   -0.74$\pm$1.00\tablenotemark{e} & IR1,  2 \\
NGC 6644 & 0.45$\pm$0.26\tablenotemark{e} &    0.54$\pm$0.28\tablenotemark{e} &    0.51$\pm$0.28\tablenotemark{e} & -0.13$\pm$0.21 &   -0.04$\pm$0.24 &   -0.07$\pm$0.24 & IR1,  4 \\
NGC 6720 & 1.14$\pm$0.31\tablenotemark{e} &    1.00$\pm$0.32\tablenotemark{e} &    0.78$\pm$0.32\tablenotemark{e} & \nodata & \nodata & \nodata & IR3, 29 \\
NGC 6741 & 0.44$\pm$0.19 &    0.26$\pm$0.22 &    0.25$\pm$0.22 & 0.14$\pm$0.19 &   -0.05$\pm$0.22 &   -0.06$\pm$0.22 & IR1, 29 \\
NGC 6751 & $<$1.65 & $<$   1.53 & $<$   1.30 & $<$1.07 & $<$   0.94 & $<$   0.71 & IR1, 26 \\
NGC 6778 & $<$0.85 & ($<$   1.02) & $<$   0.60 & $<$0.34 & ($<$   0.51) & $<$   0.08 & IR1,  1 \\
NGC 6790 & $<$-0.05 & $<$   0.21 & $<$   0.37 & -0.44$\pm$0.14 &   -0.18$\pm$0.17 &   -0.02$\pm$0.17 & IR1, 29 \\
NGC 6803 & $<$0.23 & $<$   0.19 & $<$   0.03 & -0.29$\pm$0.18 &   -0.33$\pm$0.20 &   -0.50$\pm$0.20 & IR1, 46 \\
  & -0.02$\pm$0.43\tablenotemark{e} &   -0.06$\pm$0.44\tablenotemark{e} &   -0.23$\pm$0.44\tablenotemark{e} & -0.22$\pm$0.20 &   -0.26$\pm$0.22 &   -0.43$\pm$0.22 & IR3, 46 \\
NGC 6804 & \nodata & \nodata & \nodata & $<$0.82 & $<$   0.98 & \nodata & IR1,  2 \\
NGC 6807 & $<$0.65\tablenotemark{f} & $<$   0.75 & $<$   1.35 & $<$-0.55 & $<$  -0.46 & $<$   0.14 & IR1, 46 \\
NGC 6818 & $<$0.50 & $<$   0.47 & $<$   0.24 & 0.34$\pm$0.20 &    0.31$\pm$0.22 &    0.08$\pm$0.22 & IR1, 35 \\
NGC 6826 & $<$0.37 & $<$   0.51 & $<$   0.56 & -0.49$\pm$0.22 &   -0.35$\pm$0.25 &   -0.30$\pm$0.25 & IR1, 29 \\
NGC 6833 & $<$0.26 & $<$   0.79 & $<$   0.63 & $<$-0.53 & $<$  -0.00 & $<$  -0.17 & IR1, 46 \\
NGC 6879 & $<$0.57 & $<$   0.71 & $<$   0.52 & 0.01$\pm$0.18 &    0.15$\pm$0.21 &   -0.04$\pm$0.21 & IR1, 46 \\
NGC 6881 & 0.74$\pm$0.23\tablenotemark{e} &    0.66$\pm$0.30\tablenotemark{d,e} &    0.27$\pm$0.30\tablenotemark{d,e} & 0.30$\pm$0.25 &    0.22$\pm$0.32\tablenotemark{d} &   -0.17$\pm$0.32\tablenotemark{d} & IR1,  2 \\
NGC 6884 & $<$0.43 & $<$   0.48 & $<$   0.35 & 0.02$\pm$0.18 &    0.07$\pm$0.21 &   -0.06$\pm$0.21 & IR1, 29 \\
NGC 6886 & 0.50$\pm$0.17 &    0.35$\pm$0.19 &    0.36$\pm$0.19 & 0.03$\pm$0.20 &   -0.12$\pm$0.22 &   -0.11$\pm$0.22 & IR1, 36 \\
NGC 6891 & $<$0.54 & $<$   0.59 & $<$   0.57 & $<$-0.29 & $<$  -0.23 & $<$  -0.25 & IR1, 46 \\
NGC 6905 & $<$1.54 & $<$   1.33 & $<$   1.62 & $<$1.22 & $<$   1.01 & $<$   1.30 & IR1, 26 \\
NGC 7009 & \nodata & \nodata & \nodata & -0.09$\pm$0.32\tablenotemark{e} &   -0.01$\pm$0.34\tablenotemark{e} &   -0.33$\pm$0.34\tablenotemark{e} & IR3, 17 \\
NGC 7026 & $<$0.23 & $<$   0.15 & $<$   0.02 & -0.09$\pm$0.20 &   -0.17$\pm$0.23 &   -0.30$\pm$0.23 & IR1, 46 \\
NGC 7027 & 0.57$\pm$0.16 &    0.57$\pm$0.18 &    0.45$\pm$0.18 & 0.22$\pm$0.19 &    0.22$\pm$0.22 &    0.10$\pm$0.22 & IR2, 47 \\
  & 0.70$\pm$0.28\tablenotemark{e} &    0.70$\pm$0.30\tablenotemark{e} &    0.58$\pm$0.30\tablenotemark{e} & 0.18$\pm$0.22 &    0.18$\pm$0.24 &    0.06$\pm$0.24 & IR3, 47 \\
  & 0.74$\pm$0.18 &    0.74$\pm$0.21 &    0.62$\pm$0.21 & 0.19$\pm$0.23 &    0.19$\pm$0.25 &    0.07$\pm$0.25 & IR6, 47 \\
NGC 7354 & $<$0.87 & ($<$   0.93) & $<$   0.45 & 0.46$\pm$0.24 &    (0.52$\pm$0.31) &    0.04$\pm$0.31 & IR1,  2 \\
NGC 7662 & $<$0.74 & $<$   0.87 & $<$   0.89 & 0.32$\pm$0.21\tablenotemark{e} &    0.45$\pm$0.23\tablenotemark{e} &    0.47$\pm$0.23\tablenotemark{e} & IR2, 29 \\
  & \nodata & \nodata & \nodata & 0.14$\pm$0.68 &    0.28$\pm$0.75 &    0.30$\pm$0.75 & IR3, 29 \\
SwSt 1 & 0.33$\pm$0.15\tablenotemark{f} &    0.58$\pm$0.29 & \nodata & \nodata & \nodata & \nodata & IR4, 11 \\
Vy 1-1 & \nodata & \nodata & \nodata & \nodata & \nodata & \nodata & IR1, 23 \\
Vy 1-2 & $<$0.85 & $<$   0.81 & $<$   0.78 & $<$-0.09 & $<$  -0.13 & $<$  -0.16 & IR1, 46 \\
Vy 2-2 & $<$-0.35 & $<$   0.33\tablenotemark{d} & $<$  -0.06\tablenotemark{d} & -0.93$\pm$0.18 &   -0.25$\pm$0.20\tablenotemark{d} &   -0.64$\pm$0.20\tablenotemark{d} & IR1, 46 \\
  & -0.40$\pm$0.44\tablenotemark{e} &    0.28$\pm$0.44\tablenotemark{d,e} &   -0.11$\pm$0.44\tablenotemark{d,e} & -0.89$\pm$0.31\tablenotemark{e} &   -0.21$\pm$0.31\tablenotemark{d,e} &   -0.60$\pm$0.31\tablenotemark{d,e} & IR3, 46 \\
  & $<$-0.39 & $<$   0.29\tablenotemark{d} & $<$  -0.09\tablenotemark{d} & \nodata & \nodata & \nodata & IR4, 46 \\
\tableline
\enddata
\label{newsekr}
\scriptsize
\tablenotetext{a}{The $[$Kr/O$]$ and $[$Se/O$]$ values of Type~I PNe are placed in parentheses, to indicate that O may not be a reliable metallicity indicator in these objects (see text).}
\tablenotetext{b}{Abundances are relative to the \citet{asplund05} solar composition for ease of comparison with \citetalias{sterling08}.  Using the \citet{asplund09} solar composition, the listed $[$Kr/O$]$ and $[$Se/O$]$ values would increase by 0.05 and 0.02~dex, respectively, while $[$Kr/Ar$]$ and $[$Se/Ar$]$ would increase by 0.24 and 0.21~dex.}
\tablenotetext{c}{REFERENCES (Optical/UV/Mid-infrared Data): (1) \citet{ac83}; (2) \citet{ak87}; (3) \citet{aller86}; (4)  \citet{aller88}; (5) \citet{barker78a, barker78b}; (6) \citet{bernard-salas02}; (7) \citet{bernard-salas03}; (8) \citet{costa04}; (9) \citet{cuisinier96}; (10) \citet{demarco99}; (11) \citet{demarco01a}; (12) \citet{ercolano04}; (13) \citet{feibelman94}; (14) \citet{girard07}; (15) \citet{gorny04}; (16) \citet{henry00, henry04}; (17) \citet{hyung95}; (18) \citet{hyung96}; (19)\citet{hyung93}; (20) \citet{hyung94b}; (21) \citet{hyung99}; (22) \citet{hyung01}; (23) \citet{kaler80}; (24) \citet{kaler93}; (25) \citet{kaler96}; (26) \citet{kb94}; (27) \citet{koppen91}; (28) \citet{henry00, kwitter03}; (29) \citet{liu04a, liu04b}; (30) \citet{lopez-martin02}; (31) \citet{henry00, milingo02}; (32) \citet{perinotto98}; (33) \citet{pottasch99}; (34) \citet{pottasch00}; (35) \citet{pottasch05a}; (36) \citet{pottasch05b}; (37) \citet{pottasch07}; (38) \citet{pottasch03}; (39) \citet{pottasch04}; (40) \citet{ratag97}; (41) \citet{samland92}; (42) \citet{shen04}; (43) \citet{surendiranath04}; (44) \citet{tsamis03}; (45) \citet{wesson04}; (46) \citet{wesson05}; (47) \citet{zhang05}.  REFERENCES (Near-Infrared Data): (IR1) \citetalias{sterling08}; (IR2) \citet{geballe91}; (IR3) \citet{hora99}; (IR4) \citet{likkel06}; (IR5) \citet{lumsden96}; (IR6) \citet{lumsden01}; (IR7) \citet{rudy01}.}
\tablenotetext{d}{Ar is used as a reference element.  This object exhibits N enrichments that fall just below the value for Type~I classification, and hence its O abundance may not be reliable.}
\tablenotetext{e}{Based on a weak or uncertain detection.}
\tablenotetext{f}{ICF(Kr) from Equation~\ref{kr2s2}.}
\tablenotetext{g}{IC~4997 has an uncertain O abundance, and Ar is used as the reference element for this object.}
\end{deluxetable}
\clearpage

By comparing our new analytical fits to the correlations plotted in Figure~\ref{icfs_solar} (solid lines) with those from \citetalias{sterling07} (dashed lines), it is expected that the ICFs and hence elemental Se and Kr abundances in PNe will be lower than previous estimates.  Indeed, that is exactly what is seen (for comparison, see Table~10 of \citetalias{sterling08}).  Overall, of the 33 PNe in which the Kr abundance could be determined, the average $[$Kr/(O,Ar)$]=0.82\pm0.29$, compared to the mean of 0.98$\pm$0.31 found in \citetalias{sterling08}.  Likewise, the mean $[$Se/(O,Ar)$]=0.12\pm0.27$ in the 68 PNe exhibiting $[$\ion{Se}{4}$]$ emission, down from \citetalias{sterling08}'s 0.31$\pm$0.27.  If the \citet{asplund09} solar abundances are used, the mean Se and Kr abundances in our sample would each increase by 0.05~dex.

In individual objects, the derived Se and Kr abundances are typically lower by 0.1--0.3~dex compared to previous estimates, but there are occasional exceptions.  For example, the new Kr/Kr$^{2+}$ ICFs exceed those of \citetalias{sterling07} when the S$^{2+}$ or Ar$^{2+}$ fractional abundances are sufficiently high.  In the case of BD+30$^{\rm o}$3639, the Kr abundance actually increased by 0.14~dex from that found in \citetalias{sterling08}, since the S$^{2+}$/S fraction of 0.79 \citep{bernard-salas03} lies above the point where Equation~\ref{kr2s2} intersects \citetalias{sterling07}'s Equation~2 (see Figure~\ref{icfs_solar}, top left panel).  For other PNe, the new Kr abundances do not exceed those from \citetalias{sterling08} by more than 0.01~dex.  In contrast, the Se abundance decreased in all objects of the sample.  For some objects (e.g., Cn~3-1, M~1-11, and M~3-41) the Se abundances (or upper limits) are lower by as much as an order of magnitude.  This is due to the extremely small O$^{2+}$/O fraction in these low-excitation objects, which led to a very large and uncertain ICF using \citetalias{sterling07}'s formula.  For example, the Se ICF for M~1-11 decreased from 257$\pm$130 \citepalias{sterling08} to 23.4$\pm$5.4 in the present study.  In fact, the Se ICF formula from \citetalias{sterling07} (their Equation~3) is negative and hence cannot be applied when O$^{2+}$/O~$<0.01626$.  Our Equation~\ref{se3o2} is valid for all O$^{2+}$ fractional abundances.  This is the reason that upper limits to the Se abundance are given in Table~\ref{newsekr} for He~2-459 and M~1-12 when they were not determined in \citetalias{sterling08}.

The general reduction of Se and Kr abundances in this sample of PNe decreases the number of objects that are enriched by \textit{in~situ} \emph{s}-process nucleosynthesis in their progenitor stars' AGB stage of evolution.  Since the sample is primarily composed of Galactic disk PNe, we follow \citetalias{sterling08} and assume that a PN is self-enriched in Se and Kr if its $[$Se/(O, Ar)$]$ and $[$Kr/(O, Ar)$]$ abundances exceed 0.3~dex.  This cutoff lies above the dispersion of $\sim$0.2~dex in the abundances of other light \emph{n}-capture elements (Sr, Y, and Zr) in unevolved near-solar metallicity stars \citep{burris00, travaglio04}, and therefore (statistically) accounts for star-to-star scatter in initial Se and Kr abundances.  Using this criterion, we find that 26 of 33 (79\%) objects with determined Kr abundances are enriched by \emph{s}-process nucleosynthesis in their progenitor stars, while only 14 of 68 PNe (21\%) are self-enriched in Se.  Overall, 30 of the 81 PNe (37\%) with determined Se and/or Kr abundances are \emph{s}-process enriched.\footnote{The percentages of enriched objects increase by less than 5\% when adopting the \citet{asplund09} solar abundances.}  The percentages of enriched PNe found in \citetalias{sterling08} were 85\% for Kr, 35\% for Se, and 52\% overall.

In the 22 PNe in which both Se and Kr emission lines were detected, we find an abundance ratio $[$Kr/Se$]=0.5\pm 0.2$ nearly identical to that derived in \citetalias{sterling08}.  This is in agreement (though toward the upper end of the range) with computational studies of \emph{s}-process nucleosynthesis in AGB stars \citep{gallino98, goriely00, busso01, karakas09}.  We refer the reader to \citetalias{sterling08} and \citet{karakas09} for a more extensive discussion comparing the observational results to predictions of nucleosynthesis models.

\subsection{Correlations Between Se and Kr Abundances and Other Nebular/Stellar Properties}

In \citetalias{sterling08}, we searched for correlations between Se and Kr \emph{s}-process enrichment factors and other nebular and stellar properties.  We investigate here whether the improved Se and Kr abundances affect those correlations.  We do not repeat all details of the discussion from \citetalias{sterling08}, and refer the reader to that paper for the full analysis and interpretation.

One of the most significant correlations found in \citetalias{sterling08} was that PNe with higher mass progenitor stars ($>3$--4~M$_{\odot}$; e.g., Type~I and bipolar PNe) tend to show little if any \emph{s}-process enrichment of Se and Kr, compared to nebulae ejected by lower-mass stars.  Following \citetalias{sterling08}, to improve statistics we consider a PN to be Type~I if it meets either the \citet{peimbert78} or the \citet{kb94} criteria.  Table~\ref{corr_avg} displays the average $[$Se/(O, Ar)$]$ and $[$Kr/(O, Ar)$]$ abundances for various subclasses of PNe, including Peimbert types, different morphologies, and central star types.  The average Se abundance is 0.39~dex higher in non-Type~I PNe than in Type~I objects, while the average Kr abundance is nearly an order of magnitude higher.  Indeed, no Type~I PN in this sample is enriched in either Se or Kr by more than 0.05~dex \citep[0.3~dex using the solar abundances of][]{asplund09}.  Kolmogorov-Smirnov (K-S) tests \citep[e.g.,][]{press92} indicate that the differences in \emph{s}-process enrichments in Type~I and non-Type~I PNe are statistically significant, with probabilities $p_{\rm ks}=0.025$ and 0.009 that the Se and Kr abundances (respectively) in these two subclasses are drawn from the same distribution function.  This result is robust against the metallicity reference element (Ar or O) used for non-Type~I PNe, as well as the adopted solar composition.

\begin{deluxetable}{lcccccccc}
\rotate
\tablecolumns{9}
\tablewidth{0pc}
\tabletypesize{\footnotesize}
\tablecaption{Kr and Se Abundances Vs. Nebular and Stellar Properties} 
\tablehead{
\colhead{} & \colhead{Mean} & \colhead{Mean} & \colhead{} & \colhead{Number of} & \colhead{Mean} & \colhead{Mean} & \colhead{} & \colhead{Number of} \\
\colhead{Property} & \colhead{[Se/(O,~Ar)]\tablenotemark{a}} & \colhead{[Se/(O,~Ar)]\tablenotemark{b}} & \colhead{$< \sigma >$\tablenotemark{c}} & \colhead{Se Detections} & \colhead{[Kr/(O,~Ar)]\tablenotemark{a}} & \colhead{[Kr/(O,~Ar)]\tablenotemark{b}} & \colhead{$< \sigma >$\tablenotemark{c}} & \colhead{Kr Detections}}
\startdata
Type I & $-0.22$ & $-0.01$ & 0.27 & 12 & $-0.11$ & 0.13 & 0.12 & 3 \\
Non-Type I & 0.17 & 0.20 & 0.25 & 55 & 0.85 & 0.91 & 0.25 & 30 \\
\tableline
Bipolar & 0.08 & 0.13 & 0.38 & 14 & 0.50 & 0.58 & 0.28 & 8 \\
Elliptical & 0.15 & 0.19 & 0.24 & 28 & 0.95 & 1.00 & 0.36 & 15 \\
\tableline
$[$WC$]$ & 0.19 & 0.24 & 0.25 & 16 & 0.79 & 0.84 & 0.31 & 10 \\
WELS & 0.09 & 0.12 & 0.24 & 12 & 0.45 & 0.50 & 0.16 & 3 \\
Non-[WC]/WELS & 0.10 & 0.15 & 0.28 & 39 & 0.86 & 0.92 & 0.30 & 20 \\
\tableline
Full Sample & 0.12 & 0.17 & 0.27 & 67 & 0.82 & 0.87 & 0.29 & 33 \\
\enddata
\label{corr_avg}
\tablecomments{Only PNe exhibiting Se and/or Kr emission and with determined O and Ar abundances are included.}
\tablenotetext{a}{Ar is used as the reference element for Type~I PNe, as well as M~1-31, NGC~6881, and Vy~2-2 (which have N/O ratios just below the Type~I cutoff); O is used for all other objects.  Solar abundances are taken from \citet{asplund05}.} 
\tablenotetext{b}{Solar abundances from \citet{asplund09}.} 
\tablenotetext{c}{The $< \sigma >$ are the mean absolute deviations in the Se and Kr abundances.} 
\end{deluxetable}

Bipolar PNe are believed to derive from a more massive population than other morphologies of PNe, based on their small Galactic scale height, high central star masses, and tendency to exhibit Type~I compositions \citep{corradi95, gorny97, torres-peimbert97, stanghellini02}.  The average $[$Se/(O, Ar)$]$ and $[$Kr/(O, Ar)$]$ abundances in bipolar PNe are lower than in elliptical PNe (Table~\ref{corr_avg}), though the difference is not as striking as for Type~I and non-Type~I objects.  Unlike Type~I PNe, several bipolar objects show evidence for self-enrichment in Se and Kr.  K-S tests suggest that the Se and Kr abundances in bipolar and elliptical PNe are not strongly distinct, with respective probabilities 0.42 and 0.21 that the abundances are drawn from the same distribution function.  \citetalias{sterling08} found a similar result, suggesting that morphology is not as reliable as chemical composition in gauging the masses of PN progenitor stars.

We also confirm \citetalias{sterling08}'s result that there is no significant difference in the \emph{s}-process enrichments of PNe with H-rich and H-poor ($[$WC$]$ and weak emission line stars, or WELS) central stars.  $[$WC$]$ and non-$[$WC$]$ PNe have similar average Se and Kr abundances, and K-S tests indicate that the Se and Kr abundance distributions are the same in the two subclasses ($p_{ks}$(Se)~=~0.79 and $p_{ks}$(Kr)~=~0.80).  The average enrichment factors are smaller for PNe with WELS nuclei, but K-S tests show that there is little evidence that the Se and Kr abundance distributions are distinct from those of PNe with other central star types.  These results adhere to previous findings that the nebular compositions of $[$WC$]$ and WELS PNe do not show significant differences from those with H-rich nuclei \citep{gorny95, pena01, demarco01b, girard07, gorny09}.

\begin{figure}[ht!]
\plotone{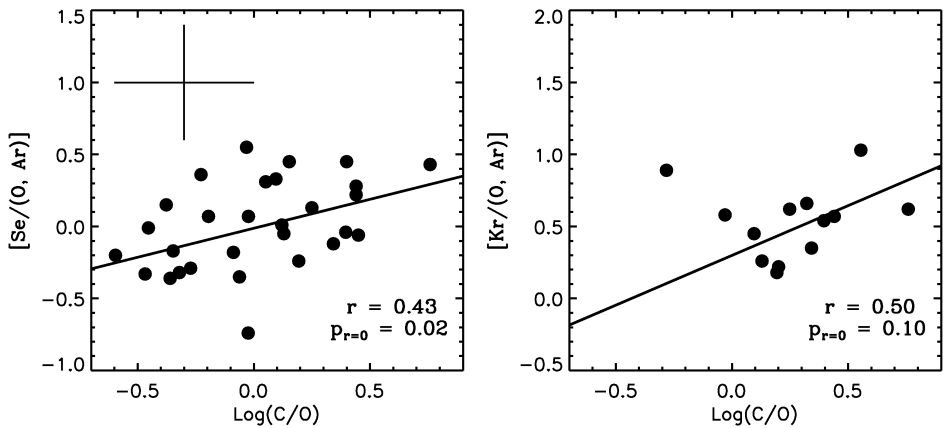}
\caption{$[$Se/(O,~Ar)$]$ (left) and $[$Kr/(O,~Ar)$]$ (right) are plotted against C/O (derived from UV forbidden lines) for PNe in the sample of \citetalias{sterling08}.  Typical error bars are shown in the left panel.  The solid lines indicate the best linear fits to the correlations.  The correlation coefficient $r$ and significance $p_{\rm r=0}$ are given in each panel.  The discrepant point in the right panel corresponds to Hb~12.}
\label{sekr_vs_c}
\end{figure}

Finally, in \citetalias{sterling08} we found that Se and Kr enrichments positively correlate with nebular C/O ratios, as is expected theoretically \citep[e.g.,][]{gallino98, busso01, karakas14} and seen for other \emph{n}-capture elements in AGB and post-AGB stars \citep{smith90, abia02, vanwinckel03}.  In Figure~\ref{sekr_vs_c}, we plot the new Se and Kr abundances against log(C/O), where the carbon abundances were determined from UV forbidden lines.  As discussed in \citetalias{sterling08}, Hb~12 is an outlier in the $[$Kr/(O, Ar)$]$ plot in that it exhibits a high Kr abundance but a low C/O ratio.  We have excluded Hb~12 from this analysis, but note that its inclusion greatly weakens the correlation between Kr and C abundances.

The correlation coefficients $r$ are given in each panel of Figure~\ref{sekr_vs_c}, along with the probability of no correlation $p_{\rm r=0}$ (i.e., the probability that the observed correlation between the same number of objects could have been found from two uncorrelated populations).  The values of $r$ indicate marginal correlations (slightly less robust than found in \citetalias{sterling08}, particularly for Kr versus C/O), likely due to the relatively large observational uncertainties in both the Se and Kr abundances and in the C/O ratios. 

We applied a least-squares fitting routine to find the best linear fit to each correlation.  These are plotted as solid lines in Figure~\ref{sekr_vs_c}, and correspond to the equations:
\begin{equation}
\mathrm{[Se/(O,~Ar)]} = (-0.01\pm0.05) + (0.40\pm0.16)\mathrm{log(C/O)},
\end{equation}
and
\begin{equation}
\mathrm{[Kr/(O,~Ar)]} = (0.30\pm0.11) + (0.69\pm0.32)\mathrm{log(C/O)},
\end{equation}
where the uncertainties are derived from the dispersion in the fits.  The fits have lower intercept values than Equations~4 and 5 of \citetalias{sterling08}, but the slopes are quite similar.  Therefore, while the correlations between Se and Kr abundances and the C/O ratio of PNe are slightly less robust than found in \citetalias{sterling08}, the abundances follow nearly the same trend.  

We have therefore verified the correlations found in \citetalias{sterling08}, in particular that Type~I PNe show little to no \emph{s}-process enrichment compared to their counterparts with less massive progenitor stars, and that there is evidence for a positive correlation between Se and Kr enrichments and nebular C/O ratios.  In addition, we confirm that there is no evidence that Se and Kr enrichment factors differ in PNe with H-rich nuclei and those with $[$WC$]$ or WELS central stars.  The results do not depend on whether Ar or O is used as the reference element in non-Type~I PNe, nor on the adopted solar composition.  Thus, the new ICFs for Se and Kr have corrected a systematic overestimation of Se and Kr abundances by 0.1--0.3~dex for most PNe, but the newly-derived abundances do not alter the overarching conclusions of \citetalias{sterling08}.

\subsection{\emph{S}-Process Enhancements for Thick Versus Thin Disk Nebulae}\label{thinthick}

In order to assess whether a PN is self-enriched in Se or Kr, and by what amount, we must adopt initial abundances for these species in the progenitor star and compare them to the observed values.  Since the initial abundances cannot be determined directly, we estimate them from abundances of measurable species and the characteristics of the Galactic population to which the progenitor star belongs.  Thus far we have used O and Ar as proxies for the overall metallicities of PN progenitor stars, and taken the initial values of Se/(O,~Ar) and Kr/(O,~Ar) to be solar.  This assumption, implicit in our previous criterion for designating a PN as self-enriched, is not necessarily valid if the sample includes objects from multiple stellar populations.

In this section, we relax the assumption that the PN progenitor stars all have scaled solar compositions, and take into account the well-established abundance patterns in the subpopulations of the Milky Way disk.  Many large spectroscopic surveys have attempted to characterize and distinguish the thin and thick disks in terms of properties such as kinematics, spatial distribution, age, and compositional profiles \citep{freeman02, bensby03, bensby04, reddy06, ramirez07, alves-brito10, feltzing13, bensby14}.  Both disks have broad, distinctly non-Gaussian metallicity ($[$Fe/H$]$) distribution functions (MDFs) that overlap heavily, but all studies find that the thick disk's MDF is skewed toward lower $[$Fe/H$]$ values than the thin disk.  The centroids of the MDFs are not always well-defined and may vary with the characteristics of the sample \citep[e.g.,][]{casagrande11, kordopatis11}, but are approximately $<[$Fe/H$]>\approx -0.6\pm0.1$~dex for the thick disk and $<[$Fe/H$]>\approx -0.1\pm 0.1$~dex for the thin disk.

In view of the large spread in $[$Fe/H$]$ within each disk component, it would be inappropriate to assume a single characteristic value for PNe of each population.  However, a robust and much tighter relation emerging from these spectroscopic surveys is the separation of the two populations in their $[\alpha$/Fe$]$ ratios, with thick disk values $\sim$0.2--0.3~dex higher than the thin disk for moderately subsolar metallicities \citep[e.g.,][]{reddy09, anders14, bensby14}.  While there is controversy over whether this quantity is essentially a step function with a break between the two populations or two parallel linear correlations separated vertically in an $[\alpha$/Fe$]$ vs.\ $[$Fe/H$]$ diagram, we will use the former approximation below for simplicity.  (Note that this phenomenon introduces an ambiguity in the use of the $\alpha$ elements O and Ar as proxies for $[$Fe/H$]$ in PNe: if the sample includes both thin and thick disk objects, $[$Fe/H$]$ cannot be unambiguously inferred from $[\alpha$/Fe$]$ because $[\alpha$/Fe$]$ is double-valued.)

In \citetalias{sterling08}, we examined correlations between Se and Kr abundances and several characteristics of the PNe in our sample, but did not attempt to distinguish between different \citet{peimbert78} classes, except to pay special attention to objects of Type~I, which descend from relatively high-mass progenitors.  Peimbert also defined two additional classes of Galactic disk PNe, Types~II and III.  Type~III PNe have radial velocities $|v_{\rm rad}| \geq$~60~km~s$^{-1}$, and in terms of kinematics represent an older stellar population than Type~II PNe, which exhibit lower radial velocities.  In their spatio-kinematic characteristics, Type~II PNe are now recognized as members of the thin disk, while Type~III PNe belong to the thick disk \citep[e.g.,][]{phillips05}.\footnote{While the modern nomenclature of thick and thin disks had not yet been developed at the time, \citet{peimbert78} proposed that in view of the similar kinematics of Type~III PNe to those of old metal-deficient stars, the PNe probably were also Fe-poor, although no means was available to verify this suggestion.}  The O, Ne, S, and Ar abundances of Type~III PNe are only slightly lower than in their Type~II counterparts \citep{perinotto04, stanghellini10}.  However, since these elements are $\alpha$~species, this does not preclude the possibility that Type~III PNe are deficient in Fe-peak elements. Measurements of $[$Zn/H$]$ in PNe \citep[][Dinerstein et al.\ 2015, in preparation]{dinerstein_geballe01, smith14}, an effective proxy for $[$Fe/H$]$, demonstrate that many high-velocity (Type~III) PNe have subsolar metallicities.  The $\alpha$-enhancement at lower $[$Fe/H$]$ values thus partially compensates for low Fe abundances, yielding similar $\alpha$-element abundances for Type~II and Type~III PNe.

We separate the full sample of \citetalias{sterling08} into two subsets, Type~II and Type~III PNe,\footnote{Both groups contain several nominally N-rich Type~I PNe.  The inclusion of these objects reduces the average abundances discussed below by 0.03--0.05~dex, but does not affect our conclusions.} which at least statistically correspond to the thin and thick disks, respectively.  To avoid circular reasoning, we base our population assignments purely on kinematical grounds rather than abundances.  This criterion is not ideal, since kinematic mixing blurs the boundaries between the thin and thick disk populations even when all three kinematic components ($U$, $V$, $W$) are known \citep{anders14, bensby14}, but for PNe radial velocities are generally the only available kinematic information.  We use the radial velocities of the Strasbourg catalog \citep{acker94}, which provides data for 109 objects in our sample, and apply Peimbert's original criterion of  $|v_{\rm rad}| \geq$~60~km~s$^{-1}$ to distinguish Type~III (thick disk) from Type~II (thin disk) PNe.  Given large uncertainties due to the absence of proper motion data and poorly-known distances, we do not attempt to correct the heliocentric velocities for solar motion or Galactic rotation.  \citep[Type~III and Type~II objects appear to be kinematically distinguishable without this correction; see Fig.\ 2 of][]{phillips05}.  This yields 39 PNe in the high-velocity Type~III group, and 70 in the Type~II group, including objects for which only upper limits to the Se and/or Kr abundances are available.

We extrapolate each PN progenitor star's elemental Fe abundance using its measured $[$Ar/H$]$ abundance and the average $[\alpha$/Fe$]$ of the disk population to which the object is kinematically associated.  We use Ar as the $\alpha$ tracer for all sources, to avoid possible systematic effects; stellar abundance studies tend to find a larger enhancement for $[$O/Fe$]$ than for heavier species such as Si, Ca, and Ti \citep[e.g.,][]{bensby14}.  While this may be partly due to systematic uncertainties caused by using different spectroscopic indicators to determine O abundances \citep[e.g.,][]{bensby04, ramirez07}, the greater enrichment of lighter $\alpha$-elements is also likely a consequence of metallicity-dependent yields in massive stars \citep[e.g.,][]{mcwilliam08}.  Our choice of $[$Ar/H$]$ also avoids possible artificial effects introduced by using different $\alpha$ tracers for different PNe.  Although there are no direct data on the abundance of Ar in stars, we take typical average values for the heavier $\alpha$~species in the metallicity regime represented in our sample: $[\alpha$/Fe$]$~=~0.25~dex for the thick disk, and 0.0 for the thin disk \citep{anders14, bensby14}.

The next step is to extrapolate, from the inferred Fe abundances, the initial abundances of Se and Kr in the PN progenitor stars.  These two elements fall in a comparative desert of stellar abundance data in the periodic table, between the high-mass tail of the iron peak at Zn and the widely-observed light-\emph{s} peak of Sr, Y, and Zr.  A few direct measurements of Se abundances in stars have recently been obtained by \citet{roederer12} and \citet{roederer14}.  These authors found a roughly constant value of $[$Se/Fe$]=0.16\pm 0.09$ for seven stars spanning $-2.8\leq [$Fe/H$]\leq -0.6$.  We adopt this value for our Type~III (thick disk) subgroup, and assume a solar Se/Fe ratio for the Type~II (thin disk) objects.  By applying the $[\alpha$/Fe$]$ and $[$Se/Fe$]$ values for the two groups to the measured $[$Ar/H$]$ abundances, we infer the initial $[$Se/H$]$.  The net effect is $[$Se/H$]=[$Ar/H$]-0.09$~dex for the thick disk and $[$Se/H$]=[$Ar/H$]$ for the thin disk.

Since Kr cannot be observed in cool stars, we approximate its behavior relative to Fe by considering the elements bracketing it in the periodic table.  In addition to their results for Se, \citet{roederer14} found a slightly higher average abundance for As than for Se: $[$As/Fe$]=0.28\pm 0.14$, and no trend with $[$Fe/H$]$.  $[$Sr/Fe$]$ is slightly elevated (0.1--0.2~dex) in unevolved metal-poor stars \citep{ishigaki13}, while Y abundances are fairly constant and slightly subsolar in disk stars with $[$Fe/H$]>-1$ \citep{mishenina13}.  Given this information, we believe that it is reasonable to adopt the same initial values for $[$Kr/Fe$]$ (and hence $[$Kr/H$]$) as for $[$Se/Fe$]$.  

The enrichment factors for each nebula are computed from the difference between the measured values of $[$(Se, Kr)/H$]$ and the initial values predicted by the procedure described above.  Se was detected in 42 of the 70 low-velocity PNe and in 23 of the 39 high-velocity objects.  We find that the average self-enrichment of Se is 0.07$\pm$0.39~dex in thin disk PNe and 0.21$\pm$0.25~dex in thick disk PNe.  In contrast, the average enrichment of Kr is considerably higher in both populations: 0.62$\pm$0.41~dex in 23 thin disk PNe and 0.75$\pm$0.33~dex in seven thick disk objects.  Note that the numbers after the $\pm$ symbols result at least in part from intrinsic dispersion in the actual range of enrichments, and should not be interpreted simply as errors in the average values.

These results suggest that thick disk PNe may exhibit larger (by 0.1--0.2~dex) Se and Kr enrichment factors on average than their thin disk counterparts.  Our assumption that the thick disk members have lower initial Se and Kr abundances than thin disk PNe may contribute to this result, but does not fully account for it.  Moreover, there is strong evidence for systematic differences in abundances between the thick and thin disks from stellar surveys, and this is supported by the significantly subsolar Se abundances in some of our observed PNe (presumably from stars that experienced little or no \emph{s}-processing).  

The possible offsets in Se and Kr self-enrichments are in qualitative agreement with theoretical calculations of \emph{s}-process yields for AGB stars that predict a metallicity dependence for the production of the light-\emph{s} species \citep[e.g.,][]{busso01, cristallo09}.  We examined this effect in the FRUITY database\footnote{http:fruity.oa-teramo.inaf.it} \citep{cristallo11}.  For stars of initial mass 1.5--2.0~M$_{\odot}$ the Se \emph{s}-process enhancement increases by 0.04--0.08~dex when going from $[$Fe/H$]=0$ ($Z=0.014$) to $[$Fe/H$]=-0.67$, and by 0.24~dex at $[$Fe/H$]=-1.15$.  For Kr the corresponding values are 0.09--0.15~dex larger at $[$Fe/H$]=-0.67$ and 0.3~dex larger at $[$Fe/H$]=-1.15$ compared to solar metallicity models.  These models also predict higher Se and Kr yields for 2.5--3.0~M$_{\odot}$ stars as metallicity decreases.  However, higher-mass stars may have initially higher rotational velocities, which results in lower \emph{s}-process production rates \citep{piersanti13}.  At lower metallicities ($[$Fe/H$]<-1$), there is large scatter in the observed (initial) abundances of light-\emph{s} species in unevolved stars, but this is mainly due to the operation of mechanisms other than the \emph{s}-process \citep{sneden08}.  Our sample is drawn from more metal-rich populations and is therefore unlikely to include a significant number of such low-metallicity objects.

While we have found tentative evidence for a metallicity dependence in the enrichments of Se and Kr, these enhancements also depend on initial stellar mass.  Additional factors such as rotation and details of convection may also play a role.  Differences in average values between different populations will therefore depend on the distribution of stellar masses as well as metallicities, leading to a natural dispersion in the enrichment factors within each population.  This expectation is supported by the large standard deviations in the average Se and Kr enhancements in thin and thick disk PNe, which indicate a substantial spread and overlap in the populations.  It would be interesting to extend this kind of comparison to additional \emph{s}-process products, for example nuclei belonging to the heavy-\emph{s} peak.  It will also be useful to develop more robust methods for determining the parent population of individual PNe, such as obtaining additional kinematic information and/or Fe-group abundances using $[$Zn/H$]$ as a tracer \citep{dinerstein_geballe01, smith14}.

\section{CONCLUSIONS}

We have conducted a study of the ionization balance of the \emph{n}-capture elements Se and Kr in ionized nebulae, utilizing recently-determined atomic data for photoionization, recombination, and charge exchange processes for relevant ions of these elements.  The new atomic parameters were incorporated into the C13 release of Cloudy \citep{ferland13}, which we used to derive new Se and Kr ionization correction factor formulae.  We applied these ICF prescriptions to our near-infrared survey of $[$\ion{Kr}{3}$]$ and $[$\ion{Se}{4}$]$ lines \citepalias{sterling08} to derive revised Se and Kr abundances in a sample of 120 PNe.  Our main conclusions are:

\begin{enumerate}

\item We tested the ability of models that incorporate the new atomic data to reproduce the observed Kr ionization balance in 15 PNe that exhibit emission lines from multiple Kr ions.  We found a systematic disagreement between the modeled and observed Kr ionization balance for all 15 PNe, regardless of stellar and nebular properties.  The $[$\ion{Kr}{3}$]$ intensities were predicted by the models to be stronger than observed, while the $[$\ion{Kr}{4}$]$ lines were predicted to be too weak.  Given that no systematic discrepancies were seen in the ionization balance of other elements, the most probable explanation is that the Kr photoionization and recombination data are inaccurate.  To correct this disagreement, we adjusted the photoionization cross sections of Kr$^+$--Kr$^{3+}$ within their stated uncertainties, and altered the DR rate coefficients for the same ions by an amount somewhat larger than the approximate uncertainties cited by \citet{sterling11c}.  Because only one Se ion has been unambiguously detected in astrophysical nebulae, a similar test of the Se atomic data is not currently possible.  

This result underlines the importance of the interplay between theoretical and experimental atomic physics and observational spectroscopy.  All three approaches were needed to best reproduce the observed Kr ionization balance in PNe.  This provides added motivation for deep, high-resolution optical and infrared spectroscopy of PNe to detect multiple ionization states of \emph{n}-capture elements.  Such observations enable more accurate \emph{n}-capture element abundances to be derived and provide critical tests to newly-determined atomic data, both of which are needed to advance the efficacy of nebular spectroscopy for the study of trans-iron element nucleosynthesis.  Our study also highlights the need for more accurate (e.g., experimental) DR measurements for low-charge ions of heavy elements.  In a forthcoming paper, we will explore the sensitivity of Se and Kr abundance determinations to uncertainties in photoionization and recombination data.  This will help to elucidate which ionic systems are most in need of further investigation.


\item We computed multiple grids of Cloudy models, spanning a wide range of central star temperatures, luminosities, nebular densities, and chemical compositions in order to derive new ionization correction schemes for unobserved Se and Kr ions.  ICF formulae were computed from correlations between Se and Kr ionic fractions and those of commonly-detected ions of lighter elements.  We determined analytical ICFs for single Se and Kr ions and for combinations of those species, which can be applied to optical and near-infrared observations of ionized nebulae.  We tested the various ICF formulae with the optical and near-infrared spectra of the 15 individually modeled PNe.  The ICF-derived Se and Kr abundances agree within the uncertainties with the modeled values in nearly all cases.  The abundances determined with ICF formulae that rely on single Kr ions show larger scatter than those computed from multiple ionization states, as is expected.

\item The new ICFs were applied to the near-infrared survey of \citetalias{sterling08} to re-evaluate Kr and Se abundances in a sample of 120~PNe.  For most objects, the revised Se and Kr abundances are 0.1--0.3~dex lower than previous estimates.  We derive new average abundances of $[$Se/(O, Ar)$]=0.12\pm 0.27$ and $[$Kr/(O, Ar)$]=0.82\pm 0.29$ in PNe in which Se and Kr emission lines were detected, compared to the respective values of $0.31\pm0.27$ and $0.98\pm0.31$ found in \citetalias{sterling08}.  These average abundances show that Se is not enriched by \textit{in~situ} \emph{s}-process nucleosynthesis in many of the PNe in which it was detected.  Indeed, according to our criterion that requires $[$(Se)/(O, Ar)$]$ and $[$Kr/(O, Ar)$]\geq 0.3$~dex to be considered self-enriched, only 21\% of the PNe exhibiting Se emission are enhanced in this element.  Many PNe have subsolar Se abundances, which may reflect their initial compositions.  In contrast, Kr abundances are enhanced relative to solar in 79\% of the PNe in which $[$\ion{Kr}{3}$]$ was detected.  

\item While the overall Se and Kr abundances in the sample of \citetalias{sterling08} decreased from their previous values, we confirm the correlations found in that paper between Se and Kr abundances and other nebular and stellar properties are unaffected.  Specifically, Type~I PNe show no evidence of significant \emph{s}-process enrichments, and bipolar nebulae tend to have lower Se and Kr abundances than elliptical PNe.  Since Type~I and bipolar PNe tend to be associated with more massive progenitor stars ($>3$--4~M$_{\odot}$), this suggests that intermediate-mass AGB stars experience smaller \emph{s}-process enrichments than lower-mass stars, at least in the case of Se and Kr.  In addition, the positive correlation between Se and Kr abundances and the C/O ratio found in \citetalias{sterling08} is seen with the newly-derived abundances.  Finally, we do not find significant differences in the \emph{s}-process enrichment of PNe with different central star types ($[$WC$]$, WELS, and H-rich). These results do not depend on the choice of the metallicity reference element for non-Type~I PNe or on the adopted solar composition.

\item We took into account the initial compositions of the PNe by dividing the sample into two subgroups based on their radial velocities: low-velocity (Peimbert Type~II) and high-velocity ($|v_{\rm rad}| \geq$~60~km~s$^{-1}$; Peimbert Type~III).  We argue that these subgroups statistically correspond to the Galactic thin and thick disk stellar populations, respectively.  To estimate the initial Se and Kr abundances of PNe in each subgroup, we utilize the $[$Se/Fe$]$ abundances found in metal-poor and near-solar metallicity stars.  The initial Se and Kr abundances were inferred using the abundance of the $\alpha$~element Ar to indicate the initial composition of each nebula, taking into account the differing $[\alpha$/Fe$]$ abundances in thin and thick disk populations.  We find that $<[$Se/H$]>=0.07\pm$0.39~dex in the thin disk sample and 0.21$\pm$0.25~dex in thick disk objects.  A similar trend is seen for the average Kr enrichments in thin and thick disk PNe: $<[$Kr/H$]>=0.62\pm$0.41 and 0.75$\pm$0.33~dex, respectively.  Therefore, when accounting for the initial compositions of the PNe in the \citetalias{sterling08} sample, the average Se and Kr enrichment factors appear to be larger by 0.1--0.2~dex in metal-poor thick disk objects than in thin disk PNe.  This offset is in qualitative agreement with models of AGB nucleosynthesis \citep[e.g.,][]{cristallo11} for 1.5--2.0~M$_{\odot}$ stars with metallicities $-1.15 \leq [$Fe/H$] \leq 0$.  To further investigate the possible offset in \emph{s}-process enrichments in thin and thick disk PNe, another method is needed to validate the association of the high-velocity PNe with the thick disk, for example Zn abundance determinations (a proxy for Fe abundances).  It would also be of interest to study the behavior of other \emph{n}-capture elements in PNe from the different disk populations.

\end{enumerate}

\acknowledgements

We thank N.\ R.\ Badnell, whose constructive feedback strengthened this paper, and C.\ Sneden for fruitful discussions.  NCS acknowledges support of this work from an NSF Astronomy and Astrophysics Postdoctoral Fellowship under award AST-0901432 and from NASA grant 06-APRA206-0049.  HLD received support from NSF award AST-0708429.  This work has made use of NASA's Astrophysics Data System, the FRUITY Database of nucleosynthetic yields from AGB stars, and of the VizieR catalogue access tool, CDS, Strasbourg, France.  The original description of the VizieR service was published in A\&AS 143, 23.

\bibliographystyle{apj}

\bibliography{paper3.bib}

\end{document}